# Survey on Instruction Selection

## An Extensive and Modern Literature Review

Gabriel S. Hjort Blindell

October 4, 2013

School of Information and Communication Technology
KTH Royal Institute of Technology
Stockholm, Sweden

Research Unit: Software and Computer Systems
School of Information and Communication Technology
KTH Royal Institute of Technology
Forum 105
164 40, Kista
Sweden

# Abstract*

Instruction selection is one of three optimization problems – the other two are instruction scheduling and register allocation – involved in code generation. The task of the instruction selector is to transform an input program from its target-independent representation into a target-specific form by making best use of the available machine instructions. Hence instruction selection is a crucial component of generating code that is both correct and runs efficiently on a specific target machine.

Despite on-going research since the late 1960s, the last comprehensive survey on this field was written more than 30 years ago. As many new approaches and techniques have appeared since its publication, there is a need for an up-to-date review of the current body of literature; this report addresses that need by presenting an extensive survey and categorization of both dated method and the state-of-the-art of instruction selection. The report thereby supersedes and extends the previous surveys, and attempts to identify where future research could be directed.

*This research was funded by the Swedish Research Council (VR) (grant 621-2011-6229). I also wish to thank Roberto Castañeda Lozano, Mats Carlsson, and Karl Johansson for their helpful feedback, and SICS (The Swedish Institute of Computer Science) for letting me use their computer equipment to compile this report.

# Contents

# List of Figures

# List of Abbreviations

| | |
|---|---|
| AI | artificial intelligence |
| APT | abstract parse tree (same as AST) |
| ASIP | application-specific instruction-set processor |
| AST | abstract syntax tree |
| BURS | bottom-up rewriting system |
| CDFG | control and data flow graph |
| CISC | complex instruction set computing |
| CNF | conjunctive normal form |
| CP | constraint programming |
| DAG | directed acyclic graph |
| DP | dynamic programming |
| DSP | digital signal processor |
| GA | genetic algorithm |
| IP | integer programming |
| IR | intermediate representation |
| ISA | instruction set architecture |
| ISE | instruction set extension |
| JIT | just-in-time |
| LL parsing | left-to-right, left-most derivation parsing |
| LR parsing | left-to-right, right-most derivation parsing |
| LALR | look-ahead LR parsing |
| MIS | maximum independent set |
| MWIS | maximum weighted independent set |
| PBQP | partitioned Boolean quadratic problem |
| PO | peephole optimisation |
| QAP | quadratic assignment problem |
| RTL | register transfer list |
| SAT problem | satisfiability problem |
| SIMD | single-instruction, multiple-data |
| SLR parsing | simple LR parsing |
| SSA | static single assignment |
| VLIW | very long instruction word |
| VLSI | very-large-scale integration |

# 1

# Introduction

Every undertaking of implementing any piece of software – be it a GUI-based Windows application written in C, a high-performance FORTRAN program for mathematical computations, or a smartphone app written in Java – necessitates a compiler in one form or another. Its core purpose is to transform the source code of an input program – written in some relatively high-level programming language – into an equivalent representation of low-level program code which is specific to a particular target machine. This program code is converted into machine code – a relatively easy task compared to compilation – which can then be executed by the designated target machine. Consequently, compilers have been, are, and most likely always will be, essential enablers for making use of modern technology; even refrigerators nowadays are controlled by tiny computers. Thus, having been under active research since the first computers started to appear in the 1940s, compilation is one of the oldest and most studied areas of computer science.

To achieve this task the compiler needs to tackle a vast range of intermediate problems. A few of these problems include syntax analysis, transformation into an intermediate representation, target-independent optimization, followed by code generation. Code generation is the process where the target-specific program code is produced and consists in turn of three subproblems – instruction selection, instruction scheduling, and register allocation – of which the first is the focus of this report.

Previous surveys that discuss instruction selection to one degree or another have been conducted by Cattell [45], Ganapathi et al. [116], Leupers [169], and Boulytchev and Lomov [33]. However, the last *extensive* survey – that of Ganapathi et al. – was published more than 30 years ago. Moreover, the classic compiler textbooks only briefly discuss instruction selection and thus provide little insight; in the combined body of over 4,600 pages that constitute the compiler textbooks [8, 15, 59, 96, 173, 190, 262], less than 160 pages – of which there is tremendous overlap and basically only discuss tree covering – are devoted to instruction selection. Hence there is a need for a new and up-to-date study of this field. This report addresses that need by describing and reviewing the existing methods – both dated as well as the state-of-the-art – of instruction selection, and thereby supersedes and extends the previous surveys.

The report is structured as follows. The rest of this chapter introduces and describes the task of instruction selection, explains common problems of comparing existing methods, and briefly covers the first papers on code generation. Chapters 2 through 5 then each discusses a fundamental principle to instruction selection: Chapter 2 introduces macro expansion; Chapter 3 discusses tree covering; which is extended into DAG and graph covering in Chapter 4 and Chapter 5. Lastly, Chapter 6 ends the report with conclusions where I attempt to identify where future research could be directed.

The report also contains several appendices. Appendix A provide a table which summarizes all approaches covered in the report, and in Appendix B there is a diagram illustrating the publication timeline. For necessary background information Appendix C provide the formal definitions regarding graphs. Appendix D contains a taxonomy of terms and concepts which are used throughout the report to facilitate the discussion and avoid confusions. Lastly, a revision log for the document is available in Appendix E.

## 1.1 What is instruction selection?

As already stated in the beginning of this chapter, the core purpose of a compiler is to transform a program written in high-level source code into low-level machine code that is tailored for the specific hardware platform on which the program will be executed. The source code is expressed in some programming language (e.g. C, Java, Haskell, etc. ), and the hardware is typically a processor of a particular model. The program under compilation will henceforth be referred to as the *input program*, and the hardware against which it is compiled will be called the *target machine* (these definitions are also available in Appendix D).

### 1.1.1 Inside the compiler

To put instruction selection into context, let us briefly examine what a compiler does. Figure 1.1 illustrates the infrastructure of most compilers. First the *frontend* parses, validates, and translates the source code of the input program into an equivalent form known as the *intermediate representation*, or IR for short. The IR code is a format used internally by the compiler which also insulates the rest of the compiler from the characteristics of a particular programming language. Hence multiple programming languages can be supported by the same compiler by simple providing a frontend for each language.

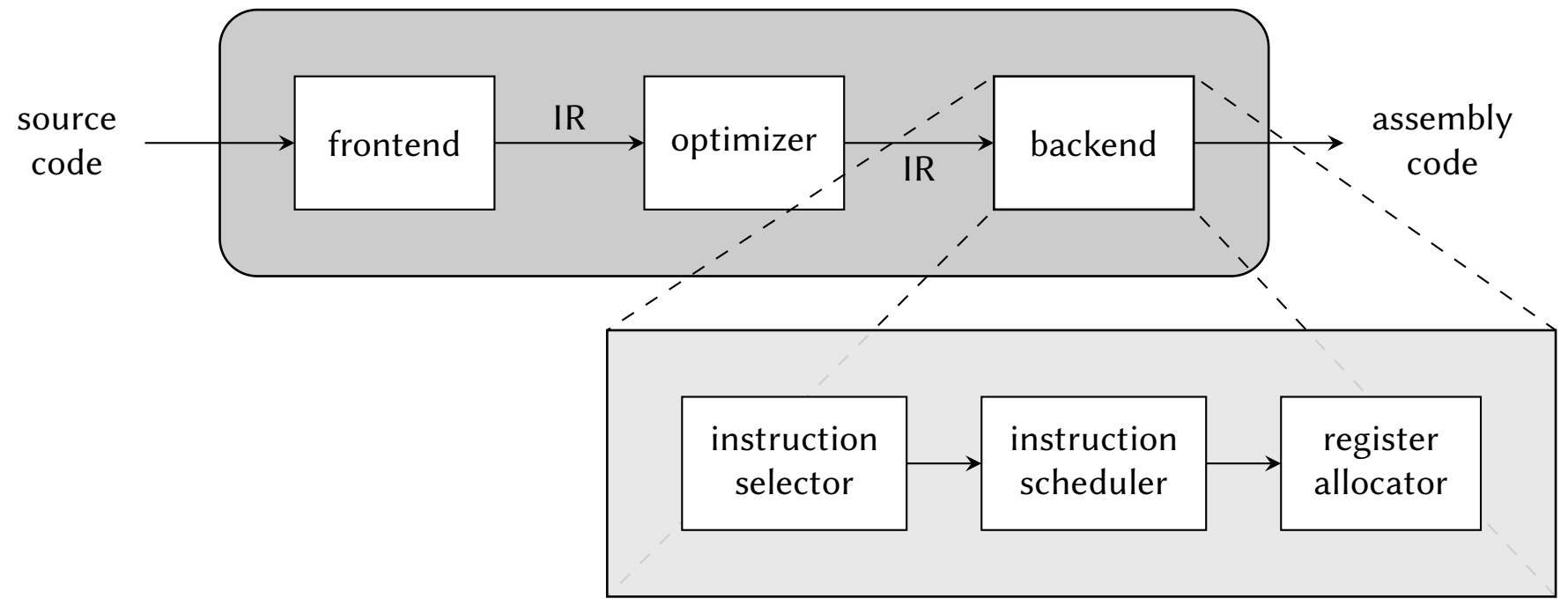

Figure 1.1: Overview of a common compiler infrastructure.

After undergoing a range of optimizations that can be performed regardless of the target machine – e.g. dead-code elimination and evaluating expressions with constant values (known as *constant folding*) – the IR code is converted by the *code generator backend* into a set of machine-specific instructions called *assembly code*; this is the component that allows the input program to be executed on a specific target machine. To do this we need to decide which machine instructions of the target machine to use for implementing the IR code; this is the responsibility of the *instruction selector*. For the selected machine instructions we also need to decide in which order they shall appear in the assembly code, which is taken care of by the *instruction scheduler*. Lastly, we must decide how to use the limited set of registers available at the target machine, which is managed by the *register allocator*. The last two tasks are out of scope for this report and will not be discussed in any greater detail.

The instruction selector must primarily select the machine instructions which implement the same expected behavior as that of the input program, and as a secondary objective it should result in efficient assembly code. We reformulate this as the following subproblems:

1. *pattern matching* – detecting when and where it is possible to use a certain machine instruction; and
2. *pattern selection* – deciding which instruction to choose in situations when multiple options exist.

The first subproblem is concerned with finding machine instruction candidates, whereas the second subproblem is concerned with selecting a subset from these candidates. Some approaches combine both subproblems into a single step, but most keep them separate and predominately differ in how they solve the pattern selection problem. In general the latter is formulated as an optimization problem where each machine instruction is assigned a cost and the goal is to minimize the total cost of the selected instructions. Depending on the optimization criteria, the cost typically reflects the number of cycles required by the target machine to execute a given machine instruction.

### 1.1.2 The interface of the target machine

The set of available machine instructions that can be used during instruction selection is dictated by the target machine's *instruction set architecture* (ISA). Most ISAs provide a rich and diverse instruction set whose complexity ranges from simple instructions that perform only a single operation – e.g. arithmetic addition – to highly complex instructions – e.g. copy the content of one memory location to another, increment the address pointers, and then repeat until a certain stop value is encountered. As the instruction set is seldom orthogonal there often exists more than one way of implementing a specific set of operations. For example, the expression $a + (b \times c)$ could be computed by first executing a `mul` instruction – which implements $d \leftarrow b \times c$ – followed by an `add` instruction implementing $e \leftarrow a + d$, or it could be computed by executing a single `mulacc` instruction that implements $a \leftarrow a + (b \times c)$.

Depending on the characteristics of the target machine, some sequences of instructions are more efficient than others in performing a specific task. This is especially true for digital signal processors (DSPs) which exhibit many customized machine instructions in order to improve the performance of certain algorithms. According to a 1994 study [254], the clock-cycle overhead of compiler-generated assembly code from C programs targeting DSPs could be as much as 1,000% compared to hand-written assembly code due to failing to exploit the full capabilities of the target machine. Consequently, the quality of the generated machine code is highly dependent on the compiler's ability to exploit hardware-specific features and resources, wherein the instruction selector is a crucial (but not the only) component in achieving this goal.

## 1.2 Problems of comparing different methods

As we saw in the previous section, the ISA can consist of many different kinds of machine instructions. Common to all contemporary instruction selection approaches, however, is that none is capable of handling every machine instruction available in the ISA. Consequently, most papers proposing a new instruction selection technique differ on the instruction set under discussion. For example, in the early literature a "complex machine instruction" refers to a memory load or store instruction that computes the memory address using schemes of varying complexity. These schemes are called *addressing modes* and using the appropriate addressing mode can reduce code size as well as increase performance. As an example, let us assume that we need to load the value at a particular position in an array of 1-byte values that reside in memory. The memory address of our wanted value is thus `@A` + `offset`, where `@A` refers to the memory address of the first element and is usually called the *base address*. A reasonable approach to fetch this value is to first execute an `add` instruction to compute `@A` + `offset` into some register $r_x$, and then execute a `load` instruction that reads the memory address from $r_x$. Such `load` instructions are said to have *absolute* or *direct* addressing modes. However, if the ISA provides a `load` with *indexed* addressing mode – where the memory address to read from is the sum a base value and offset – then we can get away with using only a single machine instruction. By contrast, in modern instruction selection approaches efficient handling of such addressing modes is for the most part trivial, and there a complex machine instruction typically denotes an instruction which produces more than one value or can only be used (or not used) in particular situations.

To mitigate these problems, and enable us to compare the different methods of instruction selection, we will introduce and define a set of *characteristics* that each refers a certain class of machine instructions.

### 1.2.1 Machine instruction characteristics

*Single-output instructions*

The simplest kind of machine instructions is the set of *single-output* instructions. These produce only a single observable output, in the sense that "observable" means a value that can be read and accessed by the program. This includes all machine instructions that implement a single operation (e.g. addition. multiplication, and bit operations like `OR` and `NOT`), but it also includes more complicated instructions that implement several operations like the memory operations just discussed with complicated addressing modes (e.g. load into register $r_d$ the value at memory location specified in base register $r_x$ plus offset specified in register $r_y$ plus an immediate value); as long as the observable output constitutes a *single* value, a single-output instruction can be arbitrarily complex.

*Multi-output instructions*

As expected, *multiple-output* instructions produce more than one observable output from the same input. Examples include `divrem` instructions that compute both the quotient as well as the remainder of two input values, as well as arithmetic instructions that, in addition to computing the result, also set *status flags*. A status flag (also known as *condition* flags or *condition codes*) is a bit that signifies additional information about the result, e.g. if there was a carry overflow or the result was `0`. For this reason such instructions are often said to have *side effects*, but in reality these bits are no more than additional output values produced by the instruction and will thus be referred to as multiple-output instructions. Load and store instructions that access a value in memory and then increment the address pointer are also considered multiple-output instructions.

*Disjoint-output instructions*

Machine instructions that produce many observable output values from many *different* input values are referred to as *disjoint-output* instructions. These are similar to multiple-output instructions with the exception that all output values in the latter originate from the same input values. Another way to put it is that if one formed the expression graphs that correspond to each output, all these graphs would be disjoint from one another; we will explain how such graphs are formed in the next chapter. Disjoint-output instructions typically includes SIMD (Single-Instruction, Multiple-Data) and vector instructions that execute the same operations simultaneously on many distinct input values.

*Internal-loop instructions*

*Internal-loop* instructions denote machine instructions whose behavior exhibits some form of internal looping. For example, the TI TMS320C55x [247] processor provides an `RPT k` instruction that repeats the immediately following instruction $k$ times, where $k$ is an immediate value given as part of the machine instruction. These machine instructions are among the hardest to exploit and are today handled either via customized optimization routines or through special support from the compiler.

*Interdependent instructions*

The last class is the set of *interdependent* machine instructions. This includes instructions that carry additional constraints that appear when the instructions are combined in certain ways. An example includes an `ADD` instruction, again from the TMS320C55x instruction set, which cannot be combined with an `RPT k` instruction if a particular addressing mode is used for the `ADD`. As we will see in this report, this is another class of machine instructions that most instruction selectors struggle with as they often fall outside the set of assumptions made by the underlying techniques. In addition, they appear only in certain situations.

### 1.2.2 Optimal instruction selection

When using the term *optimal* instruction selection, most literature assume – often implicitly – the following definition:

Definition. *An instruction selector, capable of modeling a set of machine instructions $I$ with costs $c_i$, is* optimal *if for any given input program $P$ it selects a set $S \subseteq I$ s.t. $S$ implements $P$, and there is exists no other set $S'$ that also implements $P$ and for which $\sum_{s' \in S'} c_{s'} < \sum_{s \in S} c_s$.*

In other words, if no assembly code with lower cost can be achieved using the same machine instructions that can be handled and exploited by the instruction selector, then the instruction selector is said to be optimal.

This definition has two shortcomings. First, the clause "machine instructions that can be modeled" restricts the discussion of optimality to whether the instruction selector can make best use of the instructions that it can handle; other, non-supported machine instructions that could potentially lead to more efficient assembly code are simply ignored. Although it enables comparison of instruction selectors with equal machine instruction support, the definition also allows two instruction selectors that handle different instruction sets to be both considered optimal even if one instruction selector clearly produces more efficient code than the other. Omitting the offending clause removes this problem but also renders all existing instruction selection techniques as suboptimal; excluding the simplest of ISAs, there always seems to exist some machine instruction that would improve the code quality for a particular input program but cannot be handled by the instruction selector.

Another, more pressing, problem is that two comparable instruction selectors may select different instructions, both solutions considered optimal from an instruction-selection point of view, that ultimately yield assembly code of disproportionate quality. For example, one set of instructions may enable the subsequent instruction scheduler and register allocator to do a better job compared to if other instructions had been selected. This highlights a well-known property of code generation: instruction selection, instruction scheduling, and register allocation are all interconnected with one another, forming a complex system that affects the quality of the final assembly code in complicated and often counter-intuitive ways. To produce truly optimal code, therefore, all three tasks must be solved in unison. and much research has gone into devising such systems (some of which are covered in this report).

Hence the idea of optimal instruction selection as an isolated concept is a futile sentiment, and one will be wise to pay close attention to the set of supported machine instructions if a method of instruction selection is claimed to be "optimal". Unfortunately, the notion has already been firmly established in the compiler community and permeates the literature. Due to this – and lack of better options – I will adhere to the same definition in this report but keep its use to a minimum.

## 1.3 The prehistory of instruction selection

The first papers on code generation started to appear in the early 1960s [13, 99, 193, 214] and were predominately concerned with how to compute arithmetic expressions on target machines based on accumulator registers. An *accumulator register* is a register which acts both as an input value and destination for an instruction (e.g. $a \leftarrow a + b$), and were prevalent in the early machines as the processor could be built using only a few registers. Although this simplified the hardware manufacturing process, it was not straight-forward to automatically generate assembly code that minimizes the number of transfers between the accumulator registers and main memory when evaluating the expressions.

In 1970 Sethi and Ullman [228] extended these ideas to target machines with $n$ general-purpose registers and presented an algorithm that evaluates arithmetic statements with common subexpressions and generates assembly code with as few instructions as possible. This work was later extended in 1976 by Aho and Johnson [3] who applied dynamic programming to develop a code generation algorithm that could handle target machines with more complex addressing modes such as indirection. We will revisit this method later in the report as many subsequent approaches to instruction selection are influenced by this technique.

Common among these early ideas is that the problem of instruction selection was effectively ignored or circumvented. For instance, both Sethi and Ullman's and Aho and Johnson's approaches assume that the target machines exhibit neat mathematical properties, devoid of any exotic machine instructions and multiple register classes. Since no, or very few, machines have such characteristics, these algorithms were not applicable directly in real-life settings.

Lacking formal approaches the first instruction selectors were typically made by hand and based on ad-hoc algorithms. This meant a trade-off between efficiency and retargetability: if the instruction selector was too general, the generated assembly code might not perform; if it was tailored too tightly to a particular target machine, it could constrain the compiler's support for others. Retargeting such instruction selectors therefore involved manual modifications and rewrites of the underlying algorithms. For irregular architectures, with multiple register classes and different instructions to access each class, the original instruction selector might not even be usable at all.

But even if the instruction selector was built to facilitate compiler retargeting, it was not immediately clear how to achieve this goal. In the next chapter we will examine the first methods that attacked this problem.

# 2

# Macro Expansion*

## 2.1 The principle

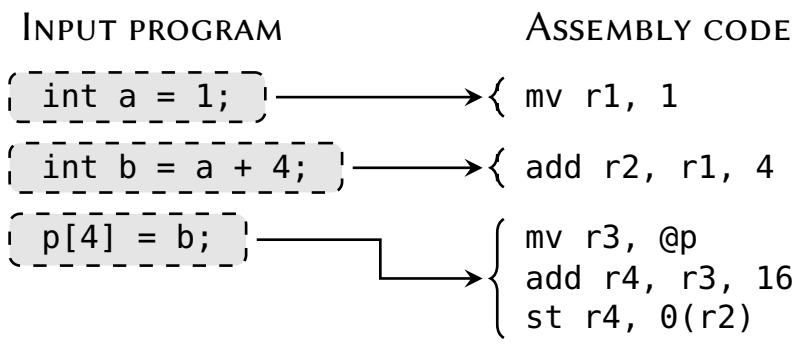

The first papers that introduced methods which dealt with instruction selection as a separate problem started to appear in the late 1960s. In these designs the instruction selection is driven by matching *templates* over the code that constitute the input program. Upon a hit the corresponding macro is executed using the matched program string as argument. Each language construct, which collectively constitutes the programming language in which the input program is expressed, typically has its own macro definition that emit appropriate assembly code for the target machine. Multiple language constructs can also be combined into a single template to make better use of the instruction set. For a given input program, the instruction selector will operate in a procedural fashion of traversing the input program, pairing the code against the templates, and executing the macros that match. If some portion of the text cannot be matched against any template then the instruction selector will fail and report an error, indicating that it is unable to produce valid assembly code for that particular input program against the specific target machine (which preferably should never happen).

The process of matching templates therefore corresponds to the pattern matching problem described in the previous chapter, and the process of selecting between multiple matching macros corresponds to the pattern selection problem. To the best of my knowledge, however, all macro-expanding instruction selectors immediately select whichever macro that matches first, thereby trivializing the latter problem.

By keeping the implementation of these macros separately from the implementation of the core – the part of the instruction selector which takes care of template matching and macro execution – the effort of retargeting the compiler against a different target machine is lessened, whereas the earlier, monolithic approaches often necessitated rewrites of the entire code generator backend.

*This chapter is primarily based on earlier surveys by Cattell [45] and Ganapathi et al. [116]. In the latter, this principle is called *interpretative code generation*.

## 2.2 Naïve macro expansion

### 2.2.1 Early approaches

We will refer to instruction selectors that directly apply the principle just described as *naïve macro expanders* – for reason that will soon become apparent. In the first such implementations the macros were either written by hand – as in the Pascal compiler developed by Ammann et al. [11, 12] – or generated automatically from a machine-specific description file. This file was typically written in some dedicated language, and many such languages and tools have appeared (and disappeared) over the years.

One example is Simcmp, a macro processor developed by Orgass and Waite [200] which was designed to facilitate bootstrapping.[1] Simcmp read its input line by line, compared the line against the templates of the available macros, and then executed the first macro that matched. An example of such a macro is given in Figure 2.1.

```
* = CAR.*.
    I = CDR('21)
    CDR('11) = CAR(I).
.X
```

(a) A macro definition.

```
A = CAR B.
```

(b) A string that matches the template.

```
I = CDR(38)
CDR(36) = CAR(I)
```

(c) Produced output.

Figure 2.1: Language transfer example using SIMCMP (from [200]).

Another example is GCL (Generate Coding Language), a language developed by Elson and Rake [78] which was used in a PL/1 compiler for generating machine code from *abstract syntax trees* (ASTs). An AST is a graph-based representation of the source code and is typically shaped like a tree (Appendix C provides an exact definition of graphs, trees, nodes, edges, and other related terms which we will encounter throughout this report). The most important feature of these trees is that only syntactically valid input programs can be transformed into an AST, which simplifies the task for the instruction selector. However, the basic principle of macro expansion remains the same.

*Using an intermediate representation instead of abstract syntax trees*

Performing instruction selection directly the source code – either in its textual form or on the AST – carries the disadvantage of tightly coupling the compiler against a particular programming language. Most compiler infrastructures therefore rely on some machine-independent *intermediate representation* (IR) which insulates the subsequent optimization routines and the backend from the details of the programming language.[2] Like the AST the IR code is often represented as a graph called the *expression graph* (or *expression tree*, depending on the shape), where each node represents a distinct operation and each directed edge represents a data dependency between two operations. Figure 2.2 illustrates an example of such an expression tree; in this report we will consistently draw trees with the root node at the top,

[1] *Bootstrapping* is the process of writing a compiler in the programming language it is intended to compile.

[2] This need was in fact recognized already in the late 1950s [57, 236].

```
t1 = a + b
c = t1 * 2
```

(a) IR code.

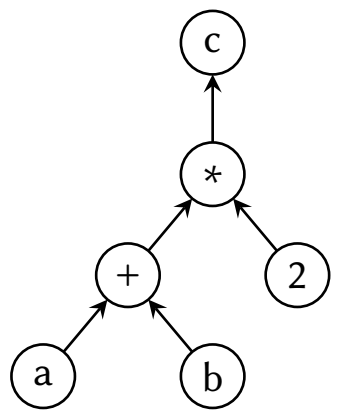

(b) Corresponding expression tree.

Figure 2.2: Example of a piece of IR code represented as an expression tree.

but note that the convention differs from one paper to another. It is also common to omit any intermediate variables and only keep those signifying the input and output values of the expression.

One of the first IR-based schemes was developed by Wilcox [261]. Implemented in a PL/C compiler, the AST is first transformed into a form of machine-independent assembly code consisting of SLM (Source Language Machine) instructions. The instruction selector then maps each SLM instruction into one or more target-specific machine instructions using macros defined in a language called ICL (Interpretive Coding Language) (see Figure 2.3).

In practice, these macros turned out to be tedious and difficult: many details, such as addressing modes and data locations, had to be dealt with manually from within the macros, and in the case of ICL the macro writer also had to keep track of which variables were part of the final assembly code and which variables were auxiliary and only used to aid the code generation process. In an attempt to simplify this task Young [269] proposed – but never implemented – a higher-level language called TEL (TEmplate Language) that would abstract away some of the implementation-oriented details; the idea was to first express the macros using TEL, and then automatically generate the lower-level ICL macros from the specification.

```
ADDB  BR   A,ADDB1      If A is in a register, jump to ADDB1
      BR   B,ADDB2      If B is in a register, jump to ADDB2
      LGPR A            Generate code to load A into register

ADDB1 BR  B,ADDB3       If B is in a register, jump to ADDB3
      GRX A,A,B         Generate A+B
      B   ADDB4         Merge

ADDB3 GRR  AR,A,B       Generate A+B
ADDB4 FREE B            Release resources assigned to B
ADDB5 POP  1            Remove B descriptor from stack
      EXIT

ADDB2 GRI  A,B,A        Generate A+B
      FREE A            Release resources assigned to A
      SET  A,B          A now designates result location
      B    ADDB5        Merge
```

Figure 2.3: Binary addition macro in ICL (from [261]).

### 2.2.2 Generating the macros automatically from a machine description

As with Wilcox's approach, many of the early macro-expanding instruction selectors depended on macros that were intricate and difficult to write, and the fact that the compiler developers often incorporated register allocation into these macros only exacerbated this problem. For example, if the target machine exhibits multiple register classes – with special machine instructions to move data from one register class to another – a record must be kept of which program values reside in which registers. Then, depending the register assignment the instruction selector will need to emit the appropriate data transfer instructions in addition to the rest of the assembly code that constitutes the input program. Due to the exponential number of possible situations, the complexity that the macro designer had to manage could be immense.

*Handling data transfers via state machines*

The first attempt to address this problem was made in 1971 by Miller [189] who developed a code generation system called Dmacs (Descriptive MACro System) that automatically infers the necessary data transfers between memory and different register classes. By providing this information in a separate file, Dmacs was also the first system to allow the details of the target machine to be declared separately instead of implicitly embedding it into the macros.

Dmacs relies on two proprietary languages: MIML (Machine-Independent Macro Language), which declares a set of procedural two-argument commands that serves as the IR format (see Figure 2.4 for an example); and a declarative language called OMML (Object Machine Macro Language) for implementing the macros that will transform each MIML command into assembly code. So far the scheme is similar to the one applied by Wilcox.

First, when adding support for a new target machine, the macro designer specifies the set of available register classes as well as the permissible transfer paths between these classes (including memory). The designer then defines the OMML macros by providing, for every macro, a list of machine instructions that each can implement the corresponding MIML command on the target machine. If necessary a sequence of machine instructions can be given to emulate the effect of the MIML command. For each machine instruction constraints are added that enforce the input and output data to reside in locations expected of that machine instruction. Figure 2.5 shows an abbreviated OMML specification for the IBM-360 machine.

Dmacs uses this information to generate a collection of *state machines* that each determines how a given set of input values can be transferred into locations that are permissible for a given OMML macro. A state machine consists of a directed graph where each node represents a

```
1:    SS      C,J
2:    IMUL    1,D
3:    IADD    2,B
4:    SS      A,I
5:    ASSG    4,3
```

Figure 2.4: An example on how an arithmetic expression $A[I] = B + C[J] * D$ can be represented as MIML commands (from [189]). The SS command is used for data referencing and the ASSG command assigns a value to a variable. The arguments to the MIML commands are referred to either by a variable symbol or by line number.

```
rclass REG:r2,r3,r4,r5,r6
rclass FREG:fr0,fr2,fr4,fr6
...
rpath WORD->REG: L REG,WORD
rpath REG->WORD: ST REG,WORD
rpath FREG-WORD: LE FREG,WORD
rpath WORD->FREG: STE FREG,WORD
...
ISUB s1,s2
from REG(s1),REG(s2) emit SR s1,s2  result REG(s1)
from REG(s1),WORD(s2) emit S s1,s2  result REG(s2)

FMUL m1,m2 (commutative)
from FREG(m1),FREG(m2) emit MER m1,m2  result FREG(m1)
from FREG(m1),WORD(m2) emit ME m1,m2   result FREG(m1)
```

Figure 2.5: Partial specification of IBM-360 in OMML (from [189]). `rclass` declared a register class and `rpath` declares a permissible transfer between a register class and memory (or vice verse), along with the machine instruction which implemented that transfer.

specific configuration of register classes and memory, some of which are marked as permissible. The edges indicate how to transition from one state to another, and are labeled with the machine instruction that will enable the transition when executed on a particular input value. During compilation the instruction selector consults the appropriate state machine as it traverses from one MIML command to the next by using the input values of the former to initialize the state machine, and then emitting the machine instructions appearing on the edges that are taken until the state machine reaches a permissible state.

The work by Miller was pioneering but limited: DMACS only handled arithmetic expressions consisting of integer and floating-point values, had limited addressing mode support, and could not model other target machine classes such as stack-based architectures. Donegan [72] later extended these ideas by proposing a new compiler generator language called CGPL (Code Generator Preprocessor Language), but the design was never implemented and still only remains on paper. Similar approaches have also been developed by Tirrell [242] and Simoneaux [232], and Ganapathi et al. [116] describe in their survey article another state machine-based compiler called UGEN, which was derived from U-CODE [204].

*Further improvements*

In 1975 Snyder [233] implemented on of the first fully operational and portable C compilers where the target machine-dependent parts could be automatically generated from a machine description. The approach is similar to that of Miller in that the frontend first transforms the input program into an equivalent representation for an abstract machine, consisting of AMOPs (Abstract Machine OPerations), which are then expanded into machine instructions via macros. The abstract machine and macros are specified in machine description language which is similar to that of Miller but handles more complex data types, addressing modes, alignment, as well as branching and function calls. If needed, more complicated macros can also be defined as customized C functions. We mention Snyder's work primarily because it was later adapted by Johnson [143] in his implementation of PCC, which we will discuss in the next chapter.

Fraser [106, 107] also recognized the need for human knowledge to guide the code generation process and implemented a system with the aim to facilitate the addition of hand-written rules when these are required. First the input program is transformed into a representation based on XL which is a form of high-level assembly code; for example, XL provides primitives for array accesses and *for* loops. Like Miller and Snyder, the machine instructions are provided via a separate description that each maps directly to a distinct XL primitive. If some portion of the input program cannot be implemented by any available instruction, the instruction selector will invoke a set of rules which rewrite the XL code until a solution is found. For example, array accesses are broken down into simpler primitives. In addition, the same rule base can be used to improve the code quality of the generated assembly code. Since these rules are provided as a separate file, they can be customized and augmented as needed to fit a particular target machine.

As we will see, this idea of "massaging" the input program until a solution can be found has been applied, in one form or another, by many instruction selection approaches that both predate and succeed Fraser's approach. In most, however, the common drawback is that if this set of rules is incomplete for a particular target machine – which is far from trivial to determine whether such is the case – the instruction selector may get stuck in an infinite loop. Moreover, such rules are often hard to reuse for other target machines.

### 2.2.3 More efficient compilation with tables

Despite its already simplistic nature, macro-expanding instructions selectors can be made even more so by representing the 1-to-1 or 1-to-$n$ mappings as a set of tables. This further emphasizes the separation between the machine-independent core of the instruction selector from the machine-dependent mappings, as well as allowing for denser implementations that require less memory and potentially reduces compilation time.

*Representing machine instructions as coding skeletons*

In 1969 Lowry and Medlock [180] presented one of the first table-driven methods for code generation. In their implementation of the Fortran H Compiler, Lowry and Medlock used *coding skeletons* for each machine instruction which consists of a bit string. The bits represent the restrictions of the machine instructions such as the permitted modes – e.g. load from memory, load from register, do not store, use this or that base register, etc. – for the operands and result. These coding skeletons are then matched against the bit strings corresponding to the input program, and an 'X' appearing in the coding skeleton means that it can match any bit.

```
L    B2,D(0,BD)    XXXXXXXX00000000
LH   B2,D(0,B2)    0000111100000000
LR   R1,R2         0000110100001101
```

However, the tables in Lowry and Medlock's compiler could only be used for the most basic of machine instructions and had to be written by hand. More extensive approaches were later developed by Tirrell [242] and Donegan [72], but these also suffered from similar disadvantages of making too many assumptions about the target machine which hindered compiler retargetability.

*Top-down macro expansion*

Later Krumme and Ackley [164] proposed a table-driven design which, unlike the earlier approaches, exhaustively enumerates all valid combinations of selectable machine instructions, schedules, and register allocations for a given expression tree. Implemented in a C compiler targeting DEC-10 machines, the method also allows code size to be factored in as an optimization goal, which was an uncommon feature at the time. Krumme and Ackley's backend applies a recursive algorithm that begins with selecting machine instructions for the root node in the expression tree, and then working its way down. In comparison, the bottom-up techniques we have examined so far all start at the leaves and then traverse upwards. We settle with this distinction for now as we will resume and deepen the discussion of bottom-up vs. top-down instruction selection in the next chapter.

Needless to say, enumerating all valid combinations in code generation leads to a combinatorial explosion and it is thus impossible to actually produce and check each and every one of them. To curb this immense search space Krumme and Ackley applied a search strategy known as *branch-and-bound*. The idea behind branch-and-bound is simple: during search, always remember the best solution found so far and then prune away all parts of the search tree which can be proven to yield a worse solution.[3] The remaining problem, however, is how to prove that a given branch in the search tree will definitely lead to solutions that are worse than what we already got (and can thus be skipped). Krumme and Ackley only partially tackled this problem by pruning away branches that for sure will eventually lead to failure (i.e. yield no solution whatsoever). Without going into any details, this is done by using not just one machine instruction table but several – one for each *mode*. In this context modes are oriented around the result of an expression, for example whether it is to be stored in a register or in memory. By having a table for each mode – which are constructed in a hierarchical manner – the instruction selector can look ahead and detect whether the current set of already-selected machine instructions will lead to a dead end.

However, with this as the only method of branch pruning the instruction selector will make many needless revisits in the search space which does not scale to larger expression trees.

### 2.2.4 Falling out of fashion

The said improvements notwithstanding, the main disadvantage of macro-expanding instruction selectors is that they can only handle macros that expand a single AST or IR node at a time and are thus limited to supporting only single-output machine instructions.[4] As this has a detrimental effect on code quality for target machines exhibiting more complicated features, such as multi-output instructions, instruction selectors based solely on naïve macro expansion were quickly replaced by newer, more powerful techniques – one of which we will discuss later in this chapter – when these started to appear in the late 1970s.

[3] In their paper Krumme and Ackley actually called this *$\alpha$-$\beta$ pruning* – which is an entirely different search strategy – but their description of it fits more the branch-and-bound approach (both are well explained in [219]).

[4] This is a truth with modification: multi-output instructions *can* be emitted by these instruction selectors, but only one of its output value will be retained in the assembly code.

*Rekindled application in the first dynamic code generation systems*

Having fallen out of fashion, naïve macro-expanding instruction selectors later made a brief reappearance in the first dynamic code generation systems that were developed in the 1980s and 1990s. In such systems the input program is first compiled into *byte code* which is a kind of target-independent machine code that can be interpreted by an underlying runtime environment. By providing an identical environment on every target machine, the same byte code can be executed on multiple systems without having to be recompiled.

Running a program in interpretive mode, however, is typically much slower than executing native machine code. By incorporating a compiler compiler into the runtime environment, the performance loss can be mitigated. Parallel to being executed, the byte code is profiled and frequently executed segments, such as inner loops, are then compiled – at runtime – into native machine code. This notion is called *just-in-time compilation* – or *JITing* for short – which allows performance to be increased while retaining the portability of the byte code. If the performance gap between running byte code instead of native machine code is great, then the compiler can afford to produce assembly code of low quality in order to decrease the overhead in the runtime environment. This was of great importance for the earliest dynamic runtime systems where hardware resources were typically scarce, which made macro-expanding instruction selection a reasonable option. A few examples include Smalltalk-80 [68], Omniware [1] (a predecessor to Java), Vcode [82], and Gburg [103] which was used within a small virtual machine.

As technology progressed, however, dynamic code generation systems also began to transitioned to more powerful techniques of instruction selection such as tree covering, which we will describe in the next chapter.

## 2.3 Improving code quality with peephole optimization

An early – and still applied – method of improving the code quality of generated assembly code is to perform a subsequent optimization step that attempts to combine and replace several machine instructions with shorter, more efficient alternatives. These routines are known as *peephole optimizers* for reason which will soon become apparent.

### 2.3.1 Peephole optimization

In 1965 McKeeman [188] advocated the use of a simple but often neglected optimization procedure which, as a post-step to code generation, inspects a small sequence of instructions in the assembly code and attempts to combine two or more adjacent instructions with a single instruction. Similar ideas were also suggested by Lowry and Medlock [180] around the same time. By doing this the code size shrinks while also improving performance as using complex machine instructions is often more efficient than implementing the same functionality using several simpler instructions. Because of its narrow window of observation, this technique became known as *peephole optimization*.

*Modeling machine instructions with register transfer lists*

Since this kind of optimization is tailored for a particular target machine, the earliest implementations were (and still often are) done ad-hoc and by hand. In 1979, however, Fraser [102] presented a first approach that allowed such programs to be generated automatically. It is also described in a longer article by Davidson and Fraser [65].

Like Miller, Fraser developed a technique that allows the semantics of the machine instructions to be described separately in a symbolic machine specification instead of implicitly embedding this knowledge into hand-written routines. This specification describes the observable effects that each machine instruction has on the target machine's registers, which Fraser called *register transfers*. Each instruction therefore has a corresponding *register transfer pattern* or *register transfer list* (RTL) – we will henceforth use the latter term. For example, assume that we have a three-address machine instruction `add` that executes an arithmetic addition using as input register $\mathtt{r_s}$ and an immediate value `imm`, stores the result in register $\mathtt{r_d}$, and sets a zero flag `Z`. Then the corresponding RTL would be expressed as

$$RTL(\mathtt{add}) = \begin{cases} \mathtt{r_d} \leftarrow \mathtt{r_s} + \mathtt{imm} \\ \mathtt{Z} \leftarrow (\mathtt{r_s} + \mathtt{imm}) \Leftrightarrow 0 \end{cases}$$

This information is then feed to a program called PO (Peephole Optimizer) that produces an optimization routine which makes two passes over the generated assembly code. The first pass runs backwards across the assembly code to determine the observable effects (i.e. the RTL) of each machine instruction in the assembly code. It runs backwards to be able to remove effects that have no impact on the program's observable behavior: for example, the value of a condition flag is unobservable if no subsequent instruction reads it and can thus be ignored. The second pass then checks whether the combined RTLs of two adjacent machine instructions are equal to that of some other machine instruction; in PO this check is done via a series of string comparisons. If such an instruction is found the pair is replaced and the routine backs up one instruction in order to check combinations of the new instruction with the following instruction in the assembly code. This way replacements can be cascaded and thus many machine instructions can be reduced into a single equivalent (provided there exists an appropriate machine instruction for each intermediate step).

Pioneering as it was PO also had several limitations. The main drawbacks were that it only handled combinations of two instructions, and that these had to be lexicographically adjacent in the assembly code and could not cross label boundaries (i.e. belong to separate basic blocks; a description of these is given in Appendix D on page 124). Davidson and Fraser [63] later removed the limitation of lexicographical adjacency by building *data flow graphs* – which are more or less the same as expression graphs – and operating on those instead of directly on the assembly code, and also extended the size of the instruction window from pairs to triples.

*Further developments*

Much research has been dedicated into improving automated approaches to peephole optimization. In 1983 Giegerich [119] presented a formal approach that eliminates the need for a fixed-size instruction window. Shortly after Kessler [155] proposed a method where RTL combination and comparison can be precomputed as the compiler is built, thus decreasing compilation time. Kessler [154] later extended his approach to incorporate $n$-size instruction window, similar to that of Giegerich, although at an exponential cost.

Another approach was developed by Massalin [187] who implemented a system called the SUPEROPTIMIZER which accepts small input programs, written in assembly code, and then exhaustively combines sequences of machine instructions to find shorter implementations that exhibit the same behavior as the input program. Granlund and Kenner [128] later adapted Massalin's ideas into a method that minimizes the number of branches used by an input program. Neither approach, however, makes any guarantees on correctness and therefore demands manual inspection of the final result. More recently Joshi et al. [145, 146] have proposed a method based on automatic theorem proving that uses SAT (Boolean satisfiability) solving to find performance-wise optimal implementations of a given input program. However, this technique exhibits exponential worst-time execution and is therefore best suited for short, performance-crucial code segments such as inner loops.

### 2.3.2 Combining naïve macro expansion with peephole optimization

In 1984 Davidson and Fraser [63] presented a technique for efficient instruction selection that combines the simplicity of naïve macro expansion with the power of peephole optimization. Similar yet unsuccessful strategies had already been proposed earlier by Auslander and Hopkins [24] and Harrison [133], but Davidson and Fraser struck the right balance between compiler retargetability and code quality which made their approach a reasonable option for production-quality compilers. This scheme has hence become known as the *Davidson-Fraser approach* and variants of it have been used in several compilers, e.g. YC [64] (Davidson and Fraser's own implementation, which also incorporates common subexpression elimination on RTL level), the ZEPHYR/VPO system [14], the Amsterdam Compiler Kit (ACK) [239], and – most famously – the GNU Compiler Collection (GCC) [157, 235].

#### *The Davidson-Fraser approach*

In the Davidson-Fraser approach the instruction selector consists of two parts: an *expander* and a *combiner* (see Figure 2.6). The task of the expander is to transform the IR code of the input program into register transfers. The transformation is done by executing simple macros that expand every IR node (assuming it is represented as such) into a corresponding RTL that describes the effects of that node. Unlike the previous macro expanders we have studied, these macros do not incorporate register allocation; instead the expander assigns each result to a virtual storage location, called a *temporary*, for which there exists an infinite amount. A subsequent register allocator then assigns each temporary to a register, potentially inserting additional code that saves some values to memory for later retrieval when the number of available register is not enough (this is called *register spilling*). But before this is done the combiner tries to merge multiple RTLs into larger RTLs in an attempt to improve code quality, using the same technique behind PO. After running the combiner and register allocator, a final procedure emits the assembly code by simply replacing each RTL with a corresponding machine instruction. For this to work, however, both the expander and the combiner must at every step of the way adhere to the *machine invariant* which dictates that every present RTL must be implementable by at least one machine instruction provided by the target machine.

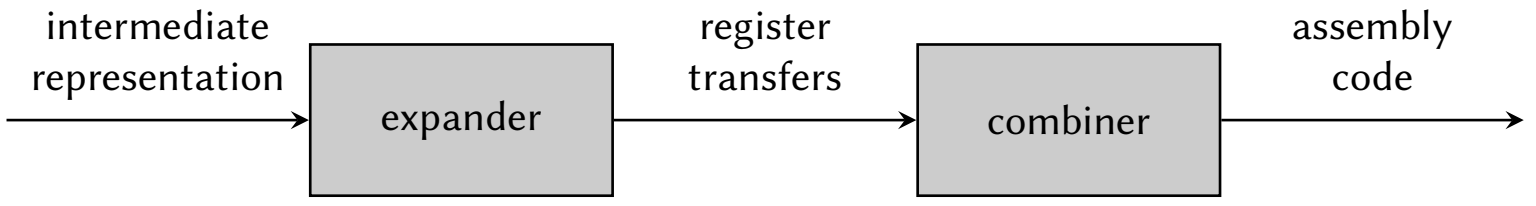

Figure 2.6: Overview of the Davidson-Fraser approach (from [63]).

By using a subsequent peephole optimizer to combine the effects of multiple RTLs, Davidson and Fraser's instruction selector can effectively extend over multiple nodes in the AST or IR tree – potentially across basic block boundaries – and its machine instruction support is in theory only restricted by the number of instructions that the peephole optimizer can compare at a time. For example, opportunities to replace three machine instructions by a single instruction will be missed if the peephole optimizer only checks combinations of pairs. Increasing the window size, however, typically incurs an exponential cost in terms of added complexity, thereby making it difficult to handle complicated machine instructions that require large instruction windows.

An improvement to Davidson and Fraser's work was later made by Fraser and Wendt [101]. In a 1988 paper Fraser and Wendt describe a method where the expander and combiner are effectively fused together into a single entity. The basic idea is to generate the instruction selector in two steps: the first step produces a naïve macro expander that is capable of expanding a single IR node at a time. Unlike Davidson and Fraser – who implemented the expander by hand – Fraser and Wendt applied a clever scheme consisting of a series of *switch* and *goto* statements, allowing theirs to be generated automatically from a machine description. Once produced the macro expander is executed on a carefully designed training set. Using function calls embedded into the instruction selector, a retargetable peephole optimizer is executed in tandem which discovers and gathers statistics on target-specific optimizations that can be done on the generated assembly code. Based on these results the code of the macro expander is then augmented to incorporate beneficial optimization decisions, thus allowing the instruction selector to effectively expand multiple IR nodes at a time which also removes the need for a separate peephole optimizer in the final compiler. Since the instruction selector only implements the optimization decisions that are deemed to be "useful", Fraser and Wendt argued, code quality is improved with minimal overhead. Wendt [258] later improved the technique by providing a more powerful machine description format – also based on RTLs – which subsequently evolved into a compact stand-alone language used for implementing code generators (see Fraser [100]).

*Enforcing the machine invariant with a recognizer*

The Davidson-Fraser approach was recently extended by Dias and Ramsey [70]. Instead of requiring each separate RTL-oriented optimization routine to abide by the machine invariant, Dias and Ramsey's design employs a *recognizer* to determine whether an optimization decision violates the restriction (see Figure 2.7). This simplifies the optimization routines as they no

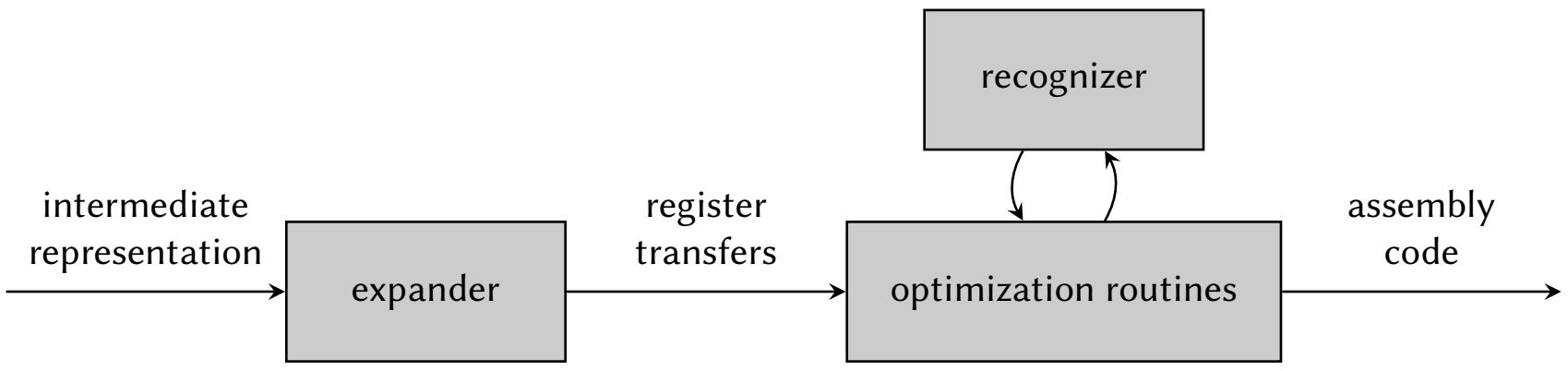

Figure 2.7: Overview of Dias and Ramsey's extension of the Davidson-Fraser approach (from [70]).

```
default attribute
  add(rd, rs1, rs2) is $r[rd] := $rs[rs1] + $r[rs2]
```

Figure 2.8: A PowerPC add instruction specified using λ-RTL (from [69]).

longer need to internalize the invariant, potentially allowing some of them to be generated automatically.

In a 2006 paper Dias and Ramsey demonstrate how the recognizer can be produced from a declarative machine description written in λ-RTL. Originally developed by Ramsey and Davidson [212], λ-RTL is a high-level functional language based on ML that raises the level of abstraction for writing RTLs (see Figure 2.8 for an example). According to Dias and Ramsey, λ-RTL-based machine descriptions are more concise and simpler to write compared to those of many other designs, including GCC. In particular, λ-RTL is precise and unambiguous which makes it suitable for automated tool generation and verification, the latter of which has been explored by Fernández and Ramsey [95] and Bailey and Davidson [25].

First the λ-RTL description is read and its RTLs transformed into *tree patterns*, which are the same as expression trees but model machine instructions instead of the input program. Using these tree patterns the task of checking whether an RTL fulfills the machine invariant is then reduced to a *tree covering problem*. Since the entire next chapter is devoted to discussing this problem, we let it be sufficient for now to say that each RTL is transformed into an expression tree, and then the tree patterns are matched on top of it. If the expression tree cannot be completely covered by any subset of patterns, then the RTL does not fulfill the machine invariant. Although this technique limits the recognizer to only reason about RTLs on a syntactical level, Dias and Ramsey argued that the need for efficiency rules out a more expensive semantic analysis (remember that the recognizer needs to be consulted for every taken optimization decision).

*"One program to expand them all"*

Shortly afterwards Dias and Ramsey [69, 213] proposed a scheme where the macro expander only needs to be implemented once per every distinct *architecture family* – e.g. register-based machines, stack-based machines, etc. – instead of once per every distinct *instruction set* – e.g. X86, PowerPC, SPARC, etc. . In other words, if two target machines belong to the same architecture family then the same expander can be reused despite the differing details of the ISAs. This is useful because its correctness only needs to be proven once, which is a difficult and time-consuming process if the expander has to be written by hand.

The idea is to have a predefined set of *tiles* that are specific for a particular architecture family. A tile represents a simple operation which is required for any target machine belonging to that architecture family. For example, stack-based machines require tiles for *push* and *pop* operations, which are not necessary on register-based machines. Then, instead of expanding each IR node in the input program into a sequence of RTLs, the expander expands it into a sequence of tiles. Since the set of tiles is identical for all target machines within the same architecture family, the expander only needs to be implemented once. After the tiles have been emitted they are replaced by the machine instructions implementing each tile, and the resulting assembly code can then be improved by the combiner.

A remaining problem is how to find machine instructions that implement the tiles for a particular target machine. In the same papers Dias and Ramsey provide a method for doing this automatically. By describing both the tiles and the machine instructions as $\lambda$-RTL, Dias and Ramsey developed a technique where the RTLs of the machine instructions are combined such that the effects equal that of a tile. In broad outline the algorithm maintains a pool of RTLs which is initialized with those found in the machine description, and then iteratively grows it using sequencing (i.e. combining several RTLs into a new RTL) and algebraic laws. This process stops when either all tiles have been implemented, or some termination criteria is reached; the latter is necessary as Dias and Ramsey proved that the general problem of finding implementations for arbitrary tiles is undecidable (i.e. there can be no algorithm which guarantees to terminate and also find an implementation for all possible input).

Although its primary aim is to facilitate compiler retargetability, Dias and Ramsey demonstrated that a prototype of their design yielded better code quality compared to the default Davidson-Fraser instruction selector of GCC. However, this required that the target-independent optimizations were disabled; when they were reactivated GCC produced better results.

### 2.3.3 Running peephole optimization before instruction selection

In the approaches just discussed, optimization routines for improving the quality of the generated assembly code are executed after code generation. In an approach by Genin et al. [118], however, a variant of this runs *before* code generation. Targeting digital signaling processors, their compilers first transforms the input program into an *internal signal flow graph* (ISFG), and then executes a pattern matcher prior to instruction selection which attempts to merge several low-level operations in the ISFG into single nodes, conveying higher-level semantic content (e.g. a multiplication followed by an addition within a loop can be combined into a product-accumulation construct).[5] Code generation is then done node by node: for each node the instruction selector invokes a rule along with the information about the current context. The invoked rule produces the assembly code appropriate for the given context, and can also insert new nodes to offload decisions that are deemed to be better handled by the rules corresponding to the inserted nodes.

According to the authors, experiments showed that their compiler generated code that was 5 to 50 times faster than that of conventional compilers at the time, and comparable with hand-written assembly code. However, the design is limited to input programs where prior knowledge about the application – in this case digital signal processing – can be encoded into specific optimization routines, which most likely has to be done manually.

[5]The paper is not clear on how this is done, but presumably the authors implemented a hand-written peephole optimizer as the intermediate format is fixed and does not change from one target machine to another.

## 2.4 Summary

In this chapter we have discussed the first techniques to address instruction selection. These were based, in one way or another, on a principle known as macro expansion, where AST or IR nodes are expanded into one or more target-specific machine instructions. Using template matching and macro invocation, these types of instruction selectors are resource-effective and straight-forward to implement.

However, naïve macro expansion assumes a 1-to-1 or 1-to-$n$ mapping between nodes in the IR (or abstract syntax) tree and the machine instructions provided by the target machine. If there exist machine instructions that can be characterized as $n$-to-1 mappings, such as multi-output instructions, then code quality yielded by these techniques will typically be low. As machine instructions are often emitted in isolation from one another, it also becomes difficult to use machine instructions that can have unintended effects on other machine instructions (like the interdependent instructions described in the previous chapter).

A remedy is to add a peephole optimizer into the code generator backend. This allows multiple machine instructions in the assembly code to be combined and replaced by more efficient alternatives, thereby amending some of the poor decisions made by the instruction selector. When incorporated into the instruction selector – an approach known as the Davidson-Fraser approach – this also enables more complex machine instructions like multi-output instructions to be expressed as part of the machine description. Because of this versatility the Davidson-Fraser approach remains one of the most powerful approaches to instruction selection to date, and a variant is still applied in GCC.

In the next chapter we will explore another principle of instruction selection that more directly addresses the problem of implementing several AST or IR nodes using a single machine instruction.

# 3

# Tree Covering

As we saw in the previous chapter, the main limitation of most instruction selector based on macro expansion is that the scope of expansion is limited to a single AST or IR node in the expression tree. This effectively excludes exploitation of many machine instructions and results in low code quality. Another problem is that macro-expanding instruction selectors typically combine pattern matching and pattern selection into a single step, thus making it impossible to consider combinations of machine instructions and then pick the one that yields the most "efficient" assembly code.

These problems can be solved by employing another principle of instruction selection called *tree covering*. This is also the principle of most instruction selection techniques found in the literature, making this the longest chapter in the report.

## 3.1 The principle

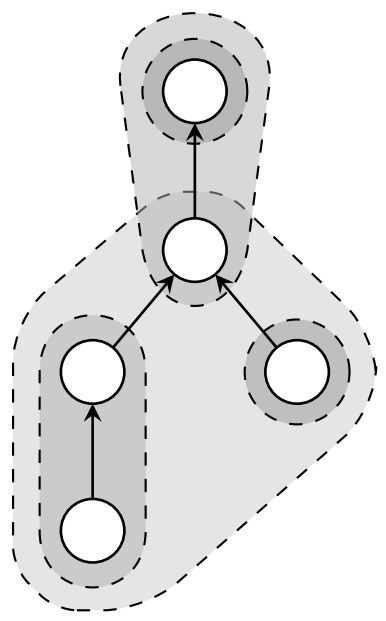

Let us assume that the input program is represented as a set of expression trees, which we are already familiar from the previous chapter (see page 10). Let us further assume that each machine instruction can be modeled similarly as another expression tree, and to distinguish between the expression trees formed from the input program and those formed from the machine instructions we will refer to the latter as *tree patterns*, or simply *patterns*. The task of instruction selection can then be reduced to a problem of finding a set of tree patterns such that every node in the expression tree is *covered* by at least one pattern. Here we see clearly how pattern matching and pattern selection constitute two separate problems in instruction selection: in the former we need to find which tree patterns are applicable for a given expression tree; and in the latter we then select from these patterns a subset that results in a valid cover of the expression tree. For most target machines there will be a tremendous amount of overlap – i.e. one tree pattern may match, either partially or fully, the nodes matched by another pattern in the expression tree – and we typically want to get away with using as few tree patterns as possible. This is for two reasons:

- Striving for the smallest number of patterns means opting for larger tree patterns over smaller ones. This in turn leads to the use of more complex machine instructions which typically results in improved code quality.
- The amount of overlap between the selected pattern is limited, meaning that the same operations in the input program will be re-implemented in multiple machine instruction only when necessary. Keeping redundant work to a minimum is another crucial factor for performance as well as for reducing code size.

As a generalization the optimal solution is typically not defined as the one that minimizes the *number* of selected patterns, but as the one that minimizes the *total cost* of the selected patterns. This allows the pattern costs to be chosen to fit the desired optimization criteria, but there is usually a strong correlation between the number of patterns and the total cost. However, as stressed in Chapter 1, what is considered an optimal solution for instruction selection need not necessarily be an optimal solution for the final assembly code. Moreover, we can only take into account the machine instructions for which there exists a suitable tree pattern.

Finding the optimal solution to a tree covering problem is not a trivial task, and it becomes even less so if only certain combinations of patterns are allowed. To be sure, most would be hard-pressed just to come up with an efficient method that finds all valid tree pattern matchings. We therefore begin with exploring the first approaches that address the pattern matching problem, but not necessarily the pattern selection problem, and then gradually transition to those that do.

## 3.2 First techniques to use tree pattern matching

In 1972 and 1973 Wasilew [255] and Weingart [256] published the first papers known to describe code generation techniques using tree pattern matching (due to lack of first-hand information we will only discuss the former of the two in this report). According to [45, 116] Weingart's design uses a single pattern tree – Weingart called this a *discrimination net* – which is automatically derived from a declarative machine description. Using a single tree, Weingart argued, allows for a compact and efficient means of representation. The process of building the AST is then extended to simultaneously push each new AST node onto a stack. As a new node appears the pattern tree is progressively traversed by comparing the nodes on the stack against the children of the current node in the pattern tree. A pattern match is found when the process reaches a leaf node in the pattern tree, whereupon the the machine instruction associated with the match is emitted.

Compared to the early macro-expanding instruction selectors (i.e. those prior to Davidson-Fraser), Weingart's approach extended the machine instruction support to include those which extend over multiple AST nodes. However, the technique suffered several problems when applied in practice. First, structuring the pattern tree to allow for efficient pattern matching proved difficult for certain target machines; it is known that Weingart struggled in particular with PDP-11. Second, it assumes that there exists at least one machine instruction on the target machine that corresponds to a particular node type of the AST, and for some programming language and target machine combinations this turned out not to be the case. Weingart partly addressed this problem by introducing additional patterns, called *conversion patterns*, that enable mismatched parts of the AST to be transformed into a form that hopefully will be matched by some pattern, but this had to be done manually and could cause the compiler to get stuck in an infinite loop. Third, like its macro-expanding predecessors the process immediately selects a pattern as soon as a match is found.

Another early pattern matching technique was developed by Johnson [143] which was implemented in PCC (Portable C Compiler), a renown system that was the first standard C compiler to be shipped with Unix. Johnson based PCC on the earlier work by Snyder [233] – which we discussed in Section 2.2.2 in the previous chapter – but replaced the naïve macro-expanding instruction selector with a technique based on tree rewriting. For each machine instruction an expression tree is formed together with a rewrite rule, subgoals, resource requirements, and an assembly string which is emitted verbatim. This information is given in a machine description format that, as seen in Figure 3.1, allows multiple, similar patterns to be condensed into a single declaration.

The pattern matching process is then relatively straight-forward: for a given node in the expression tree, the node is compared against the root node of each pattern. If these match a similar check is done for each corresponding subtree in the pattern. Once all leaf nodes in the pattern are reached, a match has been found. As this algorithm – whose pseudo-code is given in Figure 3.2 – exhibits quadratic execution time, it is desirable to minimize the number of redundant checks. This is done by maintaining a set of code generation goals which are encoded into the instruction selector as into an integer. For historical reasons this integer is called a *cookie*, and each pattern has a corresponding cookie indicating in which the situations the pattern may be useful. If both the cookies and the pattern match an attempt is made to allocate whatever resources are demanded by the pattern (e.g. a pattern may require a

```
ASG PLUS,   INAREG,
            SAREG,   TINT,
            SNAME,   TINT,
                     0,      RLEFT,
                     "       add      AL,AR\n",
...
ASG OPSIM,   INAREG|FORCC,
             SAREG,      TINT|TUNSIGNED|TPOINT,
             SAREG|SNAME|SOREG|SCON,   TINT|TUNSIGNED|TPOINT,
                         0,           RLEFT|RESCC
                         "            OI           AL,AR\n",
```

Figure 3.1: A sample of a machine description for PCC consisting of two patterns (from [144]). The first line specifies the node type of the root (`+=`, for the first pattern) together with the cookie (“result must appear in an `A`-type register”). The second and third line specify the left and right descendants, respectively, of the root node. For the first pattern the left subtree must be an `int` allocated in an `A`-type register, and the right subtree must be a `NAME` node, also of type `int`. The fourth line indicates that no registers or temporaries are required and that the match part in the expression tree is to be replaced by the left descendant of the pattern root node. The fifth and last line declares the assembly string, where lowercase letters are output verbatim and uppercase words indicate a macro invocation – `AL` stands for "Address form of Left operand"; likewise for `AR` – whose result is then put into the assembly string. In the second pattern we see that multiple restrictions can be *OR*’ed together, thus allowing multiple patterns to be expressed in a more concise manner.

FindMatchset(expression tree rooted at node $n$, set of tree patterns $P$):
initialize matchset as empty
**for each** $pinP$ **do**
    **if** Matches($n$, $p$) **then**
        add $p$ to matchset
    **end if**
**end for**
return matchset

Matches(expression tree rooted at node $n$, tree pattern rooted at node $p$):
**if** $n$ matches $p$ **and** number of children of $n$ and $p$ are equal **then**
    **for each** child $n'$ of $n$ **and** child $p'$ of $p$ **do**
        **if not** Matches($n'$, $p'$) **then**
            return **false**
        **end if**
    **end for**
**end if**
return **true**

Figure 3.2: Straight-forward pattern matching with $\mathcal{O}(nm)$ time complexity, where $n$ is the number of nodes in the expression tree and $m$ is the total number of nodes in the tree patterns.

certain number of registers). If successful the corresponding assembly string is emitted, and the matched subtree in the expression tree is replaced by a single node as specified by the rewrite rule. This process of matching and rewriting repeats until the expression tree consists of only a single node, meaning that the entire expression tree has been successfully converted into assembly code. If no pattern matches, however, the instruction selector enters a heuristic mode where the expression tree is partially rewritten until a match is found. For example, to match an `a = reg + b` pattern, an `a += b` expression could first be rewritten into `a = a + b` and then another rule could try to force operand `a` into a register.

Although successful for its time, PCC had several disadvantages. Like Weingart, Johnson's technique uses heuristic rewrite rules to handle mismatching situations. Lacking formal methods of verification there was hence always the risk that the current set of rules would be inadequate and potentially cause the compiler to never terminate for certain input programs. Reiser [215] also noted that the investigated version of PCC only supported unary and binary patterns with a maximum height of 1, thus excluding many machine instructions such as those with complex addressing modes. Lastly, PCC – and all other approaches discussed so far – still adhered to the *first-matched-first-served* principle when selecting patterns.

## 3.3 Covering trees bottom-up with LR parsing

As already noted, a common flaw among the first approaches is that they (i) apply the greediest form of pattern selection, and (ii) typically lack a formal methodology. In contrast, syntax analysis – which is the task of parsing the source code of the input program – is arguably the best understood area of compilation, and its methods also produce completely table-driven parsers that are very fast and resource-efficient.

### 3.3.1 The Glanville-Graham approach

In 1978 Glanville and Graham [121] presented a seminal paper that describes how techniques of syntax analysis can be adapted to instruction selection.[1] Subsequent experiments and evaluations showed that this design – which we will refer to as the *Glanville-Graham approach* – proved simpler and more general than contemporary approaches [9, 116, 126, 127]. Moreover, due to its table-driven nature, assembly code could be generated very rapidly (although the first implementations performed on par with other instruction selectors used at the time). Consequently the Glanville-Graham approach has been acknowledged as one of the most significant breakthroughs in this field, and these ideas have influenced many later techniques in one way or another.

*Expressing machine instructions as a grammar*

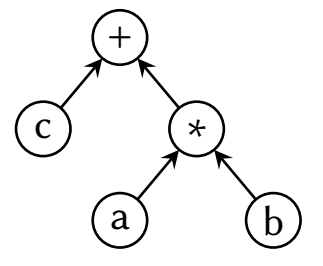

"+ c * a b"

To begin with, a well-known method of removing the need for parentheses in arithmetic expressions without making them ambiguous is to use *Polish prefix notation*. For example, 1 + (2 + 3) can be written as + 1 + 2 3 and still mean the same thing. Glanville and Graham recognized that by using this form the machine instructions can be expressed as a *context-free grammar* based on BNF (*Backus-Naur Form*). This concept is already well-described in most compiler textbooks – I recommend [8] – so we will only proceed with a brief introduction.

A context-free grammar consists of a set of *terminals* and *nonterminals*, which we refer to both as *symbols*. We will distinguish between the two by always writing terminals with an initial capital letter (e.g. Term), and nonterminals entirely in lower case (e.g. nt). For each nonterminal there exists one or more *productions* of the following form:

$$\text{lhs} \ \rightarrow \ \text{Right hand Side} \ldots$$

A production basically specifies how a particular combination of symbols (the right-hand side) can be replaced by another nonterminal (the left-hand side). Since nonterminals can appear on both sides in a production most grammars allow for infinite chains of replacements, which is one of the powerful features of context-free grammars. In terms of recursion, one can also think of nonterminals as inducing the recursive case whereas the terminals provide the base case that stops the recursion. Productions are also called *production rules* or just *rules*, and although they can typically be interchanged without causing confusion we will be consistent in this report and only use the first term as rules will hold a slightly different meaning.

To model a set of machine instructions as a context-free grammar, one would add one or more *rules* for each instruction. We call this collection of rules the *instruction set grammar*. Each rule contains a production, cost, and an action. The right-hand side of the production represents the tree pattern to be matched over an expression tree, and the left-hand side contains the nonterminal indicating the characteristics of the result of executing the machine instruction (e.g. a specific register class). The cost should be self-explanatory at this point, and the action would typically be to emit a string of assembly code. We illustrate the anatomy of a rule more succinctly with an annotated example:

[1] This had also been vaguely hinted at ten years earlier in an article by Feldman and Gries [93].

$$\underbrace{\overbrace{\overbrace{\text{reg}}^{\text{result}} \rightarrow \overbrace{+\ \text{reg}_1\ \text{reg}_2}^{\text{tree pattern}}}^{\text{production}} \qquad \overbrace{4}^{\text{cost}} \qquad \overbrace{\text{EMIT “}\texttt{add r}_1\texttt{,r}_2\text{”}}^{\text{action}}}_{\text{rule}}$$

In most literature *rules* and *patterns* usually hold the same connotations. In this report, however, in the context of grammars a rule refers to a tuple of production, cost, and action, and a pattern refers the right-hand side of the production appearing in a rule.

*Tree parsing*

The instruction set grammar provides us with formal methodology of modeling machine instructions, but it does not address the problems of pattern matching and pattern selection. For that, Glanville and Graham applied an already-known technique called *LR parsing* (Left-to-right, Right-most derivation). Because this technique is mostly associated with syntax parsing, the same application on trees is commonly referred to as *tree parsing*. We proceed with an example.

Let us assume that we have the following instruction set grammar

| | PRODUCTION | COST | ACTION |
|---|---|---|---|
| 1 | reg → + $\text{reg}_1$ $\text{reg}_2$ | 1 | EMIT "add $r_1$,$r_2$" |
| 2 | reg → * $\text{reg}_1$ $\text{reg}_2$ | 1 | EMIT "mul $r_1$,$r_2$" |
| 3 | reg → Int | 1 | EMIT "mv r,I" |

and that we want to generate assembly code for the following expression tree

`+ c * a b`

such that the result of the expression ends up in a register. If `a`, `b`, and `c` all are integers, then we can assume that each node in the expression tree is of type `Int`, *, or +; these will be our terminals.

After transforming the expression trees into sequences of terminals, we traverse each from left to right. In doing so we either *shift* the just-traversed symbol onto a stack, or replace symbols currently on the stack via a *rule reduction*. A *reduce* operation consists of two steps: first the symbols are popped according to those that appear on the right-hand side of the rule. The number and order must match exactly to be a valid rule reduction. Once popped the nonterminal appearing on the left-hand side is then pushed onto the stack, and the assembly code associated with the rule, if any, is also emitted. For a given input the performed rule reductions can also be represented as a *parse tree*, illustrating the terminals and nonterminals which were used to parse the tree. Now, turning back to our example, if we denote a *shift* by $s$ and a *reduce* by $r_x$, where $x$ is the number of the reduced rule, then a valid tree parsing of the expression tree "`+ Int(c) * Int(a) Int(b)`" could be:

$$s\ s\ s\ r_3\ s\ r_3\ r_2\ s\ r_3\ r_1$$

Verifying this is left as an exercise to the reader. For this particular tree parsing the corresponding parse tree and produced assembly code is seen below:

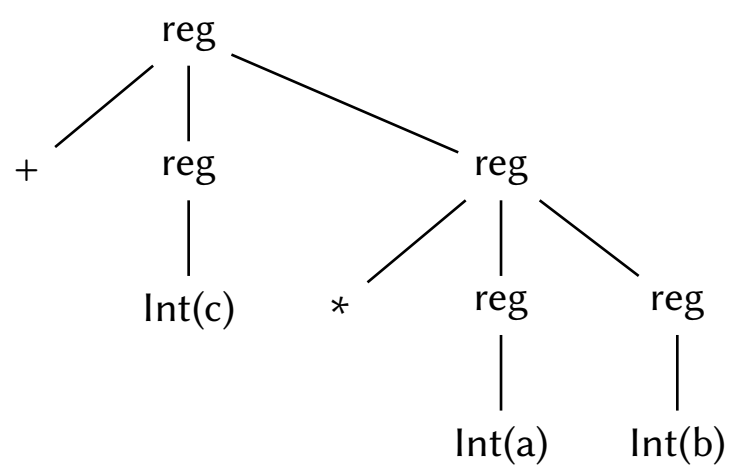

```
mv ra,a
mv rb,b
mul ra,rb
mv rc,c
add ra,rc
```

The problem that remains is how to know when to *shift* and when to *reduce*. This can be decided by consulting a *state table* which has been generated for a specific grammar. How this table is produced is out of scope for this report, but an example generated from the instruction set grammar appearing in Figure 3.3 is given in Figure 3.4. Figure 3.5 then provides a walk-through of executing an instruction selector with this state table. The subscripts that appear in some the productions in Figure 3.3 are *semantic qualifiers*, which used to express semantic restrictions that may appear in the machine instructions. For example, all two-address arithmetic machine instructions store the result in one of the register provided as input, and using semantic quantifiers this could be expressed as $r_1 \rightarrow +\ r_1\ r_2$, indicating that the destination register must be the same as that of the first operand. To make this information available during parsing, the parser pushes it onto the stacking along with its corresponding terminal or nonterminal symbol. Glanville and Graham also incorporated a register allocator into their parser, thus embodying an entire code generation backend.

*Resolving conflicts and avoiding blocking*

An instruction set grammar is typically *ambiguous*, meaning that multiple valid tree parsings may exist for the same expression tree which may lead to situations where the parser may have an option of either performing a *shift* or a *reduce*. This is known as a *shift-reduce conflict* and appear as a consequence of having non-orthogonal ISAs. To solve this kind of conflicts Glanville and Graham's state table generator always selects to *shift*. The intuition is that this will favor larger patterns over smaller ones as a *shift* postpones a decision to pattern select whilst allowing more information about the expression tree to accumulate onto the stack.[2]

Likewise there is also the possibility of *reduce-reduce conflict* where the parser has an option of choosing between two or more rules in a reduction. Glanville and Graham resolved these by selecting the rule with the longest right-hand side. If the grammar contains rules that differ only in their semantic quantifiers then there may still exist more than one rule to reduce (in Figure 3.3 rules 5 and 6 are two such rules). These are resolved at parse time by checking the semantic restrictions at parsing time in the order they appear in the grammar (see for example state 20 in Figure 3.4). However, if all rules in this set are semantically constrained then there be situations where the parser is unable to choose any rule. This is called *semantic blocking* and must resolved by always providing a default rule which can be invoked when all others fail. This fallback rule typically uses multiple, shorter machine instructions to simulate the effect of the more complex rule, and Glanville and Graham devised a clever trick to infer them automatically: for every semantically constrained rule, a tree parsing is performed over the expression tree which represents the right-hand side of that rule, and then using the machine instructions selected to implement this tree to constitute the fallback rule.

[2]The approach of always selecting the largest possible pattern is a scheme commonly known as *maximum munching*, or just *maximum munch*, which was coined by Cattell in his PhD dissertation [46].

| | PRODUCTION | ACTION |
|---|---|---|
| 1 | $r_2$ → + Ld + C $r_1$ $r_2$ | `add r₂,C,r₁` |
| 2 | $r_1$ → + $r_1$ Ld + C $r_2$ | `add r₁,C,r₂` |
| 3 | r → + Ld C r | `add r,C` |
| 4 | r → + r Ld C | `add r,C` |
| 5 | $r_1$ → + $r_1$ $r_2$ | `add r₁,r₂` |
| 6 | $r_2$ → + $r_1$ $r_2$ | `add r₂,r₁` |
| 7 | → = Ld + C $r_1$ $r_2$ | `store r₂,*C,r₁` |
| 8 | → = + C $r_1$ $r_2$ | `store r₂,C,r₁` |
| 9 | → = Ld C r | `store r,*C` |
| 10 | → = C r | `store r,C` |
| 11 | → = $r_1$ $r_2$ | `store r₂,r₁` |
| 12 | $r_2$ → Ld + C $r_1$ | `load r₂,C,r₁` |
| 13 | $r_2$ → + C $r_1$ | `load r₂,=c,r₁` |
| 14 | $r_2$ → + $r_1$ C | `load r₂,=C,r₁` |
| 15 | $r_2$ → Ld $r_1$ | `load r₂,*r₁` |
| 16 | r → Ld C | `load r,=C` |
| 17 | r → C | `mv r,C` |

Figure 3.3: An example of an instruction set grammar (from [121]). All rules are have the same unit cost. C, Ld +, = are all terminals (C stands for "const" and Ld for "load"), r is a non-terminal indicating that the result will be stored in a register, and subscripts denote the semantic qualifiers.

| | $ | r | c | + | Ld | = |
|---|---|---|---|---|---|---|
| 0 | *accept* | | | | | s1 |
| 1 | | s2 | s3 | s4 | s5 | |
| 2 | | s6 | s7 | s8 | s9 | |
| 3 | | s10 | s7 | s8 | s9 | |
| 4 | | s11 | s12 | s8 | s13 | |
| 5 | | s14 | s15 | s16 | s9 | |
| 6 | r11 | r11 | r11 | r11 | r11 | r11 |
| 7 | r17 | r17 | r17 | r17 | r17 | r17 |
| 8 | | s11 | s17 | s8 | s13 | |
| 9 | | s14 | s18 | s19 | s9 | |
| 10 | r10 | r10 | r10 | r10 | r10 | r10 |
| 11 | | s20 | s21 | s8 | s22 | |
| 12 | | s23 | s7 | s8 | s9 | |
| 13 | | s14 | s24 | s25 | s9 | |
| 14 | r15 | r15 | r15 | r15 | r15 | r15 |
| 15 | | s26 | s7 | s8 | s9 | |
| 16 | | s11 | s27 | s8 | s13 | |
| 17 | | s28 | s7 | s8 | s9 | |
| 18 | r16 | r16 | r16 | r16 | r16 | r16 |
| 19 | | s11 | s29 | s8 | s13 | |
| 20 | r5/6 | r5/6 | r5/6 | r5/6 | r5/6 | r5/6 |
| 21 | r14 | r14 | r14 | r14 | r14 | r14 |
| 22 | | s14 | s30 | s31 | s9 | |
| 23 | | s32 | s7 | s8 | s9 | |
| 24 | | s33 | s7 | s8 | s9 | |
| 25 | | s11 | s34 | s8 | s13 | |
| 26 | r9 | r9 | r9 | r9 | r9 | r9 |
| 27 | | s35 | s7 | s8 | s9 | |
| 28 | r13 | r13 | r13 | r13 | r13 | r13 |
| 29 | | s36 | s7 | s8 | s9 | |
| 30 | r4 | r4 | r4 | r4 | r4 | r4 |
| 31 | | s11 | s37 | s8 | s13 | |
| 32 | r8 | r8 | r8 | r8 | r8 | r8 |
| 33 | r3 | r3 | r3 | r3 | r3 | r3 |
| 34 | | s38 | s7 | s8 | s9 | |
| 35 | | s39 | s7 | s8 | s9 | |
| 36 | r12 | r12 | r12 | r12 | r12 | r12 |
| 37 | | s40 | s7 | s8 | s9 | |
| 38 | | s41 | s7 | s8 | s9 | |
| 39 | r7 | r7 | r7 | r7 | r7 | r7 |
| 40 | r2 | r2 | r2 | r2 | r2 | r2 |
| 41 | r1 | r1 | r1 | r1 | r1 | r1 |

Figure 3.4: State table generated from the instruction set grammar given in Figure 3.3 (from [121]). sX indicates a shift to the next state X, rX indicates the reduction of rule X, and a blank entry indicates an error state.

| | STATE STACK | SYMBOL STACK | INPUT | ACTION |
|---|---|---|---|---|
| 1 | 0 | | = + $C_a$ $r_7$ + Ld + $C_b$ Ld $r_7$ Ld $C_c$ $ | shift to 1 |
| 2 | 0 1 | = | + $C_a$ $r_7$ + Ld + $C_b$ Ld $r_7$ Ld $C_c$ $ | shift to 4 |
| 3 | 0 1 4 | = + | $C_a$ $r_7$ + Ld + $C_b$ Ld $r_7$ Ld $C_c$ $ | shift to 12 |
| 4 | 0 1 4 12 | = + $C_a$ | $r_7$ + Ld + $C_b$ Ld $r_7$ Ld $C_c$ $ | shift to 23 |
| 5 | 0 1 4 12 23 | = + $C_a$ $r_7$ | + Ld + $C_b$ Ld $r_7$ Ld $C_c$ $ | shift to 8 |
| 6 | 0 1 4 12 23 8 | = + $C_a$ $r_7$ + | Ld + $C_b$ Ld $r_7$ Ld $C_c$ $ | shift to 13 |
| 7 | 0 1 4 12 23 8 13 | = + $C_a$ $r_7$ + Ld | + $C_b$ Ld $r_7$ Ld $C_c$$ | shift to 25 |
| 8 | 0 1 4 12 23 8 13 25 | = + $C_a$ $r_7$ + Ld + | $C_b$ Ld $r_7$ Ld $C_c$ $ | shift to 34 |
| 9 | 0 1 4 12 23 8 13 25 34 | = + $C_a$ $r_7$ + Ld + $C_b$ | Ld $r_7$ Ld $C_c$ $ | shift to 9 |
| 10 | 0 1 4 12 23 8 13 25 34 9 | = + $C_a$ $r_7$ + Ld + $C_b$ Ld | $r_7$ Ld $C_c$ $ | shift to 14 |
| 11 | 0 1 4 12 23 8 13 25 34 9 14 | = + $C_a$ $r_7$ + Ld + $C_b$ Ld $r_7$ | Ld $C_c$ $ | reduce rule 15 ($r_2 \rightarrow$ Ld $r_1$)<br>assign result to r8<br>emit "`load r8,*r7`"<br>shift to 38 |
| 12 | 0 1 4 12 23 8 13 25 34 38 | = + $C_a$ $r_7$ + Ld + $C_b r_8$ | Ld $C_c$ $ | shift to 9 |
| 13 | 0 1 4 12 23 8 13 25 34 38 9 | = + $C_a$ $r_7$ + Ld + $C_b$ $r_8$ Ld | $C_c$ $ | shift to 18 |
| 14 | 0 1 4 12 23 8 13 25 34 38 9 18 | = + $C_a$ $r_7$ + Ld + $C_b$ $r_8$ Ld $C_c$ | $ | reduce rule 16 (r $\rightarrow$ Ld C)<br>assign result to r9<br>emit "`load r9,C`"<br>shift to 41 |
| 15 | 0 1 4 12 23 8 13 25 34 38 41 | = + $C_a$ $r_7$ + Ld + $C_b$ $r_8$ $r_9$ | $ | reduce rule 1 ($r_2 \rightarrow$ + Ld + C $r_1$ $r_2$)<br>emit "`add r9,B,r8`"<br>shift to 32 |
| 16 | 0 1 4 12 23 32 | = + $C_a$ $r_7$ $r_2$ | $ | reduce rule 8 ( $\rightarrow$ = + C $r_1$ $r_2$)<br>emit "`store r9,A,r7`" |
| 17 | 0 | | $ | accept |

Figure 3.5: A walk-through of executing Glanville and Graham's parser on an expression $a = b + c$, where $a$, $b$ and $c$ are constants residing in memory, which yielded the IR code "= + $C_a$ $r_7$ + Ld + $C_b$ Ld $r_7$ Ld $C_c$", where $r_7$ is the base address register (from [121]). The execution is based on the precomputed table from Figure 3.4. The proceedings of the reduction steps may need some explanation. A rule reduction may involve two operations: the mandatory *reduce* operation; followed by an optional operation which may be a *shift* or another rule reduction. Let us examine step 11. First a *reduce* is executed using rule 15 which pops Ld $r_7$ from the symbol stack. This is followed by pushing the result of the rule, $r_8$, on top. At the same time, 9 and 14 are popped from the state stack which leaves state 34 on top. By now consulting the table using the top elements of both stacks, the next, additional action (if any) can be inferred. In this case, input symbol $r_8$ at state 34 leads to a *shift* to state 38.

*Advantages*

By relying on a state table Glanville-Graham style instruction selectors are completely table-driven and implemented by a core that basically consists of a series of table lookups.[3] As a consequence, the time it takes for the instruction selector to produce the assembly code is linearly proportional to the size of the expression tree. Although the idea of table-driven code generation was not novel in itself – we have seen a number of these in the previous chapter – earlier attempts had all failed to provide an automated procedure for producing the tables. In addition, many decisions regarding pattern selection are precomputed by resolving *shift-reduce* and *reduce-reduce* conflicts at the time that the state table is generated which leads to faster compilation.

Another advantage of the Glanville-Graham approach is its formal foundation which enables means of automatic verification. For instance, Emmelmann [80] presented one of the first approaches to prove the completeness of an instruction set grammar. The intuition of Emmelmann's automatic prover is to find all expression trees that can appear in an input program but cannot be handled by the instruction selector. Let us denote an instruction set grammar as $\mathcal{G}$ and the grammar which describes the expression trees as $\mathcal{T}$. If we further use $L(x)$ to represent the set of all trees accepted by a grammar $x$, we can then determine whether the instruction set grammar is incomplete by checking if $L(\mathcal{T})\backslash L(\mathcal{G})$ yields a non-empty set. Emmelmann recognized that this intersection can be computed by creating a *product automaton* which essentially implements the language that accepts only the trees in this set of counterexamples. Brandner [34] recently extended this method to handle productions that contain dynamic checks called *predicates* – we will discuss shortly when exploring attribute grammars – by splitting terminals to expose these otherwise-hidden characteristics.

*Disadvantages*

Although it addressed several of the problems with contemporary instruction selection techniques, the Glanville-Graham approach had disadvantages of its own. First, since the parser can only reason on syntax any restrictions regarding specific values or ranges must be given its own nonterminal. In conjunction with the limitation that each production can match only a single tree pattern, this typically meant that rules for versatile machine instructions with several addressing or operand modes had to be duplicated for each such combination. For most target machines, however, this turned out to be impracticable; in the case of the VAX machine – a CISC-based architecture from the 1980s, where each machine instruction accepted a multitude of operand modes [48] – such an instruction set grammar would contain over eight million rules [127]. Refactoring – the task of combining similar rules by expressing shared features via nonterminals – brought this number down to about thousand rules, but this had to be done carefully to not have a negative impact on code quality. Second, since the parser traverses from left to right without backtracking, assembly code regarding one operand has to be emitted before any other operand can be observed. This can potentially lead to poor decisions which later have to be undone by emitting additional code. Hence to design a compact instruction set grammar – which also yielded good code quality – the grammar writer had to possess extensive knowledge about the inner workings of the instruction selector.

[3] Pennello [203] developed a technique to express the state table directly as assembly code, thus eliminating even the need to perform table lookups. This was reported to improve the rate of LR parsing by six to ten times.

### 3.3.2 Extending grammars with semantic handling

In purely context-free grammars there is simply no way of handling semantic information. For example, the exact register represented by a reg nonterminal is not available. Glanville and Graham worked around this limitation by pushing the information onto the stack, but even then their modified LR parser could reason upon it using simple equality comparisons. Ganapathi and Fischer [112, 113, 114, 115] addressed this problem by replacing the use of traditional, context-free grammars with a more powerful set of grammars known as *attribute grammars*. Like with the Glanville-Graham approach we will only discuss how it works at a high level; readers interested of the details are advised to consult the referenced papers.

#### *Attribute grammars*

Attribute grammars – or *affix grammars* as they are also called – were introduced by Knuth in 1968 who extended context-free grammars with *attributes*. Attributes are used to store, manipulate, and propagate additional information about individual terminals and nonterminals during parsing, and an attribute is either *synthesized* or *inherited*. Using parse trees as a point of reference, a node with a synthesized attribute forms its value from the attributes of its children, and a node with an inherited attribute copies the value from the parent. Consequently, information of synthesized attributes flows *upwards* along the tree while information of inherited attributes flows *downwards*. We therefore distinguish between synthesized and inherited attributes by a ↑ or ↓, respectively, which will be prefixed to the attribute of a particular symbol (e.g. the synthesized `x` attribute of a reg nonterminal is written as reg↑`x`).

The attributes are then used within *predicates* and *actions*. Predicates are used for checking the applicability of a rule, and actions are used to produce new synthesized attributes. Hence, when modeling machine instructions we use can predicates to express the constraints, and actions to indicate effects such as code emission and what the register the result will be stored in. Let us look at an example.

In Figure 3.6 we see a set of rules for modeling three byte-adding machine instructions: an increment version `incb` (increments a register by 1, modeled by rules 1 and 2); a two-address version `add2b` (adds two registers and stores the result in one of the operands, modeled by rules 3 and 4); and a three-address version `add3b` (the result can be stored elsewhere, modeled by rule 5). Naturally, the `incb` instruction can only be used when one of the operands is a constant of value 1, which is enforced by the predicate *IsOne*. In addition, since this instruction destroys the previous value of the register, it can only be used when no subsequent operation uses the old value; this is enforced by the predicate *NotBusy*. The EMIT action then emits the corresponding assembly code. Since addition is commutative, we require two rules to make the machine instruction applicable in both cases. Similarly we have two rules for the `add2b` instruction, but the predicates have been replaced by a *TwoOp* which checks if one of the operands is the target of assignment or if the value is not needed afterwards (i.e. is not busy). Since the last rule does not have any predicates it also acts as default rule, thus preventing situations of semantic blocking which we discussed when covering the Glanville-Graham approach.

| | PRODUCTION | PREDICATES | ACTIONS |
|---|---|---|---|
| 1 | byte↑r → + byte↑a byte↑r | *IsOne*(↓a), *NotBusy*(↓r) | EMIT "`incb ↓r`" |
| 2 | byte↑r → + byte↑r byte↑a | *IsOne*(↓a), *NotBusy*(↓r) | EMIT "`incb ↓r`" |
| 3 | byte↑r → + byte↑a byte↑r | *TwoOp*(↓a, ↓r) | EMIT "`addb2 ↓a,↓r`" |
| 4 | byte↑r → + byte↑r byte↑a | *TwoOp*(↓a, ↓r) | EMIT "`addb2 ↓a,↓r`" |
| 5 | byte↑r → + byte↑a byte↑b | | GETREG(↑r)<br>EMIT "`addb3 ↓r,↓a,↓b`" |

Figure 3.6: An instruction set expressed as an affix grammar (from [114]). Terminals and nonterminals are typeset in normal font with initial capital, attributes are prefixed by either a ↑ or ↓, predicates are typeset in italics, and actions are written entirely using capitals.

*Advantages and disadvantages*

The use of predicates removes the need of introducing new nonterminals for expressing specific values and ranges which leads to a more concise instruction set grammar. In the case of the VAX machine, the use of attributes lead to a grammar half the size of that required for the Glanville-Graham approach, without applying any extensive refactoring [113]. Attribute grammars also facilitate incremental development of the machine descriptions: one can start with implementing the most general rules to achieve an instruction set grammar that produces correct but inefficient code. Rules for handling more complex machine instructions can then can be added incrementally, thus making it possible to balance implementation effort against code quality. Another neat feature is that other optimization routines, such as constant folding,[4] can be expressed as part of the grammar instead of as a separate component.

To permit attributes to be used together with LR parsing, however, the properties of the instruction set grammar must be restricted. First, only synthesized attributes may appear in nonterminals. This is because an LR parser constructs the parse tree bottom-up and left-to-right, starting from the leaves and working its way up towards the root. Hence an inherited value to be used in guiding the construction of subtrees only becomes available once the subtree has already been created. Second, since predicates may render a rule as invalid for rule reduction all actions must appear at the right-most end of the production; otherwise these may lead to incorrect effects that cannot be undone. Third, like with the Glanville-Graham approach the parser has to make take decisions regarding one subtree without taking any consideration of sibling subtrees that may appear to the right. This can therefore lead to assembly code which could have been improved if all subtrees had been available beforehand, and this is again a limitation due to the use of LR parsing. Ganapathi [111] later made an attempt to resolve this problem by implementing a instruction selector in Prolog – a logic-based programming language – but this incurred exponential worst-time execution.

[4] *Constant folding* is the task of precomputing arithmetic expressions appearing in the input program that operates on static values, such as constants.

### 3.3.3 Maintaining multiple parse trees for better code quality

Since LR parsers make a single pass over the expression trees – and thus only produce one out of many possible parse trees – the quality of the produced assembly code is heavily dependent on the instruction set grammar to guide the parser in finding "good" parse trees.

Christopher et al. [52] attempted to address this concern by using the concepts of the Glanville-Graham approach but extending the parser to produce *all* parse trees, and then select the one which yields the best assembly code. This was achieved by replacing the original LR(1) parser with an implementation of Earley's algorithm [74], and although this approach certainly improves code quality – at least in theory – it does so at the cost of enumerating all parse trees which is often too expensive in practice.

In 2000 Madhavan et al. [181] extended the Glanville-Graham approach to achieve optimal selection of patterns while allegedly retaining the linear time complexity of LR parsing. By incorporating an new version of LR parsing [230], reductions that were previously executed directly as matching rules were found are now allowed to be postponed by an arbitrary number of steps. Hence the instruction selector essentially keeps track of multiple parse trees, allowing it to gather enough information about the input program before committing to a decision which could turn out to be suboptimal. In other words, like Christopher et al. the design by Madhavan et al. also covers all parse trees, but immediately discards those which are determined to result in less efficient assembly code. To do this efficiently the approach also incorporates *offline cost analysis*, which we will explore later in Section 3.6.3. More recently Yang [266] proposed a similar technique involving the use of *parser cactuses* [sic] where deviating parse trees are branched off a common trunk to reduce space requirements. However, the underlying principle of both still prohibits the modeling of many typical target-machine features of such as multi-output instructions.

## 3.4 Covering trees top-down with recursion

The tree covering approaches examined so far – in particular those based on LR parsing – all operate *bottom-up*: the instruction selector begins to cover the leaf nodes in the expression tree, and based on these decisions it then subsequently covers the remaining subtrees until it reaches the root, continually matching and selecting applicable patterns along the way. This is by no means the only method of covering as it can also be done *top-down*. In such approaches the instruction selector covers the expression tree starting from the root node, and then recursively works its way downwards. Consequently, the flow of semantic information (e.g. the particular register in which a result will be stored) is also different: a bottom-up instruction selector lets this information trickle upwards along the expression tree – either via auxiliary data structures or through tree rewriting – whereas a top-down implementation decides this beforehand and pushes it downwards. The latter are therefore said to be *goal-driven* as pattern selection is guided by a set of additional requirements which must be fulfilled by the selected pattern. Since this in turn will incur new requirements for the subtrees, most top-down techniques are implemented using recursion which also provides backtracking. This is needed because certain selection of patterns can cause the lower parts of the expression tree to become uncoverable.

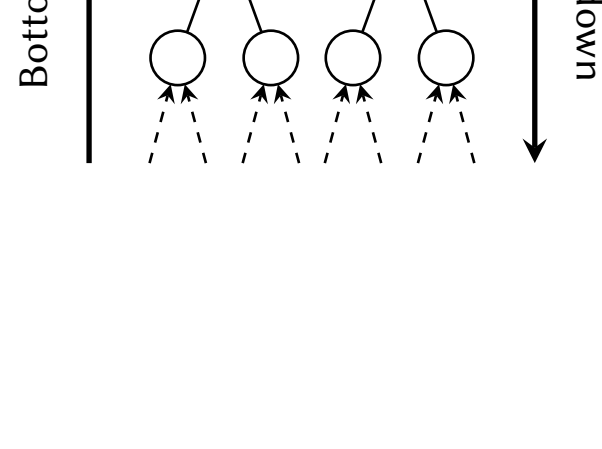

### 3.4.1 First approaches

*Instruction selection guided by means-end analysis*

To the best of my knowledge, Newcomer [194] was the first to develop a method which uses top-down tree covering to address instruction selection. In his 1975 PhD dissertation Newcomer proposes a design that exhaustively finds all combinations of patterns that cover a given expression tree, and then selects the one with the lowest cost. Cattell [45] also describes this in his survey paper, which is the main source for this discussion.

The machine instructions are modeled as *T-operators* which are basically tree patterns with costs and attributes attached. The attributes describe various restrictions such as which registers can be used for the operands. There is also a set of T-operators that the instruction selector uses to perform necessary transformations of the input program – this need will become clear as we continue the discussion. With this toolbox the scheme takes an AST as expected input and then covers it following to the aforementioned top-down approach: the instruction selector attempts to find all matching patterns for the root node of the AST, and then proceeds to recursively cover the remaining subtrees for each match. Pattern matching is done using a straight-forward technique that we know from before (see Figure 3.2 on page 25), and for efficiency all patterns are indexed according to the type of their root node. The result of this procedure is thus a set of pattern sequences which each covers the entire AST. Afterwards each sequence is checked whether the attributes of every pattern is equal to a *preferred attribute set* (PAS), which thus corresponds to a goal. If not the instruction selector will attempt to rewrite the subtree using the transformation T-operators until the attributes match. To guide this process Newcomer applied a heuristic search strategy known as *means-end analysis*, which was introduced by Newell and Simon [195] in 1959. The intuition behind means-end analysis is to recursively minimize the quantitative difference – how this is calculated is not mentioned in [45] – between the current state (i.e. what the subtree looks like *now*) and a goal state (what it *should* look like). To avoid infinite looping the transformation process stops once it reaches a certain depth in the search space. If successful the applied transformations are inserted into the pattern sequence; if not the sequence is dropped. From the found pattern sequences the one with the lowest total cost is selected, followed by code emission.

Newcomer's approach was pioneering as the application of means-end analysis made it possible to guide the process of modifying the input program until it could be implemented on the target machine without have to resort to target-specific mechanisms. However, the design was not without flaws. First, the design had little practical application as it only handles arithmetic expressions. Second, the T-operators used for modeling the machine instructions as well as transformations had to be constructed by hand, which was non-trivial and hindered compiler retargetability. Second, the process of transforming the input program could end prematurely due to the search space cut-off, causing the instruction selector to fail to find generate any assembly code whatsoever. Lastly, the search strategy proved much too expensive to be usable in practice except for the smallest of input trees.

*Making means-end analysis work in practice*

Cattell et al. [44, 47, 176] later improved and extended Newcomer's work into a more practical framework which was implemented in the Production Quality Compiler-Compiler (PQCC), a derivation of the Bliss-11 compiler originally written by Wulf et al. [264]. Instead of performing the means-end analysis as the input program is compiled, their approach does it as a preprocessing step when generating the compiler itself – much like with the Glanville-Graham approach.

The patterns are expressed a set of templates which are formed using recursive composition, and are thus similar to the productions found in instruction set grammars. Unlike Glanville and Graham's and Ganapathi and Fischer's designs, however, the templates are not written by hand but automatically derived from a target-specific machine description. Each machine instruction is described as a set of *machine operations* which describe the effects of the instruction, and are thus akin to the RTLs introduced by Fraser [102] in the previous chapter. These effects are then used by a separate tool called CGG (Code-Generator Generator) to create the templates which will be used by the instruction selector. In addition to producing the trivial templates that correspond directly with a machine instruction, CGG also produces a set of single-node patterns as well as a set of larger patterns that combine several machine instructions. The former ensures that the instruction selector can produce assembly code for all input programs as any expression tree can thereby always be trivially covered, while the latter reduces compilation time as it is quicker to match a large pattern than many smaller ones. To do this CGG uses a combination of means-end analysis and heuristic rules which apply a set of axioms (e.g. $\neg\neg E \Leftrightarrow E$, $E + 0 \Leftrightarrow E$, $\neg(E_1 \geq E_2) \Leftrightarrow E_1 < E_2$, etc.) to manipulate and combine existing tree patterns into new ones. There are, however, no guarantees that these "interesting" tree patterns will ever be applicable. Once generated, instruction selection is then performed in a greedy, top-down fashion which always selects the template with the lowest cost that matches at the current node in the expression tree (pattern matching is done using a scheme identical to Newcomer). If there is a draw the instruction selector picks the template with the least number of memory loads and stores.

Compared to the LR parsing-based approaches discussed previously, Cattell's et al. approach has both advantages and disadvantages. The main advantage is that regardless of the provided templates, the generated instruction selectors are always capable of generating valid assembly code for all possible input trees.[5] At least in Ganapathi and Fischer's design this correctness has to be assured by the grammar designer. The disadvantage is that it is considerably slower:

[5]There is of course the possibility that the set of predefined templates is insufficient to ensure this property, but CGG is at least capable of detecting such situations and issue a warning.

whereas the tree parsing-based instruction selectors exhibit linear time complexity, both for pattern matching and selection, Cattell's et al. instruction selector has to match each template individually which could take quadratic time in the worst case.

*Recent techniques*

To the best of my knowledge, the only recent approach (i.e. less than 20 years old) to use this kind of recursive top-down methodology for tree covering is that of Nymeyer et al. [197, 198]. In two 1996 and 1997 papers Nymeyer et al. describe a method where A* search – another strategy for controlling the search space (see [219]) – is combined with *bottom-up rewriting system* theory, or BURS theory for short. We will discuss BURS theory in more detail later in this chapter – the anxious reader can skip directly to Section 3.6.3 – but for now let it be sufficient to mention that grammars based on BURS allow transformation rules, such as rewriting $X + Y$ into $Y + X$, to be included, which potentially simplifies and reduces the number of rules required for expressing the machine instructions. However, the authors did not publish any experimental results, making it difficult to judge whether the A*-BURS theory combination would be an applicable technique in practice.

## 3.5 A note on tree rewriting and tree covering

At this point some readers may feel that tree *rewriting* – where patterns are iteratively selected for rewriting the expression tree until it consists of a single node of some goal type – is something entirely different compared to tree *covering* – where compatible patterns are selected for covering all nodes in the expression tree. Indeed there appears to be a subtle difference, but a valid solution to a problem expressed using tree writing is also a valid solution to the equivalent problem expressed using tree covering, and vice versa. It could therefore be argued that the two are interchangeable, but I regard tree rewriting as one of the *means* to solving the tree covering problem which I regard as a fundamental *principle*.

## 3.6 Separating pattern matching and pattern selection

In the previously discussed tree covering approaches, the tasks of pattern matching and pattern selection are unified into a single step. Although this enables single-pass code generation – a trend which long since has fallen out of fashion [227] – it typically also prevents the instruction selector from considering the impact of combinations of patterns. By keeping these two concerns separate, and allowing the instruction selector to make multiple passes over the expression tree, it can gather enough information about all applicable patterns before having to commit to premature decisions. However, excluding LR parsing the pattern matchers we have seen so far have all been implementations of algorithms with quadratic time complexity. Fortunately, we can do better.

### 3.6.1 Algorithms for linear-time tree pattern matching

Over the years many algorithms have been discovered for finding all occurrences where a set of tree patterns match within another given tree (see for example [51, 55, 73, 140, 149, 210, 211, 229, 257, 265]. In tree covering, however, most pattern matching algorithms have been derived from methods of string pattern matching. This was first discovered by Karp et al. [149] in 1972, and Hoffmann and O'Donnell [140] then extended these ideas to form the algorithms most applied by tree-based instruction selection techniques. So in order to understand pattern matching with trees, let us first explore how this is done with strings.

*Matching trees is the same as matching strings*

The algorithms most commonly used for string matching were introduced by Aho and Corasick [5] and Knuth et al. [160]. Independently discovered from one another and published in 1975 and 1977, respectively, both algorithms operate in the same fashion and are thus nearly identical in their approach. The intuition is that when a partial match of a substring which has repetitive pattern fails, the matcher does not need to return all the way to the input character where the matching initially started. This is illustrated in the inlined table where the substring `abcabd` is matched against the input string `abcabcabd`. The arrow indicates the current character under consideration. At first, the substring matches the beginning of the input string up until the last character (position 5). When this fails, instead of returning to position 1 and restarting the matching from scratch, the matcher remembers that the first three characters of the substring (`abc`) have already been matched at this point and therefore continues to position 6, attempting to match the fourth character in the substring. Consequently all occurrences of the substring can be found in linear time. We continue our discussion with Aho and Corasick's approach as it is capable of matching multiple substrings whereas the algorithm of Knuth et al. only considers a single substring (although it can easily be extended to handle multiple substring as well).

| | 0 | 1 | 2 | 3 | 4 | 5 | 6 | 7 | 8 |
|---|---|---|---|---|---|---|---|---|---|
| Input string | a | b | c | a | b | c | a | b | d |
| Substring | a | b | c | a | b | d | | | |
| | | | | | | ↑ | | | |
| | | | | a | b | c | a | b | d |
| | | | | | | | ↑ | | |

Aho and Corasick's algorithm relies on three functions – *goto*, *failure*, and *output* – where the first implements a state machine and the latter two are simple lookup tables. How these are constructed is out of scope for our purpose – the interested reader can consult the referenced paper – but we will illustrate how it works with an example. In Figure 3.7 we see the corresponding functions for matching the strings `he`, `she`, `his`, and `hers`. As a character is read from an input string, say `shis`, it is first given as argument to the *goto* function whose state machine has been initialized to state `0`. With `s` as input it transitions to state 3, and after `h` it ends up in state 4. For the next input character `i`, however, there exists no corresponding edge from the current state. At this point the *failure* function is consulted which dictates that the state machine should fall back to state 1. We retry with `i` in that state, which takes us to state 6, and with the last input character with transition to state 7. For this state the *output* function also indicates that we have found a match with pattern string `his`.

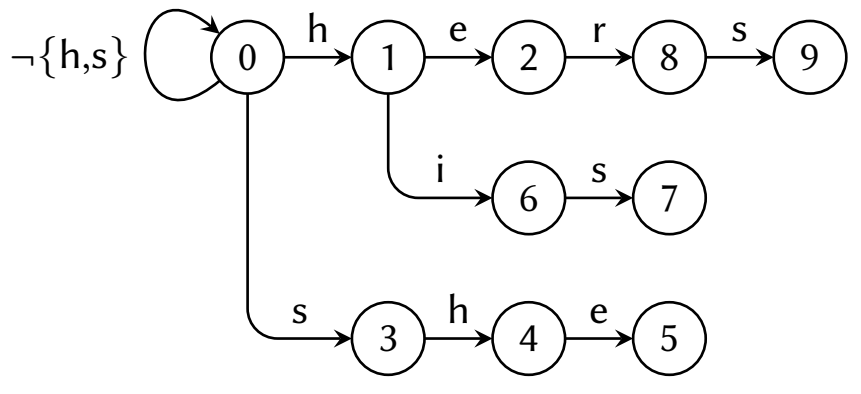

(a) Goto function.

| $i$ | 1 | 2 | 3 | 4 | 5 | 6 | 7 | 8 | 9 |
|---|---|---|---|---|---|---|---|---|---|
| $f(i)$ | 0 | 0 | 0 | 1 | 2 | 0 | 3 | 0 | 3 |

(b) Failure function.

| $i$ | output($i$) |
|---|---|
| 2 | {he} |
| 5 | {she, he} |
| 7 | {his} |
| 9 | {hers} |

(c) Output function.

Figure 3.7: String matching state machine (from [5]).

### *The Hoffmann-O'Donnell algorithm*

Hoffmann and O'Donnell [140] developed two algorithms which incorporate the ideas of Aho and Corasick and Knuth et al. In their 1982 paper they first present an $\mathcal{O}(np)$ algorithm which matches tree patterns in a top-down fashion, and then a $\mathcal{O}(n + m)$ bottom-up algorithm which trades linear-time pattern matching for longer preprocessing times ($n$ is the size of the subject tree, $p$ is the number of tree patterns, and $m$ is the number of matches found).

Pattern matching for the latter – which is the most applied due to its linear runtime behavior – is simple and outlined in Figure 3.8: starting at the leaves, each node is labeled with an identifier denoting the set of patterns that match the subtree rooted of that node. We call this set the *matchset*. The label to assign a particular node is retrieved by using the labels of the child nodes as indices to a table which is specific to the type of the current node. For example, label lookups for nodes representing addition are done using one table while lookups for nodes representing subtraction are done using another table. The dimension of the table is equal to the number of children that the node may have, e.g. binary operation nodes have 2-dimensional tables while nodes representing constant values have a 0-dimensional table, thus consisting of just a single value. An fully labeled example is also available in (g) of Figure 3.9 appearing on page 41. The matchsets are then retrieved via a subsequent top-down traversal.

Since the bottom-up algorithm introduced by Hoffmann and O'Donnell has had an historical impact on instruction selection, we will spend some time to discuss the details on how the lookup tables used for labeling are generated.

```
LabelTree(expression tree rooted at node n):
  for each child node m_i of n do
      LabelTree(m_i)
  end for
  label n with T_n[labels of m_1, ..., m_k]
```

Figure 3.8: Hoffmann-O'Donnell's bottom-up algorithm for labeling expression trees (from [140]).

*Definitions*

We begin with introducing some definitions, and to our aid we will use the two tree patterns I and II appearing in (a) and (b), respectively, of Figure 3.9. First, we say that patterns I and II constitute our *pattern forest*, consisting of nodes with symbols $a$, $b$, $c$, or $v$, where an $a$ node has exactly two children and $b$, $c$, and $v$ nodes have none. The $v$ symbol is a special *nullary* symbol as such nodes represent placeholders that can match any subtree. We say that these symbols collectively constitute the *alphabet* of the pattern forest, which we will denote by $\Sigma$. The alphabet needs to be finite and *ranked*, meaning that each symbol in $\Sigma$ has a *ranking function* that gives the number of children for a given symbol. Hence, in our case $rank(a) = 2$ and $rank(b) = rank(c) = rank(v) = 0$. Following the terminology of Hoffmann and O'Donnell's paper, we also introduce the notion $\Sigma$*-term* and define it as follows:

1. For all $i \in \Sigma$ where $rank(i) = 0$, $i$ is a $\Sigma$-term.
2. If $i(t_1, \ldots, t_m) \in \Sigma$ where $m = rank(i) > 0$, then $i(t_1, \ldots, t_m)$ is a $\Sigma$-term provided that every $t_i$ is a $\Sigma$-term.
3. Nothing else is a $\Sigma$-term.

A tree pattern is therefore a $\Sigma$-term, allowing us to write patterns I and II in as $a(a(v, v), b)$ and $a(b, v)$, respectively. $\Sigma$-terms are also ordered – e.g. $a(b, v)$ is different from $a(v, b)$ – meaning that commutative operations such as addition must be handled through pattern duplication, like in the Glanville-Graham approach.

We also introduce the following definitions: a pattern $p$ is said to *subsume* another pattern $q$ (written $p \geq q$) if any matchset that includes $p$ always also includes $q$. For example, given two pattern $a(b, b)$ and $a(v, v)$ we have that $a(b, b) \geq a(v, v)$ since the $v$ nodes must obviously also match whenever the $b$ nodes match. By this definition all patterns also subsume themselves. A pattern $p$ *strictly subsumes* another pattern $q$ (written $p > q$) if and only if $p \geq q$ and $p \neq q$. Lastly, a pattern $p$ *immediately subsumes* another pattern $q$ (written $p >_i q$) if and only if there exists no other pattern $r$ such that $p > r$ and $r > q$. For completeness, two patterns $p$ and $q$ are said to be *inconsistent* if no pattern appears in the matchset of the other, and *independent* if there exists a tree for which $p$ matches but not $q$, and another tree for which $q$ matches but not $p$. Pattern forests which contain no independent patterns are known as *simple pattern forests*, and our pattern forest is simple since there exists no tree for which both patterns I and II match.

*Generating the lookup tables for simple pattern forests*

The notion of simple pattern forests is important. In general the size of each lookup table is exponential with the size of the pattern forest, as is the time to generate these tables. However, Hoffmann and O'Donnell recognized that for simple pattern forests the number of possible matchsets is equal to the number of patterns, and that each such matchset can be represented by a single tree pattern – namely the largest pattern appearing in the matchset. The intuition is as follows: if the pattern forest is simple – meaning it is devoid of independent patterns – then for every pair of patterns in every matchset it must hold that one pattern subsumes the other. Consequently, for every matchset we can form an order of subsumption between the patterns

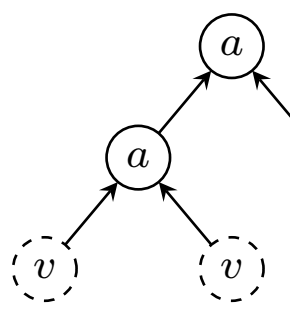

(a) Pattern I.

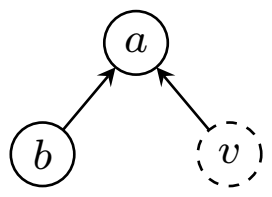

(b) Pattern II.

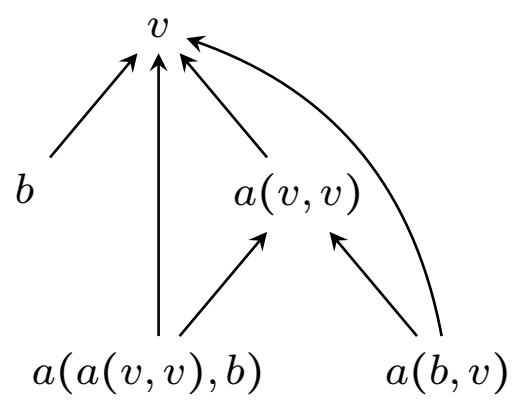

(c) Subsumption graph, loop edges excluded.

| $T_a$ 1 / 0 | 0 | 1 | 2 | 3 | 4 |
|---|---|---|---|---|---|
| 0 | 2 | 2 | 2 | 2 | 2 |
| 1 | 3 | 3 | 3 | 3 | 3 |
| 2 | 2 | 4 | 2 | 2 | 2 |
| 3 | 2 | 4 | 2 | 2 | 2 |
| 4 | 2 | 4 | 2 | 2 | 2 |

(d) Table for symbol $a$.

| $T_b$ | 1 |
|---|---|

(e) Table for symbol $b$.

| $T_c$ | 0 |
|---|---|

(f) Table for symbol $c$.

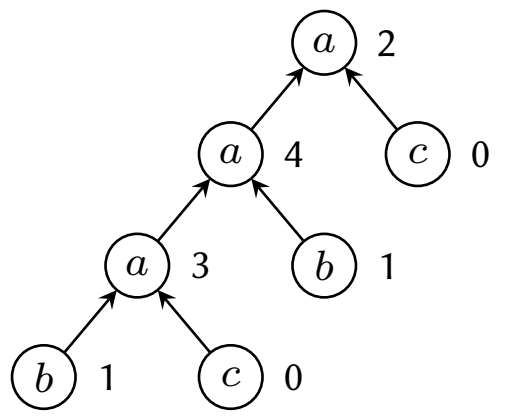

(g) A labeled example.

Figure 3.9: Tree pattern matching using Hoffmann-O'Donnell (from [140]). Nullary nodes $v$ are accentuated with a dashed border. The subpatterns have been labeled as follows: $v$ (0), $b$ (1), $a(v, v)$ (2), $a(b, v)$ (3), and $a(a(v, v), b)$ (4).

```
BuildSubsumptionGraph(set of subpatterns S):
initialize G̅_S with one node and loop edge per s ∈ S
for each s = a(s_1, ..., s_m) ∈ S in increasing order of height do
    for each s' ∈ S s.t. height of s' ≤ height of s do
        if s' = v or s' = a(s'_1, ..., s'_m) s.t. for 1 ≤ i ≤ m, edge s_i → s'_i ∈ G̅_S then
            add edge s → s' to G̅_S
        end if
    end for
end for
```

(a) Algorithm for producing the subsumption graph.

```
GenerateTable(set of subpatterns S, subsumption graph G̅_S, symbol a ∈ Σ):
do topological sort on G̅_S
initialize all entries in T_a with v ∈ S
for each s = a(s_1, ..., s_m) ∈ S in increasing order of subsumption do
    for each m-tuple ⟨s'_1, ..., s'_m⟩ s.t. for 1 ≤ i ≤ m, s'_i ≥ s_i do
        T_a[s'_1, ..., s'm] = s
    end for
end for
```

(b) Algorithm for producing the lookup tables.

Figure 3.10: Preprocessing algorithms of Hoffmann-O'Donnell (from [140]).

appearing in that matchset. For example, if we have a matchset containing $m$ patterns we can arrange these patterns such that $p_1 > p_2 > \ldots > p_m$. Since the pattern with "highest" order of subsumption (i.e. the pattern which subsumes all other patterns in the matchset) can only appear in one matchset – otherwise there exists two patterns in the forest which are pairwise independent – we can uniquely identify the entire matchset using the label of that pattern. Remember also that an index along one dimension in the table will be the label assigned to the child of the current node to be labeled in the expression tree. Consequently, if we assume that labels 1 and 2 represent the subpattern $b$ and $a(v, v)$, respectively, then a table lookup $T_a[1, 2]$ effectively means that we are currently standing at the root of a tree $t$ where the largest pattern that can possibly match is $a(b, a(v, v))$, and that the label to assign this node is the label of the largest pattern that is subsumed by $a(b, a(v, v))$. If we know the order of subsumption then we can easily retrieve this pattern.

To find the order of subsumption we first enumerate all unique subtrees that appear in our pattern forest. In our case this includes $v$, $b$, $a(v, v)$, $a(b, v)$, and $a(a(v, v), b)$, which we will represent with $S$. We then assign each subpattern a sequential number, starting from 0. These represent the labels which will be used during pattern matching, so the order in which these are assigned is not important.

Next we form the *subsumption graph* of $S$, denoted as $\overline{G}_S$, where each node $n_i \in \overline{G}_S$ represents a subpattern $s_i \in S$, and each directed edge $n_i \to n_j$ indicates that $s_i$ subsumes $s_j$. For our pattern set we end up with the subsumption graph illustrated in (c) of Figure 3.9,[6] which we produce using the algorithm given in (a) of Figure 3.10. The algorithm basically works as follows: a pattern $p$ only subsumes another pattern $q$ if and only if the roots of $p$ and $q$ are of the same symbol, and every child tree of $p$ subsumes the corresponding child tree of $q$. To check whether one tree subsumes another we simply check the presence of an edge between the corresponding nodes in $\overline{G}_S$. Since every pattern subsumes itself, we start by adding all loop edges, check whether one pattern subsumes another, adds the corresponding edge if this is the case, and then repeat until $\overline{G}_S$ stabilizes. Fortunately we can minimize the number of checks by first ordering the subpatterns in $S$ by increasing order of height and then comparing the subpatterns in that order. Once we have $\overline{G}_S$, we can find the order of subsumption for all patterns by making a *topological sort* of the nodes in $\overline{G}_S$, for which there are known methods (a definition of topological sort is available in Appendix C).

Now we generate the lookup tables by following the algorithm outlined in (b) of Figure 3.10; in our example this gives us the tables shown in (d), (e), and (f) of Figure 3.9. Starting with a table initialized using the nullary pattern, the algorithm then incrementally updates each entry with the label of the next larger pattern that matches the corresponding tree. Since we iterate over the patterns in increasing order of subsumption, the last assignment to each entry will be that of the largest pattern that matches in our pattern forest.

As already stated, since patterns are required to be ordered we need to duplicate patterns containing commutative operations by swapping the subtrees of the operands. Doing this, however, yields patterns which are pairwise independent, thus destroying the property of the pattern forest being simple. In such cases the algorithm is still able to produce usable lookup tables, but a choice will be forced between the matchsets which include one commutative pattern but not the other (it comes down to whichever subpattern in $S$ is used last during table generation). Consequently, not all matching instances will be found during pattern matching, which may in turn prevent optimal pattern selection.

[6] There is also a corresponding *immediate subsumption graph* $G_S$ which in general is shaped like a *directed acyclic graph* (DAG), but for simple pattern forests always results in a tree. However, we do not need $G_S$ for generating the lookup tables.

*Compressing the lookup tables*

Chase [49] further advanced Hoffmann and O'Donnell's table generation technique by developing an algorithm that compresses the final lookup tables. The key insight is that the lookup tables often contain redundant information as many rows and columns are duplicates. For example, this can be seen clearly in $T_a$ from our previous example, which is also available in (a) of Figure 3.11. By introducing a set of *index maps* the duplicates can be removed by mapping identical columns or rows in the index map to the same row or column in the lookup table. The lookup table can then be reduce to contain only the minimal amount of information, as seen in (b) of Figure 3.11. By denoting the compressed version of $T_a$ by $\tau_a$, and the corresponding index maps as $\mu_{a,0}$ and $\mu_{a,1}$, we then replace the previous lookups – which were done via $T_i[l_0, \ldots, l_m]$ – with $\tau_i[\mu_{i,0}[l_0], \ldots, \mu_{i,m}[l_m]]$.

| $T_a$ 0 \ 1 | 0 | 1 | 2 | 3 | 4 |
|---|---|---|---|---|---|
| 0 | 2 | 2 | 2 | 2 | 2 |
| 1 | 3 | 3 | 3 | 3 | 3 |
| 2 | 2 | 4 | 2 | 2 | 2 |
| 3 | 2 | 4 | 2 | 2 | 2 |
| 4 | 2 | 4 | 2 | 2 | 2 |

(a) Uncompressed table.

| $\mu_{a,1}$ | 0 | 1 | 2 | 3 | 4 |
|---|---|---|---|---|---|
| | 0 | 1 | 0 | 0 | 0 |

| $\mu_{a,0}$ | |
|---|---|
| 0 | 0 |
| 1 | 1 |
| 2 | 2 |
| 3 | 2 |
| 4 | 2 |

| $\tau_a$ 0 \ 1 | 0 | 1 |
|---|---|---|
| 0 | 2 | 2 |
| 1 | 3 | 3 |
| 2 | 2 | 4 |

(b) Compressed table.

Figure 3.11: Compression of lookup tables (from [49]).

Table compression also provide another benefit in that for some pattern forests, the lookup tables can be so large that they cannot even be constructed in the first place. Fortunately Chase discovered that they can be compressed as the are generated, thus pushing the limit on how large lookup tables can be produced. Cai et al. [40] later improved the asymptotic bounds of Chase's algorithm.

### 3.6.2 Optimal pattern selection with dynamic programming

Once it became possible to find all matchsets for the entire expression tree in linear time, techniques started to appear that tackled the problem of optimal pattern selection in linear time. According to the literature Ripken [216] was the first to propose a viable approach for optimal instruction selection, which is described in a 1977 technical report. Ripken based his method on the dynamic programming algorithm by Aho and Johnson – which was mentioned in Chapter 1 – and later extended it to handle more realistic instruction sets with multiple register classes and addressing modes. For brevity we will henceforth abbreviate dynamic programming as DP.

Although Ripken appears to have been the first to propose a design of an optimal DP-based instruction selector, it only remained at that – a proposal. The first *practical* attempt at such a system was instead made by Aho et al. [6, 7, 245].

### *Twig*

In 1985 Aho and Ganapathi presented a language called CGL (Code Generator Language) where tree patterns are described using attribute grammars (see Figure 3.12 for examples), which are then processed by a compiler generator called Twig. Like Ripken, the generated instruction selector selects the optimal set of patterns using a version of Aho and Johnson's DP algorithm, and makes three passes over the expression tree: a top-down pass, followed by a bottom-up pass, and lastly another top-down pass. In the first pass all matchsets are found using an implementation of the Aho-Corasick string matching algorithm [5]. The second pass applies the DP algorithm for selecting the patterns, and the last pass then emits the corresponding assembly code. We proceed with examining the DP algorithm as we are already quite familiar with how to do tree pattern matching.

The DP algorithm is outlined in Figure 3.13. For each node in the expression tree we maintain a data structure for remembering the lowest cost of reducing the node to a particular nonterminal, as well as the rule which enables this reduction. Remember that a pattern found in the matchset corresponds to the right-hand side of a production appearing in the instruction set grammar (which we discussed in Section 3.3). The cost of applying a rule is the sum of the cost of the rule itself and the costs of reducing the child nodes to whatever nonterminals used by the rule. Costs therefore increase monotonically. It can also happen that not just one rule is applied for the current node, but several. This occurs when the grammar contains

```
node const mem assign plus ind;
label reg no_value;
reg:const                                   /* Rule 1 */
  {  cost = 2; }
  ={ NODEPTR regnode = getreg( );
     emit(''MOV'', $1$, regnode, 0);
     return(regnode);
  };
no_value: assign(mem, reg)                  /* Rule 3 */
  {  cost = 2+$%1$->cost; }
  ={ emit(''MOV'', $2$, $1$, 0);
     return(NULL);
  };
reg: plus(reg, ind(plus(const, reg)))       /* Rule 6 */
  {  cost = 2+$%1$->cost+$%2$->cost; }
  ={ emit(''ADD'', $2$, $1$, 0);
     return($1$);
  };
```

Figure 3.12: Rule samples from TWIG (from [7]).

```
ComputeCosts(expression tree rooted at node n):
for each child n_i of n do
    ComputeCosts(n_i)
end for
initialize costs array for n with ∞
for each rule r_i in matchset of n do
    c_i ← cost of applying r_i at n
    l_i ← left-hand nonterminal of r_i
    if c_i < costs[l_i] then
        costs[l_i] ← c_i
        set r_i as cheapest rule for reducing n to l_i
    end if
end for
```

Figure 3.13: Computing the costs with dynamic programming (from [7]).

```
SelectAndEmit(expression tree rooted at node n,
              goal-nonterminal g):
r ← cheapest rule that reduces n to g
for each nonterminal l that appears on the right-hand side of r do
    SelectAndEmit(node to reduced to l, l)
end for
emit the assembly code associated with r
```

Figure 3.14: DP-driven instruction selection algorithm with optimal pattern selection and code emission (from [7]).

*chain rules* where the right-hand side of the production consists of a single nonterminal. For example, a reduction of a terminal A to c can either be done via a c → A rule, or via a series of rule invocations, e.g.:

$$\begin{aligned} \mathrm{b} &\rightarrow \mathrm{A} \\ \mathrm{b}' &\rightarrow \mathrm{b} \\ \mathrm{c} &\rightarrow \mathrm{b}' \end{aligned}$$

The algorithm must therefore take these chain rules and check them before any other rule that could depend on them. This is called to compute the *transitive closure*.

Once the costs have been computed for the root node we can then select the cheapest set of rules that reduces the entire expression tree to the desired nonterminal. The selection algorithm is outlined in Figure 3.14: for a given goal, represented by a nonterminal, the cheapest applicable rule to reduce the current node is selected. The same is then done for each nonterminal which appears on the right-hand side in the selected rule, thus acting as the goal for the corresponding subtree. This algorithm also correctly applies the chain rules as the use of such a rule causes SelectAndEmit to be invoked on the same node but with a different goal. Note, however, that since patterns can be arbitrarily large one must take measures to select the correct subtree which can appear several levels down from the current node in the expression tree.

### *DP versus LR parsing*

This DP scheme has several advantages over those based on LR parsing. First, reduction conflicts are automatically handled by the cost computing algorithm, thus removing the need of ordering the rules which affects code quality for LR parsers. Second, rule cycles that cause LR parsers to get stuck in an infinite loop no longer need to be explicitly broken. Third, machine descriptions can be made more concise as rules differing only in costs can be combined into a single rule; again taking the VAX machine as an example, Aho et al. reported that the entire TWIG specification could be implemented using only 115 rules, which is about half the size of Ganapathi and Fischer's attribute-based instruction set grammar for the same machine.

However, the approach requires that the problem of code generation exhibits properties of optimal substructure – i.e. that optimal code can be generated by solving each of its subproblems optimally. This is not always the case as some set of patterns, whose total sum is greater compared to another set of selected patterns, can actually lead to better assembly code. For example, let us assume that we need to select patterns for implementing a set of operations that can be executed independently from one another. The instruction set provides two options: either use two machine instructions, each taking 2 cycles; or use a single machine instruction which takes 3 cycles. If the two machine instructions in the first option are executed sequentially then the second option is more efficient. However, if the instruction scheduler is capable of parallelizing the instructions then the first option is better. This further highlights the problem of considering optimal instruction selection in isolation, as we discussed in Chapter 1.

*Further improvements*

Several improvements of TWIG were later made by Yates and Schwartz [267] and Emmelmann et al. [79]. Yates and Schwartz improved the rate of pattern matching by replacing the TWIG's top-down approach with the faster bottom-up algorithm proposed by Hoffmann and O'Donnell, and also extended the attribute support to allow for more powerful predicates. Emmelmann et al. modified the DP algorithm to be run as the IR trees are built by the frontend which also inlines the code of auxiliary functions directly into the DP algorithm to reduce the overhead. Emmelmann et al. implemented their improvements in a system called BEG (Back End Generator), and a modified version of this is currently used in the CoSy compiler [23].

Fraser et al. [104] made similar improvements in a system called IBURG that is both simpler and faster than TWIG – IBURG requires only 950 lines of code compared to TWIG's 3,000 lines of C, and generates assembly code of comparable quality at a rate that is 25x faster – and has been used in several compilers (e.g. RECORD [174, 184] and REDACO [162]). Gough and Ledermann [124, 125] later made some minor improvements of IBURG in an implementation called MBURG. Both IBURG and MBURG have later been reimplemented in various programming languages, such as the Java-based JBURG [243], and OCAMLBURG [246] which is implemented in C--.

According to [42, 173], Tjiang [244] later merged the ideas of TWIG and IBURG into a new implementation called OLIVE – the name is a spin-off of TWIG – and made several additional improvements such as allowing rules to use arbitrary cost functions instead of fixed, numeric values. This allows for more versatile instruction selection as rules can be dynamically deactivated by setting an infinite costs, which can be controlled from the current context. OLIVE is used in the implementation of SPAM [237] – a fixed-point DSP compiler – and Araujo and Malik [16] employed it in an attempt to integrate instruction selection with scheduling and register allocation.

### 3.6.3 Faster pattern selection with offline cost analysis

In the dynamic programming approach just discussed, the rule costs needed for selecting the patterns are dynamically computed while the pattern matcher is completely table-driven. It was later discovered that these calculations can also be done beforehand and represented as tables, thus improving the speed of the pattern selector as it did for pattern matching. We will refer to this aspect as *offline cost analysis*, which means that the cost computations are precomputed as part of generating the compiler instead of during input program compilation.

*Extending matchset labels with costs*

To make use of offline cost analysis we need to extend the labels to not only represent matchsets, but also the information about which pattern will lead to the lowest covering cost given a specific goal. To distinguish between the two we refer to this extended form of labels as *states*. A state is essentially a representation a specific combination of goals, patterns, and costs, where each possible goal $g$ is associated with a pattern $p$ and a relative cost $c$. A goal in this context typically dictates where the result of an expression must appear (e.g. a particular register class, memory, etc. ), and in grammar terms this means that each nonterminal is associated with a rule and a cost. This combination is such that

1. for any expression tree whose root node has been labeled with a particular state,
2. if the goal of the root node must be $g$,
3. then the entire expression tree can be covered with minimal cost by selecting pattern $p$ at the root. The relative cost of this covering, compared to if the goal had been something else, is equal to $c$.

A key point to understand here is that a state does not necessarily need to carry information about how to optimally cover the *entire* expression tree – indeed, such attempts would require an infinite number of states. Instead the states only convey enough information about how to cover the distinct *key shapes* that can appear in any expression tree. To explain this further, let us observe how most (if not all) target machines operate: between the execution of two machine instructions, the data is synchronized by storing it in registers or in memory. The manner in which some data came to appear in a particular location has in general no impact on the execution of the subsequent machine instructions. Consequently, depending on the available machine instructions one can often break an expression tree at certain key places without compromising code quality. This yields a forest of many, smaller expression trees, each with a specific goal at the root, which then can be optimally covered in isolation. In other words, the set of states only need to collectively represent enough information to communicate where these cuts can be made for all possible expression trees. This does not mean, however, that the expression tree is *actually* chopped into smaller pieces before pattern selection, but thinking about it in this way helps to understand why we can get away with a finite number of states and still get optimal pattern selection.

Since a state is simply an extended form of a label, the process of labeling an expression tree with states is exactly the same as before (see Figure 3.8 on page 39) as we simply need to replace the lookup tables. Pattern selection and code emission is then done as described in

```
SelectAndEmit(expression tree rooted at node n,
              goal-nonterminal g):
  rule r ← RT[state assigned to n, g]
  for each nonterminal l that appears on the right-hand side of r do
      SelectAndEmit(node to reduced to l, l)
  end for
  emit the assembly code associated with r
```

Figure 3.15: Table-driven algorithm for performing optimal pattern selection and code emission (from [208]). *RT* stands for *rule table* which specifies the rule to select given a certain state and goal for optimal covering.

Figure 3.15, which is more or less identical to the algorithm used in conjunction with dynamic programming (compare with Figure 3.14 on page 45). However, we have yet to describe how to go about computing these states.

*First technique to apply offline cost analysis*

According to [26] the idea of offline cost analysis for pattern selection was first introduced by Henry [138] in his 1984 PhD dissertation, and Hatcher and Christopher [134] then appears to have been the pioneers in actually attempting it. However, this approach appears to be fairly unknown as it is little cited in other literature.

Described in their 1986 paper, Hatcher and Christopher's approach is an extension of the work by Hoffmann and O'Donnell. The intuition is to find which rule to apply for some expression tree, whose root node has been assigned a label $l$, such that the entire tree can be reduced to a given nonterminal nt at the lowest cost. Hatcher and Christopher argued that for optimal pattern selection we can consider each pair of a label $l$ and nonterminal nt, and then always apply the rule that will reduce the largest expression tree $T_l$, which is representative of $l$, into nt at the lowest cost. In Hoffmann and O'Donnell's design, where there is only one nullary symbol that may match any subtree, $T_l$ is equal to the largest pattern appearing in the matchset, but to accommodate instruction set grammars Hatcher and Christopher's version includes one nullary symbol per nonterminal. This means that $T_l$ has to be found by overlapping all patterns appearing in the matchset. We then calculate the cost of transforming a larger pattern $p$ into a subsuming, smaller pattern $q$ (i.e. $p > q$) for every pair of patterns. This cost, which is later annotated to the subsumption graph, is calculated by recursively rewriting $p$ using other patterns until it is equal to $q$, making the cost of this transformation equal to the sum of all applied patterns. We represent this cost with a function $\mathit{reducecost}(p \xrightarrow{*} q)$. With this information we retrieve the rule that leads to the lowest-cost reduction of $T_l$ into a goal g by finding the rule $r$ for which

$$\mathit{reducecost}(T_l \xrightarrow{*} \mathrm{g}) = \mathit{reducecost}(T_l \xrightarrow{*} \text{tree pattern of } r) + \text{cost of } r.$$

This will either select the largest pattern appearing in the matchset of $l$, or if one exists a smaller pattern that in combination with others has a lower cost. We have of course glossed over many details, but this covers the main idea of Hatcher and Christopher's approach.

By encoding the selected rules into an additional table to be consulted during pattern matching, we achieve a completely table-driven instruction selector which also performs optimal pattern selection. Hatcher and Christopher also augmented the algorithm for retrieving the matchsets to include the duplicated patterns with commutative operations which were thus omitted from some matchsets. However, if there exist patterns that are truly independent then Hatcher and Christopher's design does not always guarantee that the expression trees can be optimally covered. In addition, it is not clear whether optimal pattern selection for the largest expression trees representative of the labels is an accurate approximation for optimal pattern selection for *all* expression trees.

### *Generating the states using BURS theory*

A different and more well-known method for generating the states was developed by Pelegrí-Llopart and Graham [202]. In their seminal 1988 paper Pelegrí-Llopart and Graham prove that the manners of tree rewriting can always arranged such that all rewrites occur at the leaves of the tree, resulting in a *bottom-up rewriting system* (abbreviated BURS). We say that a collection of such rules constitute a *BURS grammar*, which is similar to the grammars already seen with the exception that BURS grammars allow multiple symbols – including terminals – to appear on the right-hand side of a production. An example of such a grammar is given in (a) of Figure 3.16, and Dold et al. [71, 273] later developed a method for proving the correctness of BURS grammars using abstract state machines.

Using this theory Pelegrí-Llopart and Graham developed an algorithm that computes the tables needed for optimal pattern selection based on a given BURS grammar. The idea is as follows: for a given expression tree $E$, a *local rewrite (LR) graph* is formed where a node represents a specific subtree appearing in $E$, and an edge indicates the application of a particular rewrite rule on that subtree. An example of such a graph is given in (b) of Figure 3.16. Setting some nodes as goals (i.e. desired results of tree rewriting), a subgraph called the *uniquely invertable LR (UI LR) graph* is then selected from the LR graph such that the number of rewrite possibilities is minimized. Finding this UI LR graph is an NP-complete problem, so Pelegrí-Llopart and Graham applied a heuristic that iteratively removes nodes that are deemed as "useless". Each UI LR graph then corresponds to a state, and by generating all LR graphs for all possible expression tree that can be given as input we can find all the necessary states.

| | PATTERN |
|---|---|
| 1 | r → op a a |
| 2 | r → R |
| 3 | r → a |
| 4 | a → r |
| 5 | a → C |
| 6 | a → + C r |
| 7 | C → 0 |
| 8 | x → + x 0 |
| 9 | + y x → + x y |
| 10 | op x y → + x y |

(a) BURS grammar.

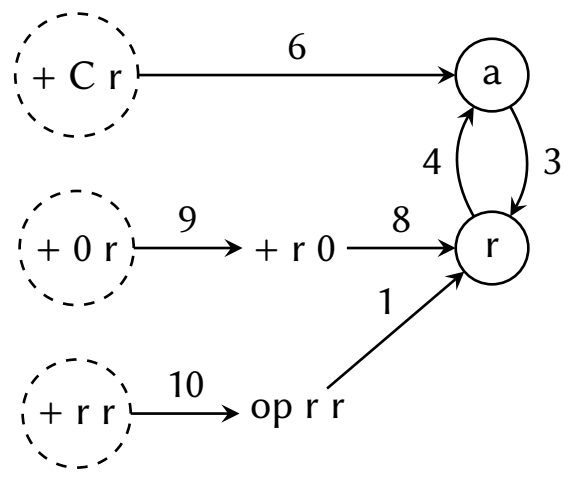

(b) Example of an LR graph based on the expression tree + 0 + C C and the grammar shown in (a). Dashed nodes represent subtrees of the expression tree and fully drawn nodes represent goals. Edges indicate rule applications, with the number of the applied rule appearing next to it.

Figure 3.16: BURS example (from [202]).

|   | PATTERN | COST |
|---|---|---|
| 1 | r → Int | 1 |
| 2 | r → Ld r | 1 |

(a) A rewrite system that may lead to an unbounded number of states. Ld stands for "load".

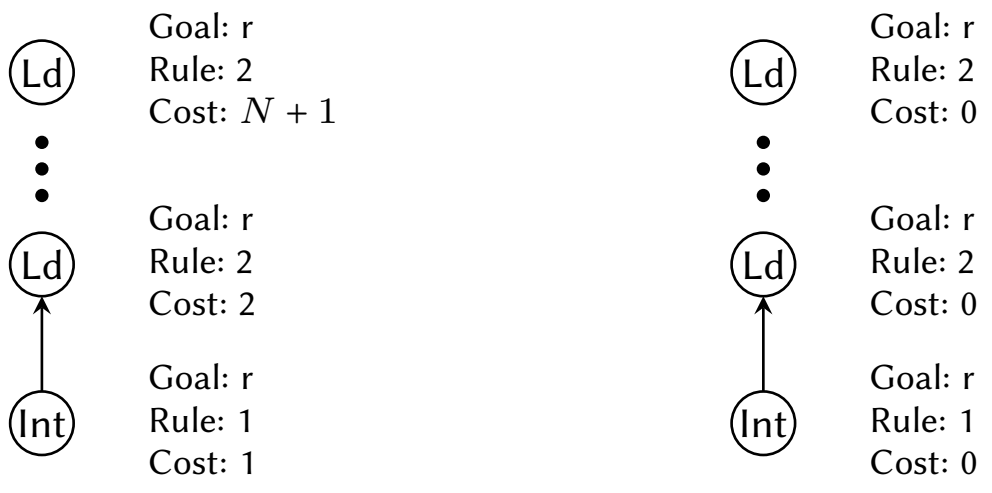

(b) Input tree labeled with states that incorporates the full cost (requires an infinite number of states).

(c) Input tree labeled with states that only incorporates the delta costs (only requires 2 states).

Figure 3.17: Incorporating costs into states (from [202]).

*Achieving a bounded number of states*

To achieve optimal pattern selection the LR graphs are augmented such that each node no longer represents a tree pattern but a $(p, c)$ pair, where $c$ denotes the minimal cost of reaching pattern $p$. This is the information embodied by the states as we discussed earlier. A naïve approach would be to include the *full* cost of reaching a particular pattern into the state, but depending on the rewrite system this may require an infinite number of states. An example where this occurs is given in (b) of Figure 3.17. A better method is to instead account for the *relative* cost of a selected pattern. This is achieved by computing $c$ as the difference between the cost of $p$ and the smallest cost associated with any other pattern appearing in the LR graph. This yields the same optimal pattern selection but the number of needed states is bounded, as seen in (c) of Figure 3.17. This cost is called the *delta cost* and the augmented LR graph is thus known as the *$\delta$-LR graph*. To limit the data size when when generating the $\delta$-LR graphs, Pelegrí-Llopart and Graham used an extension of Chase's table compression algorithm [49] which we discussed on page 43.

During testing Pelegrí-Llopart and Graham reported that their implementation yielded state tables only slightly larger than those produced by LR parsing, and generated assembly code of comparable quality compared to Twig, but at a rate that was about 5x faster.

*BURS $\not\Leftrightarrow$ offline cost analysis*

Since its publication many subsequent papers mistakenly associate to the idea of offline cost analysis with BURS, typically using terms like *BURS states*, when these two aspects are in fact orthogonal to each other. Although the work by Pelegrí-Llopart and Graham undoubtedly led to making offline cost analysis an established aspect of modern instruction selection, the application of BURS theory is only *one* means to achieving optimal pattern selection using tables.

Shortly after Pelegrí-Llopart and Graham introduced BURS, Balachandran et al. [26] published a paper in 1990 that describes an alternative method for generating the states which is both simpler and more efficient. At its heart their algorithm iteratively creates new states using those already committed to appear in the state tables. Remember that each state represents a combination of nonterminals, rules, and costs, where the costs have been normalized such that the lowest cost of any rule appearing in that state is 0. Hence two states are thus identical if the rules selected for all nonterminals and costs are the same. As a new state is created it is checked whether it has already been seen – if so it is dropped, otherwise it is added to the set of committed states – and the process repeats until no new states can be created. We will go into more details shortly.

Compared to Pelegrí-Llopart and Graham, this algorithm is less complicated and also faster as it directly generates a smaller set of states instead of first enumerating all possible states and then reducing them. In addition, Balachandran et al. expressed the machine instructions as a more traditional instruction set grammar – like those used in the Glanville-Graham approach – instead of as a BURS grammar.

*Linear-form grammars*

In the same paper Balachandran et al. also introduce the idea of *linear-form* grammars, which means that right-hand side of every production appearing in the grammar is restricted to one of the following shapes:

1. lhs → Op $n_1 \ldots n_k$: Op is a terminal representing an operator where $rank(k) > 0$, and $n_i$ are all nonterminals. This called a *base rule*.
2. lhs → T: T is a terminal. This is also a base rule.
3. lhs → nt: nt is a nonterminal. This is called a *chain rule*.

A non-linear grammar can easily be rewritten into linear form by introducing new nonterminals and rules for making the necessary transitions, e.g.:

| PRODUCTION | COST | | PRODUCTION | COST |
|---|---|---|---|---|
| reg → Ld + Const Const | 2 | | reg → Ld $n_1$ | 2 |
| | | | $n_1$ → + $n_2$ $n_2$ | 0 |
| | | | $n_2$ → Const | 0 |
| Original grammar | | ⇒ | Linear-form grammar | |

The advantage of this kind of grammars is that the pattern matching problem is then reduced to simply comparing the operator of the root node in the tree pattern with the operator of a node in the expression tree. This also means that the productions appearing in the rules become more uniform, which greatly simplifies the task of generating the states.

*A simpler work-queue approach for state table generation*

Another state generating algorithm similar to Balachandran et al. was proposed by Proebsting [207, 208]. This was also implemented by Fraser et al. [105] in a renowned code generation system called Burg,[7] which since its publication in 1992 has sparked a naming convention within the compiler community that I have chosen to call the *BURGer phenomenon*.[8] Although Balachandran et al. were first, we will continue with studying Proebsting's algorithm as it is better documented.

[7]The keen reader will notice that Fraser et al. also implemented the DP-based system Iburg which was introduced on page 46. The connection between the two is that Iburg began as a testbench for the grammar specification which was to be used as input to Burg. Fraser et al. later recognized that some of the ideas for the testbench showed some merit, and therefore improved and extended them into a stand-alone generator. However, the authors neglected to say in their papers what these acronyms stand for.

The algorithm is centered around a work queue that contains a backlog of states under consideration. Like Balachandran et al. it is assumed that the instruction set grammar is in linear form. The queue is first initialized with the states that can be generated from all possible leaf nodes. A state is then popped from the queue and used in combination with other already-visited states in an attempt to produce new states. This is done by effectively simulating what would happen if a set of nodes, appearing as children to some operator symbol, are labeled with some combination of states which includes the state that was just popped: if the operator node can be labeled using an already existing state then nothing happens; if not then a new appropriate state is created, making sure that all applicable chain rules have been applied to the state as this can affect the costs, and append it to the queue. This is checked for every possible combination of states and operator symbols, and the algorithm terminates when the queue becomes empty, which indicates that all states necessary for the instruction set grammar have been generated.

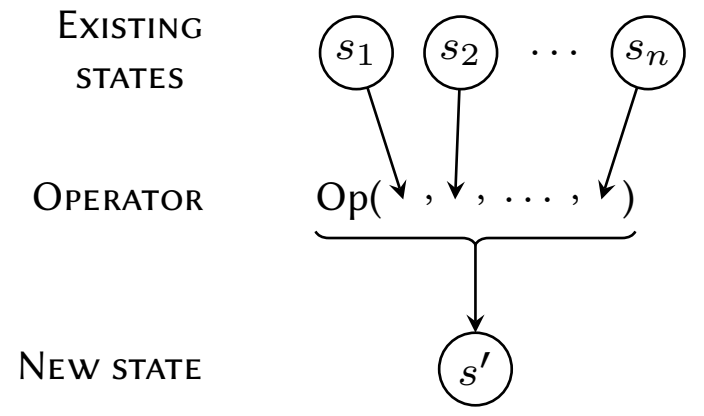

*Further improvements*

The time required to generate the state tables can be decreased if the number of committed states can be minimized. According to [207] the first attempts to do this were made by Henry [137], whose methods were later improved and generalized by Proebsting [207, 208]. Proebsting developed two methods for reducing the number of generated states: *state trimming*, which extends and generalizes the ideas of Henry; and a new technique called *chain rule trimming*. Without going into any details, state trimming increases the likelihood that two created states will be identical by removing the information about nonterminals which can be proven to never take part in a least-cost covering. Chain rule trimming then further minimizes the number of states by attempting to use the same rules whenever possible. This technique was later improved by Kang and Choe [147, 148] who exploited properties of common machine descriptions to decrease the amount of redundant state testing.

*More applications*

The approach of extending pattern selection with offline cost analysis has been applied in numerous compiler-related systems. Some notable applications that we have not already mentioned include: UNH-CODEGEN [136], DCG [83], LBURG [132], and WBURG [209]. BURG has also been made available as a Haskell clone called HBURG [240], and was adapted by Boulytchev [32] to assist instruction set selection, and LBURG is used LCC (Little C compiler)

[8] During the research for this survey I came across the following systems, all with equally creative naming schemes: BURG [105], CBURG [224], DBURG [85], GBURG [103], IBURG [104], JBURG [243], HBURG [240], LBURG [132], MBURG [124, 125], OCAMLBURG [246], and WBURG [209].

[132], and was also adopted by Brandner et al. [35] for developing an architecture description language from which the machine instructions can automatically be inferred.

### 3.6.4 Lazy state table generation

The two main approaches in achieving optimal pattern selection – those who dynamically compute the costs as the input program is compiled, and those who rely on statically computed costs via state tables – both have their respective advantages and drawbacks. The former approaches have the advantage of being able to support dynamic costs, i.e. that the cost of a pattern depends on context rather than remains static, but they are also considerably slower than their purely table-driven counterparts. These latter techniques, however, tend to result in larger instruction selectors due to the use of state tables, which are very time-consuming to generate – for pathological grammars this may even be infeasible – and they only support grammar rules where the costs are fixed.

*Combining the best of state tables and DP*

In 2006 Ertl et al. [86] developed a method that allows the state tables to be generated lazily and on demand. The intuition is that instead of generating the states for *all* possible expression trees in advance, one can get away with only generating the states needed for the expression trees that actually appear in the input program.

The approach can be outlined as follows: as the instruction selector traverses an expression tree, the states required for covering its subtrees are created using dynamic programming. Once the states have been generated the subtree is labeled and patterns selected using the familiar table-driven techniques. Then, if an identical subtree appears elsewhere – either in the same expression tree or in another of the input program – the same states can be reused. This allows the cost of state generation to be amortized as the subtree can now be optimally covered faster compared to if it had been processed using a purely DP-based pattern selector. Ertl et al. reported the overhead of state reuse was minimal compared to purely table-driven implementations, and the time required to first compute the states and then label the expression trees was on-par with selecting patterns using ordinary DP-based approaches. Moreover, by generating the states lazily it is possible to handle larger and more complex instruction set grammars which otherwise would require an unmanageable number of states.

Ertl et al. also extended this approach to handle dynamic costs by recomputing and storing the states in hash tables whenever the root costs differ. This incurs an additional overhead, the authors reported, but still performed faster when compared to a purely DP-based instruction selector.

## 3.7 Other techniques for covering trees

So far in this chapter we have discussed the conventional approaches for covering trees: LR parsing, top-down recursion, dynamic programming, and the use of state tables. In this section we will look other techniques which are also founded on this principle, but solve it using more unorthodox methods.

### 3.7.1 Approaches based on formal frameworks

*Homomorphisms and inversion of derivors*

In a 1988 paper Giegerich and Schmal [120] propose an algebraic framework intended to allow all aspects of code generation – i.e. not only instruction selection but also instruction scheduling and register allocation – to be formally described in order to simplify machine descriptions and enable formal verification. In brief terms, Giegerich and Schmal reformulated the problem of instruction selection into a "problem of a hierarchic derivor", which essentially entails the specification and implementation of a mechanism

$$\gamma : T(Q) \to T(Z)$$

where $T(Q)$ and $T(Z)$ denote the term algebras for expressing programs in an intermediate respectively target machine language. Hence $\gamma$ can be viewed as the resultant instruction selector. However, machine descriptions typically contain rules where each machine instruction in $Z$ is expressed in terms of $Q$. We therefore view the machine specification as a *homomorphism*

$$\delta : T(Z) \to T(Q)$$

and the task of an instruction selection-generator is thus to derive $\gamma$ by inverting $\delta$. Usually this is achieved by resorting to pattern matching. For optimal instruction selection the generator must also interleave the construction of the inverse of $\delta$ (i.e. $\delta^{-1}$) with a *choice function* $\xi$ whenever some $q \in T(Q)$ has several $z \in T(Z)$ such that $\delta(q) = z$. Conceptually, this gives us the following functionality:

$$T(Q) \xrightarrow{\delta^{-1}} 2^{T(Z)} \xrightarrow{\xi} T(Z)$$

In the same paper Giegerich and Schmal also demonstrate how some other approaches, such as tree parsing, can be expressed using this framework. A similar approach, based on rewriting techniques, was later proposed by Despland et al. [66, 67] in an implementation called Pagode [41]. Readers interested in more details are advised to consult the referenced papers.

*Equational logic*

In 1991 Hatcher [135] developed an approach similar to that of Pelegrí-Llopart and Graham but relies on equational logic [199] instead of BURS theory. The two are closely related in that both apply a set of predefined rules – i.e. those derived from the machine instructions in combination with axiomatic transformation – to rewrite an input program into a single-goal term. However, in an equational specification has the advantage that all rules are based on a set of *built-in operations*, which each has a cost and implicit semantics expressed as code emission. The cost of a rule is then equal to the sum of all built-in operations applied in the rule, thus removing the need to set these manually. In addition, no built-in operations are predefined but are instead given as part of the equational specification which allows for a very general mechanism for describing the target machines. Experimental results from selected problems, with an implementation called UCG – the paper does not say what this acronym stands for – indicated that assembly code of comparable quality could be generated in less time compared to contemporary techniques.

### 3.7.2 Tree automata and series transducers

In the same manner that the Glanville-Graham approach adopts techniques derived from syntax parsing, Ferdinand et al. [94] recognized that the mature theories of finite tree automata [117] can be used to address instruction selection. In a 1994 paper Ferdinand et al. describe how the problems of pattern matching and pattern selection – with or without optimal solutions – can be solved using tree automata, and also present algorithms for generating them as well as an automaton which incorporates both. An experimental implementation demonstrated the feasibility of this approach, but the results were not compared to those of other techniques. A similar design was later proposed by Borchardt [31] who made use of tree series transducers [81] instead of tree automata.

### 3.7.3 Rewriting strategies

In 2002 Bravenboer and Visser presented a design where rule-based program transformation systems [252] are adapted to instruction selection. Through a system called Stratego [253], a machine description can be augmented to contain the pattern selector itself, allowing it to be tailored to the specific instruction set of the target machine. Bravenboer and Visser refer to this as providing a *rewriting strategy*, and their system allows the modeling of several strategies such as exhaustive search, maximum munch, and dynamic programming. Purely table-driven techniques, however, are apparently not supported, thus excluding the application of offline cost analysis. Bravenboer and Visser argue that this setup allows several pattern selection techniques to be combined, but their paper does not contain an example where this would be beneficial.

### 3.7.4 Genetic algorithms

Shu et al. employed the theories of *genetic algorithms* (GA) [122, 219] – which attempt to emulate natural evolution – to solve the pattern selection problem. Described in a 1996 paper, the idea is to formulate a solution as a DNA string – called a *chromosome* – and then split, merge, and mutate them in order to hopefully end up with better solutions. For a given expression tree – whose matchsets have been found using an $\mathcal{O}(nm)$ pattern matcher – Shu et al. formulated each chromosome as a binary bit string where a 1 indicates the selection of a particular pattern. Likewise, a 0 indicates that the pattern is not used in the tree covering. The length of a chromosome is therefore equal to the sum of the number of patterns appearing in all matchsets. The objective is then to find the chromosome which maximizes a *fitness function* $f$, which Shu et al. defined as

$$f(c) = \frac{1}{k * p_c + n_c}$$

where $k$ is a tweakable constant greater than 1, $p_c$ is the number of selected patterns in the chromosome $c$, and $n_c$ is the number of nodes in $c$ which are covered by more than one pattern. First fixed number of chromosomes are randomly generated and evaluated. The best ones are kept and subjected standard GA operations, such as fitness-proportionate reproduction, single-point crossover, and one-bit mutations, in order to produce new chromosomes. The process then repeats until some termination criteria is hit. When applied to medium-sized expression trees (at most 50 nodes), the authors claimed to be able to find optimal tree coverings in "reasonable" time.

### 3.7.5 Trellis diagrams

The last approach that we will examine in this chapter is a rather unusual approach by Wess [259, 260]. Specifically targeting digital signal processors, Wess's design integrates instruction selection with register allocation through the use of *trellis diagrams*.

A trellis diagram is a graph where each node consists of an *optimal value array* (OVA). An element in the OVA represents either that the data is stored in memory (m) or in a particular register ($r_x$), and its value indicates the lowest accumulated cost from the leaves to this node. This is computed similarly as in the dynamic programming technique. If the data is stored in a register, the OVA element also indicates which other registers are available for that operation. For example, a target machine with two registers – $r_1$ and $r_2$ – yields the following OVA:

| $i$ | 0 | 1 | 2 | 3 | 4 |
|---|---|---|---|---|---|
| OVA | | | | | |
| $\mathrm{TL}_i$ | m | $r_1$ | $r_1$ | $r_2$ | $r_2$ |
| $\mathrm{AL}_i$ | $\{r_1, r_2\}$ | $\{r_1\}$ | $\{r_1, r_2\}$ | $\{r_2\}$ | $\{r_1, r_2\}$ |

where $\mathrm{TL}_i$ represents the target location of the data, and $\mathrm{AL}_i$ represents the available locations.

We create the trellis diagrams using the following scheme. For each node in the expression tree a new node representing an OVA is added to the trellis diagram. Two trellis diagram nodes are added for nodes that appear as leaves in the expression tree in order to handle situations where the input values first need to be transferred to another location before being used (this is needed if the input value resides in memory). Let us now denote $e(i, n)$ as the OVA element $i$ at a node $n$. For unary operation nodes, an edge between $e(i, n)$ and $e(j, m)$ is added to the trellis diagram if there exists an instruction that takes the value stored in the location indicated by $j$ in $m$, executes the operation of $n$, and stores the result in the location indicated by $i$. Similarly, for a binary operation node $o$ an edge pair is drawn from $e(i, n)$ and $e(j, m)$ to $e(k, o)$ if there exists such a corresponding instruction which stores the result in $k$. This can be generalized to $n$-ary operations, and a full example is given in Figure 3.18.

The edges in the trellis diagram thus correspond to the possible combinations of machine instructions and register allocations that implement a particular operation in the expression tree, and a path from every leaf node in the trellis diagram to its root represents a selection of such combinations which collectively implement the entire expression tree. By keeping track of the costs, we can get the optimal instruction sequence by selecting the path which ends the OVA element of the trellis diagram root with the lowest cost.

The strength of Wess's approach is that target machines with asymmetric register files (i.e. where different instructions are needed for accessing different registers) are easily handled as instruction selection and register allocation is done simultaneously. However, the number of nodes in the trellis diagram is exponential to the number of registers. This problem was mitigated by Fröhlich et al. [108] who augmented the algorithm to build the trellis diagram in

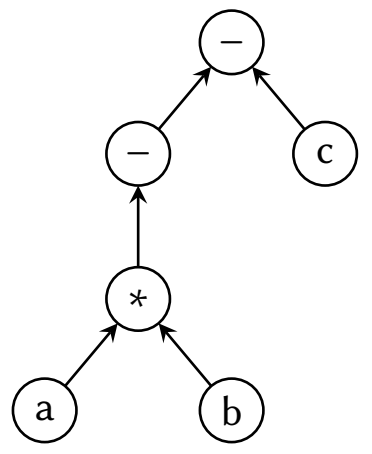

(a) Expression tree.

| | INSTRUCTION |
|---|---|
| 1 | $r_1 * r_2 \rightarrow r_1$ |
| 2 | $r_1 * m \rightarrow r_1$ |
| 3 | $r_1 - m \rightarrow r_1$ |
| 4 | $-r_1 \rightarrow r_1$ |
| 5 | $m \rightarrow r_1$ |
| 6 | $m \rightarrow r_2$ |
| 7 | $r_1 \rightarrow m$ |
| 8 | $r_2 \rightarrow m$ |
| 9 | $r_1 \rightarrow r_2$ |
| 10 | $r_2 \rightarrow r_1$ |

(b) Instruction set. Note that these are not grammar productions as the symbols refer to explicit registers and not register classes. All instructions are assumed have equal cost.

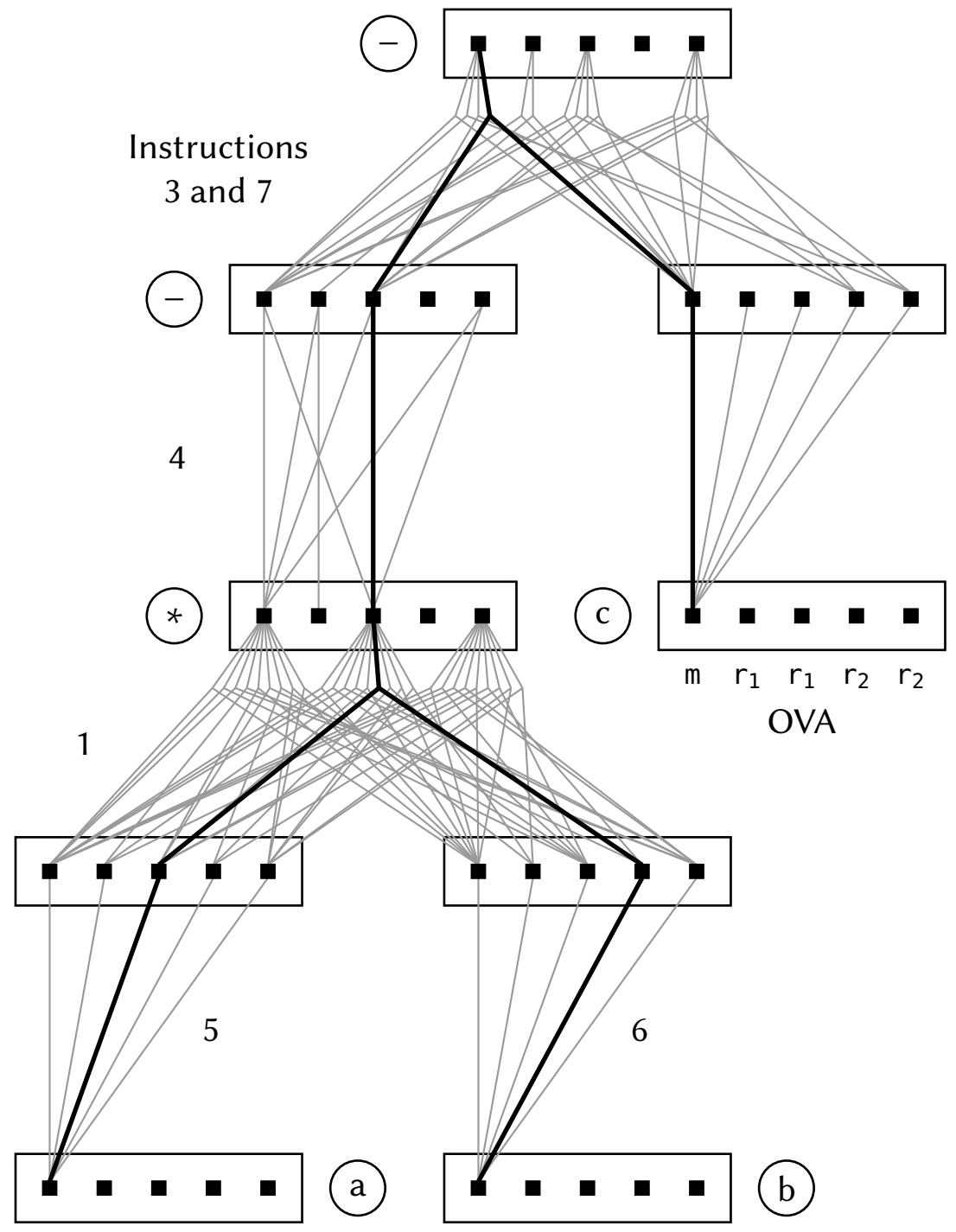

(c) Trellis diagram. Gray edges represent available paths, and black edges indicate the (selected) optimal path.

Figure 3.18: A trellis diagram applied on an expression $-(a * b) - c$ for a the two-register target machine described in the text. The variables $a$, $b$, and $c$ are assumed to be initially stored in memory, which is handled by inserting additional OVAs into the trellis diagram to allow selection of necessary data transfer instructions.

a lazy fashion. However, these approaches require a one-to-one mapping between expression nodes and machine instructions in order to be effective.[9]

## 3.8 Summary

In this chapter we have looked at numerous approaches which are based on the principle of tree covering. In contrast to macro expansion, these approaches allow usage of more complex patterns which in turn yields more efficient instruction selectors; by applying dynamic programming, optimal code can be generated in linear time. Several techniques also incorporate offline cost analysis into the table generator which further speeds up code generation. In other words, these approaches are very fast and efficient while supporting code generation for a wide array of target machines. Consequently, tree covering has become the most known – if perhaps no longer the most applied – principle of instruction selection. Restricting oneself to trees, however, comes with several inherent disadvantages.

[9]This, in combination of how instructions are selected, makes one wonder whether these approaches actually conform the principles of tree and DAG covering; I certainly struggled with deciding how to categorize them. Finally, though, I opted against putting it in separate category as that would become a very short one indeed.

### 3.8.1 Restrictions that come with trees

The first disadvantage of trees has to do with expression modeling: due to the definitions of trees, common subexpressions cannot be properly modeled in an expression tree. For example, the inlined code cannot be modeled directly without applying one of the following work-around:

```
x = a + b;
y = x + x;
```

1. Repeating the shared operations, which in Polish notation results in

$$= y + + a\ b + a\ b$$

2. Splitting the expression into a forest, resulting in

$$= x + a\ b$$
$$= y + x\ x$$

The first technique leads to additional machine instructions in the assembly code, while the second approach hinders the use of more complex machine instructions; in both cases code quality is compromised.

Since trees only allowed a single root node, multi-output machine instructions cannot be represented as tree patterns since such instructions require multiple root nodes. Consequently, such features – which appear in many ISAs as this also includes instructions that set condition flags – cannot be exploited during instruction selection. Even disjoint-output instructions whose each individual operation can be modeled as trees can also not be selected as the instruction selector can only consider a single tree pattern at a time.

Another disadvantage is that expression trees typically cannot model control flow. For example, a `for` statement requires a loop edge between basic blocks, which violating the definition of trees. For this reason the scope of tree-based instruction selectors as they are limited to selecting instructions for a single expression tree at a time, and handle code emission for control flow using a separate component. This in turn excludes matching and selection of internal-loop instructions, whose behavior incorporate some kind of control flow

To summarize, although the principle of tree covering greatly improve code quality over the principle of pure macro expansion (i.e. ignoring peephole optimization), the inherent restrictions of trees prevents full exploitation of the machine instructions provided by most target machines. In the next chapter we will look at a more general principle that addresses some of these issues.

# 4

# DAG Covering

As we saw in the previous chapter the principle of tree covering comes with several disadvantages: common subexpressions cannot be properly expressed, and many machine characteristics such as multi-output instructions cannot be modeled. Since these restrictions are due to relying solely on trees, we can lift them by extending tree covering to *DAG covering*.

## 4.1 The principle

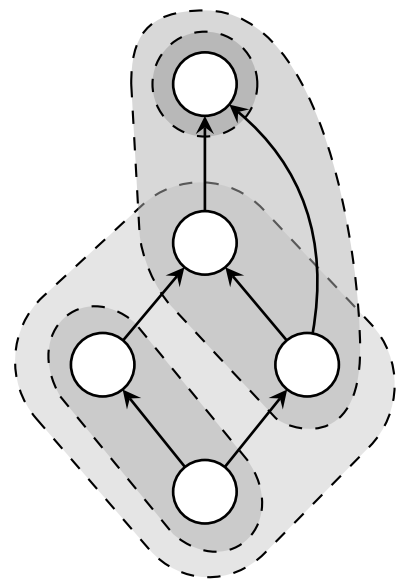

If we remove the restriction that every pair of nodes in a tree must have exactly one path, we get a new graph shape called a *directed acyclic graph*, commonly known as DAGs. As nodes are now allowed multiple outgoing edges, intermediate values of an expression can now be shared and reused within the same graph, as well as contain multiple root nodes. To distinguish between expression modeled as trees from those modeled as DAGs, we refer to the latter as *expression DAGs*. Once this has been formed instruction selection proceeds with applying the same concepts of pattern matching and pattern selection that we learned from the previous chapter. As patterns are allowed multiple root nodes, we can model and exploit a larger set of machine instructions, such as multi-output instructions like `divmod` which compute the quotient and remainder simultaneously.

## 4.2 Optimal pattern selection on DAGs is NP-complete

Unfortunately, the cost of this gain in generality and modeling capabilities that DAGs give us is a substantial increase in complexity. While selecting an optimal set of patterns over trees can be done in linear time, doing the same for DAGs is an NP-complete problem. Bruno and Sethi [38] and Aho et al. [4] proved this already in 1976, although both were most concerned with the optimality of instruction scheduling and register allocation. In 1995 Proebsting [206] gave a very concise proof for optimal instruction selection, which was later reformulated by Koes and Goldstein [161] in 2008. We will paraphrase the proof of Koes and Goldstein in this report. Note that this does not necessarily involve the task of pattern matching – as we will see, this can still be done in polynomial time if the patterns consist of trees.

### 4.2.1 The proof

The idea is to reduce the SAT (Boolean satisfiability) problem to an optimal (i.e. least-cost) DAG covering problem. The SAT problem is to decide whether a Boolean formula in *conjunctive normal form* (CNF) can be satisfied. A CNF expression is a formula in the form

$$(x_{11} \vee x_{12} \vee \ldots) \wedge (x_{21} \vee x_{22} \vee \ldots) \wedge \ldots$$

A variable $x$ can also be negated via $\neg x$. Since this problem has proven to be NP-complete [62], all polynomial-time reductions of SAT to any other problem $\mathcal{X}$ must also render $\mathcal{X}$ as NP-complete.

*Modeling SAT as a covering problem*

First we transform a SAT problem instance $S$ into a DAG. The intuition is that if we can cover the DAG with unit cost patterns such that the total cost is equal to the number of nodes (assuming that each variable and operator incurs a new node) then there exists a truth assignment to the variables for which the formula evaluates to $T$. For this purpose, the nodes in the DAG can be of types $\{\vee, \wedge, \neg, v, \square, \bigcirc\}$. We define *type*$(n)$ as the type of a node $n$ and refer to nodes of type $\square$ and $\bigcirc$ as *box nodes* and *stop nodes*, respectively. We also define *children*$(n)$ as the set of child nodes to $n$. For every Boolean variable $x \in S$ we create two nodes $n_1$ and $n_2$ such that *type*$(n_1) = v$ and *type*$(n_2) = \square$, and a directed edge $n_1 \rightarrow n_2$. We do the same for every binary Boolean operator $op \in S$ by creating two nodes $n'_1$ and $n'_2$ such that *type*$(n'_1) = op$ and *type*$(n'_2) = \square$, along with an edge $n'_1 \rightarrow n'_2$. We connect the operation with its input operands $x$ and $y$ via the box nodes of the operands, i.e. with two edges $n_x \rightarrow n'_1$ and $n_y \rightarrow n'_1$ where *type*$(n_x)$ = *type*$(n_y) = \square$. For the unary operation $\neg$ we obviously only need one such edge, and since Boolean *or* and *and* are commutative it does not matter in which order the edges are arranged with respect to the operator node. Consequently, only box nodes will have more than one outgoing edge.

By simply traversing over the Boolean formula we can construct the corresponding DAG in linear time. An example of a SAT problem translated to a DAG covering problem is given in (b) of Figure 4.1.

*Boolean operations as patterns*

We then construct a set of tree patterns $P_{\text{SAT}}$ which will allow us to deduce how the variables in $S$ should be set in order to satisfy the expression (see (a) of Figure 4.1). The patterns are formed such that it has a box node as leaf if that value is assumed to be set to $T$ (true). Moreover, the pattern has a box node as root if the result of the operation evaluates to $F$ (false). One way of looking at it is that if an operator in a pattern *consumes* a box node then that value must be set to $T$, and if the operator *produces* a box node then the result must evaluate to $F$. To force the entire expression to evaluate to $T$, the only pattern containing a stop node (i.e. of type $\bigcirc$) also consumes a box node.

In addition to the node types that can appear in the expression DAG, pattern nodes may also use an additional node type $\bullet$ which we will refer to as *anchor nodes*. We now say that a

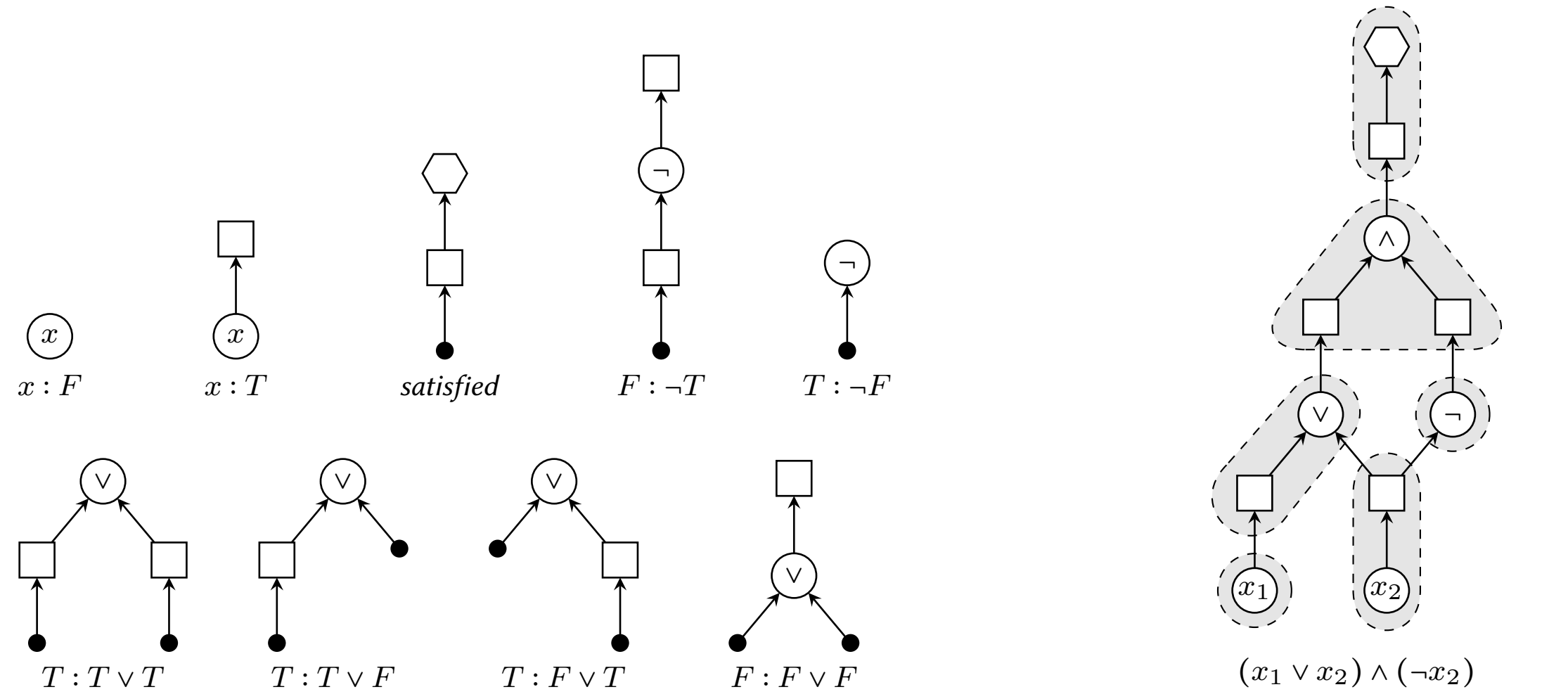

(a) SAT patterns. All patterns have the same unit cost. For brevity, the patterns for the ∧ operation are omitted (but can be easily inferred from the ∨ patterns).

(b) Example of a SAT problem represented as a DAG covering problem.

Figure 4.1: Reducing SAT to DAG covering (from [161]).

tree pattern with root $p_r$ *matches* a node $v \in V$, where $V$ is the set of nodes in the expression DAG $D = (V, E)$, if and only if:

1. $\mathit{type}(v) = \mathit{type}(p_r)$,
2. $|\mathit{children}(v)| = |\mathit{children}(p_r)|$, and
3. $\forall c_v \in \mathit{children}(v), c_r \in \mathit{children}(p_r) : (\mathit{type}(c_r) = \bullet) \vee (c_r \text{ matches } c_v)$.

In other words, the structure of the tree pattern must correspond to the structure of the matched subgraph, with the exception of anchor nodes that may match at any node. We also introduce two new definitions: *matchset*$(v)$, which is the set of patterns in $P_{\text{SAT}}$ that match $v$; and $m_{p_i \rightarrow v}(v_p)$, which is the DAG node $v \in V$ that corresponds to the pattern node $v_p$ in a matched pattern $p_i$. Lastly, we say that a DAG $D = (V, E)$ is *covered* by a mapping function $f : V \rightarrow 2^T$ from DAG nodes to patterns if and only if $\forall v \in V$:

1. $p \in f(v) \Rightarrow p$ matches $v$ and $p \in P_{\text{SAT}}$,
2. $\mathit{in\text{-}degree}(v) = 0 \Rightarrow |f(v)| > 0$, and
3. $\forall p \in f(v), \forall v_p \in p$ s.t. $\mathit{type}(v_p) = \bullet \Rightarrow |f(m_{p_i \rightarrow v}(v_p))| > 0$.

The first constraint enforces that each selected pattern matches and is an actual pattern. The second constraint enforces matching of the stop node, and the third constraint enforces matching on the rest of the DAG. An *optimal cover* of $D = (V, E)$ is thus a mapping $f$ that covers $D$ and minimizes

$$\sum_{v \in V} \sum_{p \in f(v)} \mathit{cost}(p)$$

where $\mathit{cost}(t)$ is the cost of pattern $p$.

*Optimal solution to DAG covering ⇒ solution to SAT*

We now postulate that if the optimal cover has a total cost equal to the number of non-boxed nodes in the DAG, then the corresponding SAT problem is satisfiable. Since all patterns in $P_{\mathrm{SAT}}$ cover exactly one non-box node and have equal unit cost, then if every non-box node in the DAG is covered by exactly one pattern then we can easily infer the truth assignments to the Boolean variables simply by looking at which patterns were selected to cover the variable nodes.

We have thereby shown that an instance of the SAT problem can be solved by reducing it, in polynomial time, to an instance of the optimal DAG covering problem. Hence optimal DAG covering – and therefore also optimal instruction selection based on DAG covering – is NP-complete. □

## 4.3 Covering DAGs with tree patterns

Aho et al. [4] were among the first in literature to provide code generation approaches that operate on DAGs. In their paper, published in 1976, Aho et al. propose some simple, greedy heuristics as well as an optimal code generator that produces code for a commutative one-register machine. However, their methods assume a one-to-one mapping between expression nodes and machine instructions and thus effectively ignores the problem of instruction selection.

Since instruction selection on DAGs with optimal pattern selection is computationally difficult, the first proper approaches based on this principle use DAGs to model the expressions of the input program but keep the patterns as trees. By doing so the developers could adopt already-known, linear-time methods from tree covering and extend them to DAG covering. The expression DAGs are then transformed into trees by resolving common subexpression according to the means we discussed in the summary of the previous chapter. We refer to this as *undagging*.

### 4.3.1 Undagging the expression DAG into a tree

An expression DAG can be undagged into a tree in two ways. The first method is to split the edges from the shared nodes (i.e. where reuse occurs), resulting in a set of disconnected expression trees which can then be covered individually using common tree covering techniques. An implicit connection is maintained between the trees by forcing the result of the expression tree rooted at the previously-shared node to be written to a specific location (typically main memory). This is then used by the other expression trees, which may need to be augmented to reflect this new source of input. An example of an implementation that uses this approach is Dagon, a technology binder developed by Keutzer [156] which maps technology-independent descriptions onto circuits. The second method is to duplicate the nodes where reuse occurs, which potentially also results many trees. Both concepts are illustrated in Figure 4.2.

While undagging allows instruction selection to be done on DAGs with tree-based covering techniques, it does so at a cost of efficiency loss: too aggressive edge splitting inhibits the application of large patterns as it produces many small trees; and too aggressive node duplication leads to inefficient assembly code as many operations are needlessly re-executed in the final program. Moreover, the intermediate results of a split DAG must be forcibly stored, which can be troublesome for heterogeneous memory-register architectures. This last problem was examined by Araujo et al. [17].

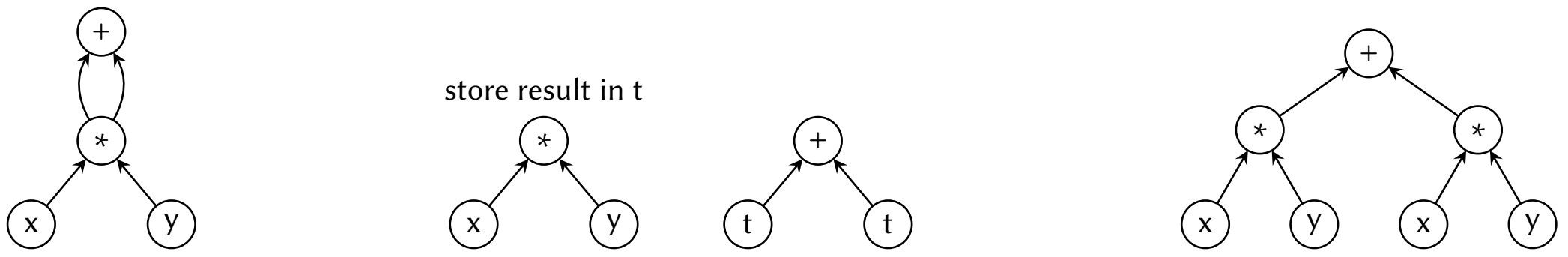

(a) DAG. (b) Result of edge splitting. (c) Result of node duplication.

Figure 4.2: Undagging an expression DAG with a common subexpression.

*Balancing splitting and duplication*

Fauth et al. [92, 191] proposed in 1994 an approach that tries to mitigate the deficiency of undagging by balancing node duplication and edge splitting. Implemented in CBC (Common Bus Compiler), the instruction selector applies an heuristic algorithm that favors duplication until it is deemed too costly; when this happens the algorithm resorts to splitting. The decision between whether to duplicate or split is taken by comparing the cost between the two solutions and selecting the cheapest one. The cost is calculated as a weighted sum between the number of nodes in the expression DAG and the expected number of nodes executed along each execution path (a rough estimate of code size and execution time, respectively). Once this is done each resulting expression tree is covered by an improved version of Iburg (see page 46) with extended match condition support. However, the experimental data is too limited to judge how efficient this technique is compared to if the expression DAGs had been transformed into trees using only one method and not the other.

*Selecting instructions with the goal to assist register allocation*

In 2001 Sarkar et al. [222] developed a greedy approach to instruction selection with the aim of reducing *register pressure* – i.e. the number of vacant registers demanded by the assembly code – in order to facilitate the subsequent instruction scheduling and register allocation. Hence, instead of the usual number of execution cycles, the cost of each machine instruction reflects the amount of register pressure that the instruction incurs if selected (the paper does not go into any details on how these costs are formulated). Instruction selection is done by first undagging the expression DAG – which has been augmented into a graph with additional data dependencies – into a forest using splitting, and then applying conventional tree covering methods on each expression tree. A heuristic is applied on deciding where to perform these cuts. Once patterns have been selected, the nodes which are covered by patterns consisting of more than one node are reduced to *super nodes*. The algorithm then checks whether the resultant graph contains any cycles; if it does then the covering is illegal as there exist cyclic data dependencies. The cuts are then revised and the process is repeated until a valid covering is achieved.

Sarkar et al. implemented their register-sensitive approach in Jalapeño – a register-based Java virtual machine developed by IBM – and achieved for a small set of problems a 10% performance increase due to the use of fewer spill instructions compared to the default instruction selector.

### 4.3.2 Extending dynamic programming to DAGs

In a 1999 paper Ertl [85] presents an approach where conventional tree covering methods are applied to DAGs without having to first break them up into trees. The idea is to first make a bottom-up pass over the DAG *as if* it was a tree, and compute the costs using the conventional dynamic programming method of pattern selection which we discussed in the previous chapter in Section 3.6.2. Each node is thus labeled with the same costs as if the DAG had first been reduced to a tree through node duplication. However, Ertl recognized that if several patterns reduce the same node to the same nonterminal, then the reduction to that nonterminal can be shared between rules where this symbol appears on the right-hand side of the production. This is then exploited during code emission by emitting assembly code only once for such nonterminals, which decreases code size while improving performance as the number of duplicated operations is reduced. An example illustrating this situation is given in Figure 4.3. There we see that the addition operation will be implemented twice as it it covered by two separate patterns that each reduces the subtree to a different nonterminal. The Reg node, on the other hand, is reduced twice to the same nonterminal (i.e. reg), and can thus be shared between the rules where this symbol appears in the patterns. With appropriate data structures this can be done in linear time.

For certain instruction set grammars Ertl's technique guarantees optimal pattern selection, and Ertl devised a checker called Dburg which detects when the grammar does not belong into this category. The basic idea behind Dburg is to check whether every locally optimal decision is also globally optimal, and perform inductive proofs over the set of all possible input DAGs. To do this efficiently Ertl implemented Dburg using ideas of Burg (hence the name).

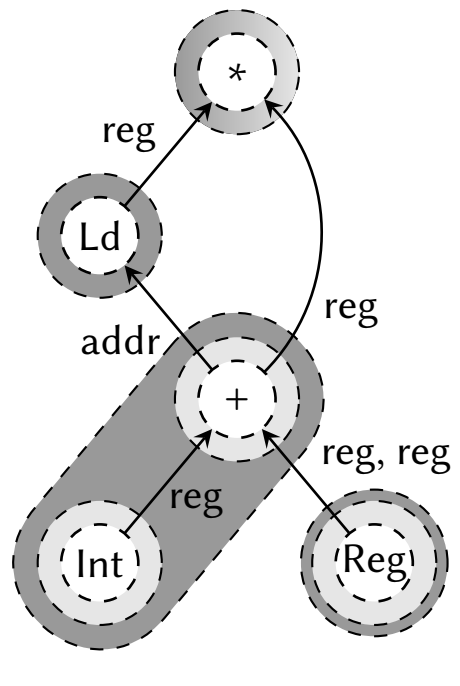

Figure 4.3: A tree covering of a DAG, where the optimal patterns have been selected. The two shades indicate the relation between rules, and the text along the edges indicate the nonterminals to which each pattern is reduced to. Note that the Reg node is covered by two patterns (as indicated by the double dash pattern) which both reduce to the same nonterminal which can be shared.

*Further developments*

Koes and Goldstein [161] recently extended Ertl's ideas by providing a heuristic similar to Fauth et al. that splits the expression DAG at points where duplication is estimated to have a detrimental effect on code quality. Like Ertl, Koes and Goldstein's algorithm first selects optimal patterns by performing a tree-like bottom-up pass, ignoring the fact that the input is a DAG. Then, at points where multiple patterns overlap two costs are calculated: *overlap-cost*, which is an estimate of the cost of letting the patterns overlap and thus incur duplication of operations in the final assembly code; and *cse-cost*, which is an estimate of the cost of instead splitting the edges at such points. If *cse-cost* is lower than *overlap-cost* then the node where overlapping occurs is marked as *fixed*. Once all such nodes have been processed, a second bottom-up DP pass is performed on the expression DAG, but this time no pattern is allowed to span over fixed nodes (i.e. a fixed node can only be matched at the root of a pattern). Lastly, a top-down pass emits the assembly code. Koes and Goldstein also compared their own implementation, called NOLTIS, against an *integer programming* (IP) implementation, and found that NOLTIS achieved optimal pattern selection in 99.7% of the test cases.

### 4.3.3 MATCHING TREE PATTERNS DIRECTLY ON DAGS

Unlike the approaches above – which first convert the expression DAGs into trees before covering – there also exist techniques which match the tree patterns directly on the DAGs.

*Using Weingart-like pattern trees*

In a highly influential 1994 paper Liem et al. [179, 201, 205] presented a design which they implemented in CODESYN, a well-known code synthesis system which in turn is part of an embedded systems development environment called FLEXWARE. The approach uses the same pattern matching technique as Weingart [256] – which we discussed in the previous chapter – by combining all available tree patterns into a single pattern tree. Using an $\mathcal{O}(nm)$ pattern matcher, all matchsets are found by traversing the pattern tree in tandem with the expression DAG. Pattern selection is then done by the dynamic programming technique which we know from before (although how they went about adapting it to DAGs is not stated). However, due to the use of a single pattern tree, it is doubtful that the approach can be extended to handle DAG patterns.

*LLVM*

Another direct trees-on-DAGs approach is applied in the well-known LLVM [167] compiler infrastructure.[1] According to a blog entry by Bendersky [30] – which at the time of writing provided the only documentation except for the LLVM code itself – the instruction selector is basically a greedy DAG-to-DAG rewriter, where machine-independent DAG representations of basic blocks are rewritten into machine-dependent DAG representations. Operations for branches and jumps are included into the expression DAG and pattern-matched just like any other node. The patterns, which are limited to trees, are expressed in a machine description format that allows common features to be factored out into abstract instructions. Using a tool called TableGen, the machine description is then expanded into complete tree patterns which are processed by a matcher generator. The matcher generator first performs a lexicographical sort on the patterns: first by decreasing complexity, which is the sum of the pattern size and a constant (this can be tweaked to give higher priority for particular machine instructions); then by increasing cost; and lastly by increasing size of the output pattern. Once sorted each pattern is converted into a recursive matcher which is essentially a small program that checks whether the pattern matches at a given node in the expression DAG. The generated matchers are then compiled into a byte-code format and assembled into a matcher table which is consulted during instruction selection. The table is arranged such that the patterns are checked in the order of the lexicographical sort. As a match is found the pattern is greedily selected and the matched subgraph is replaced with the output (usually a single node) of the matched pattern.

[1] It is also very dissimilar to GCC which applies macro expansion combined with peephole optimization (see Section 2.3.2).

Although powerful and in extensive use, LLVM's instruction selector has several drawbacks. The main disadvantage is that any pattern that cannot be handled by TableGen has to be handled manually through custom C functions. Since patterns are restricted to tree shapes this includes all multiple-output patterns. In addition, the greedy scheme compromises code quality.

We see that a common and recurring disadvantage of the discussed approaches is that they are all isolated to handling only patterns shaped as trees. We therefore proceed with techniques where also patterns are allowed the form of DAGs.

### 4.3.4 Selecting patterns with integer programming

In an attempt to achieve an integrated approach to code generation – i.e. where instruction selection, instruction scheduling, and register allocation is done simultaneously – Wilson et al. [263] formulated these problems as an *integer programming* (IP) model. Integer programming [50, 226] is a method for solving combinatorial optimization problems, and with their 1994 paper Wilson et al. appear to have been the first to use IP for code generation.

Valid pattern selection can be expressed as the following linear inequality

$$\forall n \in G : \sum_{p \in P_i} z_p \leq 1$$

which reads: for every node $n$ in the expression DAG $G$, at most one pattern $p$ from the matchset of $n$ ($P_i$) may be selected ($z_p$, which is a Boolean variable).[2] Similar linear inequalities can be formulated for expressing the constraints of instruction scheduling and register allocation – which were also incorporated into the IP model by Wilson et al. – but these are out of scope for this report. In addition, the model can be extended with whatever additional constraints are demanded by the target machine simply by adding more linear inequalities, making it suitable for code generation for target machines with irregular architectures as well as supporting interdependent machine instructions.

[2]The more common constraint is that *exactly one* pattern must be selected, but Wilson et al. allowed nodes become inactive and thus need not be covered.

Solving this monolithic IP model typically requires considerably more time compared to conventional instruction selection techniques, but the ability to extend the model with additional constraints is a much-valued feature for handling complicated target machines, which anyway cannot be properly modeled using traditional, heuristic approaches. In addition, since the IP model incorporates all aspects of code generation, Wilson et al. reported that, for a set of test cases, the generated assembly code was of comparable code quality to that of hand-optimized assembly code, which means that an optimal solution to the IP problem can be regarded as an optimal solution to code generation. Consequently, several later techniques which we will examine in this chapter are founded on this pioneering work.

### 4.3.5 Modeling entire processors as DAGs

In an approach for compiling input programs into microcode,[3] Marwedel [185] developed a retargetable system called MSS (Mimola Software System). Mimola [272], which stands for "Machine Independent MicrOprogramming LAnguage", is a description language for modeling the entire data paths of the processor, instead of just the instruction set as we have commonly seen. This is commonly used for DSPs where the processor is small but highly irregular. MSS consists of several tools, but we will concentrate on the MSSQ compiler as its purpose is most aligned with instruction selection. MSSQ was developed by Leupers and Marwedel [172] as a faster version of MSSC [196], which in turn is an extension of the tree-based MSSV [186].

[3]Microcode is essentially the hardware language which CPUs use internally in order to execute the machine instructions of the compiled input programs. For example, the microcode controls how the registers and program counter should be updated.

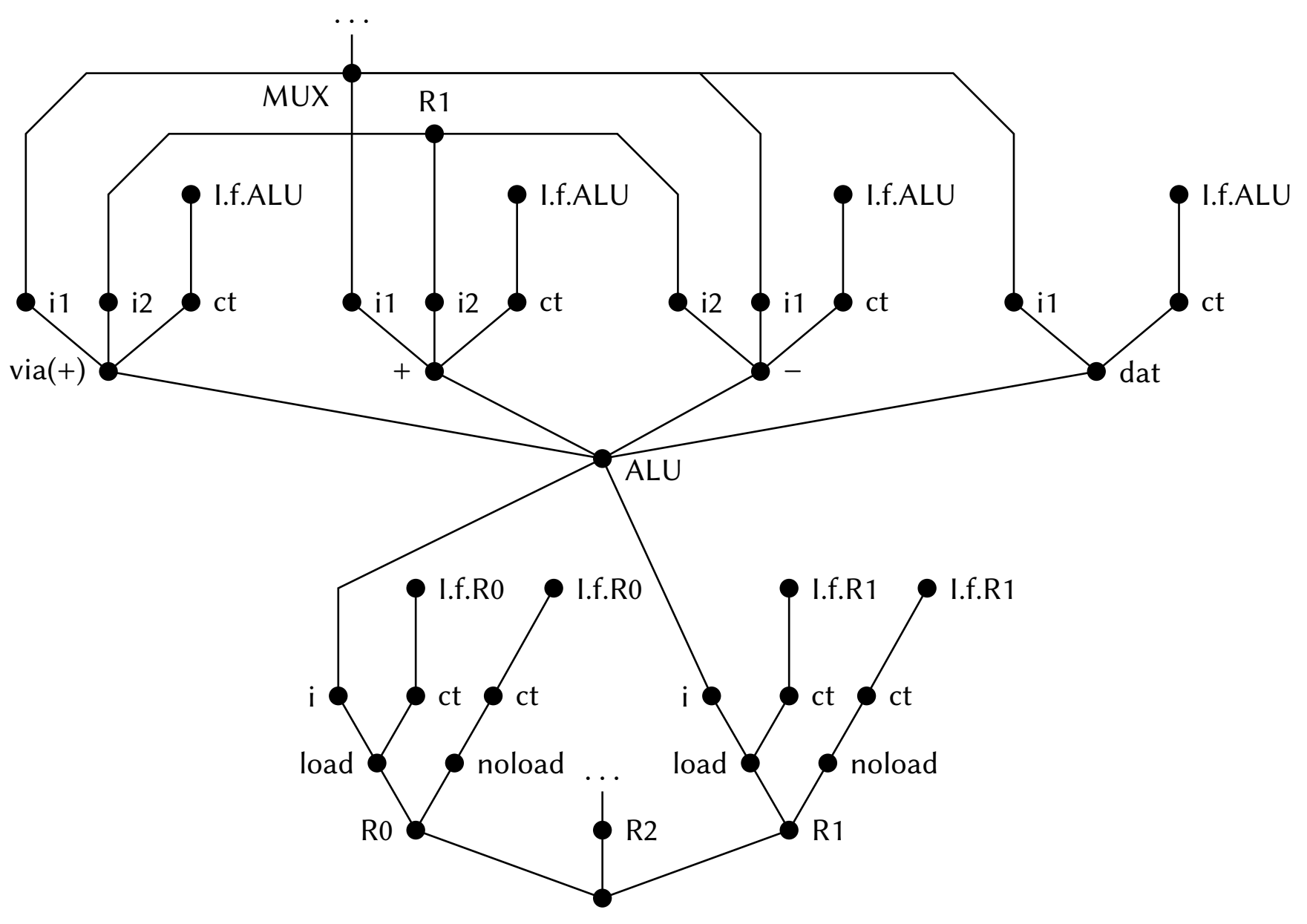

Figure 4.4: The CO graph of a simple processor, containing an ALU, two data registers, a program counter, and a control store.

From the Mimola specification – which contain the processor registers as well as all the operations that can be performed on these registers within a single cycle – a hardware DAG called the *connection-operation graph* (CO graphs) is automatically derived. An example is given in Figure 4.4. A pattern matcher then attempts to find subgraphs within the CO graph that will cover the expression trees of the input program. Because the CO graph contains explicit nodes for every register, a pattern match found on this graph – which is called a *version* – is also an assignment of program variables (and temporaries) to registers. If match cannot be found, due to lack of registers, the expression tree will be rewritten by splitting assignments and inserting temporaries. The process then backtracks and repeats in a recursive fashion until the entire expression tree is covered. A subsequent process then selects a specific version from each matchset and tries to schedule them so that they can be *compacted* into *bundles* for parallel execution.

Although microcode generation is at a lower hardware level than assembly code generation – which is usually what we refer to with instruction selection – we see several similarities between the problems that must be solved in each, and that is why it is included in this report (further examples include [27, 165, 182]). In the next chapter we will see another approach which also models the entire processor but applies a more powerful technique.

## 4.4 Covering DAGs with DAG patterns

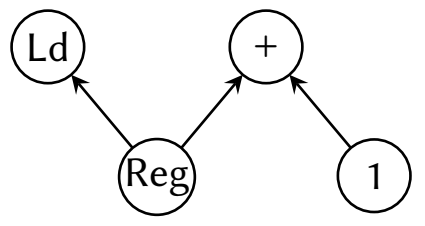

Many machine instructions can only be properly modeled as DAGs. For example, the inlined figure illustrates a load-and-increment instruction which loads a value from memory at the location specified by an address register, and simultaneously increments the address by 1. Such instructions are useful when iterating over arrays as both the fetch and increment routines can be implemented by a single machine instruction.

Conceptually, pattern matching with DAG patterns can be done in two ways: either by splitting the patterns into trees, matching them individually, and then attempting to recombine them to their original DAG form; or by matching them directly; in general this can be done in quadratic time. However, I have been unable to find a single approach which performs direct DAG-on-DAG matching *whilst also* being limited to only DAGs. The reason is, presumably, that performing DAG matching is equally or near-equally as complex as subgraph isomorphism, and since the latter is more powerful it would be an unnecessary and unreasonable restriction to limit such matchers to DAGs. Indeed, as we will later see in this chapter, several DAG-oriented approaches apply general subgraph isomorphism algorithms for tackling the problem of pattern matching.

### 4.4.1 Splitting DAG patterns into trees

Leupers and Marwedel [168, 175] developed one of the first instruction selection techniques to handle multiple-output machine instructions. In a 1996 paper, Leupers and Marwedel describe an approach where the DAG patterns of multi-output instructions – Leupers and Marwedel refer to these as *complex patterns* – are first split by their *register transfers* (RTs) into multiple tree patterns. Register transfers are akin to a Fraser's *register transfer lists* [102] which were introduced in Chapter 2 on page 16, and essentially means that each observable effect gets its own tree pattern. Each individual RT may in turn correspond to one or more machine instructions, but this is not strictly necessary for the approach.

First, assuming a prior procedure has been run that undags the expression DAG, each expression tree is optimally covered using the RTs as patterns. The grammar used by IBURG is not written manually but is automatically generated from a MIMOLA hardware description. Once RTs have been selected, the expression tree is reduced to a tree of *super nodes*, where each super node represents a set of nodes covered by some RT which have been collapsed into a single node. Since multiple-output and disjoint-output machine instructions implement more than one RT, the goal is now to cover the super node graph using the patterns which are formed when the machine instructions are modeled as RTs. Leupers and Marwedel addressed this problem by applying and extending the IP model of Wilson et al. which we saw earlier.

However, because the step of selecting RTs to cover the expression tree is separate from the step which implement them with machine instructions, the assembly code is not necessarily optimal for the whole expression tree. To achieve this property the covering of RTs and selection of machine instructions must be done in tandem.

*Extending the IP model to support SIMD instructions*

Leupers [170] later extended the aforementioned IP model to handle SIMD instructions. Presented in paper from 2000, Leupers's approach assumes every SIMD instruction performs two operations which each takes a disjoint set of input operands. This is collectively called a *SIMD pair*, and the idea can easily be extended to $n$-input SIMD instructions. The linear inequalities describing the constraints for selecting SIMD pairs are then as follows:

$$\sum_{r_j \in R(n_i)} x_{ij} = 1 \tag{4.1}$$

$$x_{ij} \leq \sum_{r_l \in R_m(n_k)} x_{kl} \tag{4.2}$$

$$\sum_{j:(n_i;n_j) \in P} y_{ij} = \sum_{r_k \in R_{hi}(n_i)} x_{ik} \tag{4.3}$$

$$\sum_{j:(n_j;n_i) \in P} y_{ji} = \sum_{r_k \in R_{lo}(n_i)} x_{ik} \tag{4.4}$$

Equations 4.1 and 4.2 ensure valid pattern selection in general, while equations 4.3 and 4.4 are specific for the SIMD instructions. We know Equation 4.1 from the discussion of Wilson et al., which enforces that every node in the expression DAG is covered by some rule; $R(n_i)$ represents the set of rules whose patterns match at node $n_i$, and $x_{ij}$ is a Boolean variable indicating whether node $n_i$ is covered by rule $r_j$. Equation 4.2 – which is new for us – enforces that a rule selected for a child of a node $n_i$ reduces to the same nonterminal required by the rule selected for $n_i$; $n_k$ is the $m$th child node of parent node $n_i$, and $R_m(n_k)$ represents the set of applicable rules that reduce the $m$th child node to the $m$th nonterminal of rule $r_j$. Equation 4.3 and Equation 4.4 enforce that the application of a SIMD instruction actually covers a SIMD pair, and that any node can be covered by at most one such instruction; $y_{ij}$

is a Boolean variable indicating whether two nodes $n_i$ and $n_j$ can be packed into a SIMD instruction, and $R_{hi}(n_i)$ and $R_{lo}(n_i)$ represent the set of rules applicable for a node $n_i$ that operate on the higher and lower portion of a register, respectively.

The goal is then to maximize the use of SIMD instructions, which is achieved by maximizing the following objective function:

$$f = \sum_{n_i \in N} \sum_{r_j \in S(n_i)} x_{ij}$$

where $N$ is the set of nodes in the expression DAG, and

$$S(n_i) = R_{hi}(n_i) \cup R_{lo}(n_i) \subset R(n_i).$$

The paper contains some experimental data which suggests that usage of SIMD instructions reduces code size by up to 75% for selected test cases and target machines. However, the approach assumes that each individual operation of the SIMD instructions can be expressed as a single node in the input DAG. Hence it is unclear whether the method can be extended to more complex SIMD instructions, and whether it scales for larger input programs. Tanaka et al. [238] later extended this work for selecting SIMD instructions while also taking the cost of data transfers into account by introducing transfer nodes and transfer patterns.

*Attacking both pattern matching and pattern selection with IP*

In 2006 Bednarski and Kessler [29] developed an integrated approach where both pattern matching and pattern selection are solved using integer programming. The design – which later was applied by Eriksson et al. [84] – is an extension of their earlier work where instruction selection had previously more or less been ignored (see [152, 153]).

In broad outline, the IP model assumes that a sufficient number of pattern instances has been generated for a given expression DAG $G$. This is done using a pattern matching heuristic which computes an upper bound. For each pattern instance $p$, the model contains solution variables that:

- map a pattern node in $p$ to an IR node in $G$;
- map a pattern edge in $p$ to an IR edge in $G$; and
- decide whether $p$ is used in the solution. Remember that we may have an excess of pattern instances so they cannot all be selected.

Hence, in addition to the typical linear inequality equations we have seen previously for enforcing coverage, this IP model also includes equations to ensure valid matchings of the pattern instances. Bednarski and Kessler implemented this approach in a framework called Optimist, and then used IBM CPLEX Optimizer [241] to solve the IP model.

When compared against an implementation using an integrated DP approach (also developed by the same authors; see [152]), Bednarski and Kessler found that the Optimist substantially reduced code generation time while retaining code quality. However, for several test cases – the largest expression DAG containing only 33 nodes – Optimist failed to generate any assembly code whatsoever within the set time limit. One reasonable cause could be that the IP model also attempts to solve pattern matching – a problem which we have seen can be solved externally – and thus further exacerbates an already computationally difficult problem.

*Modeling instruction selection with constraint programming*

Although integer programming allows auxiliary constraints to be included to the model, they may be cumbersome to express as linear inequalities. Bashford and Leupers [28] therefore developed a model using *constraint programming* (CP) [217], which is another method for tackling combinatorial optimization problems but allows a more direct form of expressing the constraints (this is also discussed in [171] which is a longer version that combines both [170] and [28]). In brief terms a CP model consists of a set of *domain variables* – which each has an initial set of values that it can assume – and a set of *constraints* that essentially specify the valid combinations of values for a subset of the domain variables. A *solution* to the CP model is then an assignment of all domain variables – i.e. each domain variable takes exactly one value – which is valid for all the constraints.

Like Wilson et al., Bashford and Leupers's approach is an integrated design which focuses on code generation for DSPs with highly irregular architectures. The instruction set of the target machine is broken down into a set of register transfers which are used to cover individual nodes in the expression DAG. As each RT concerns specific registers on the target machine, the covering problem basically also incorporates register allocation. The goal is then to minimize the cost of covering by combining multiple RTs that can be executed in parallel via some machine instruction.

For each node in the expression DAG a *factorized register transfer* (FRT) is introduced, which essentially embodies all RTs that match at that node. An FRT is formally defined as

$$\langle Op, D, [U_1, \ldots, U_n], F, C, T, CS \rangle.$$

*Op* is the operation of the node. $D$ and $U_1, \ldots, U_n$ are domain variables representing the *storage locations* of the result and the respective inputs to the operation. These are typically the registers that can be used for the operation, but also include *virtual storage locations* which convey that the value is produced as an intermediate result in a chain of operations (e.g. the multiplication term in a multiply-accumulate instruction is such a result). A set of constraints then ensure that, if the storage locations of $D$ and $U_i$ differ between two operations which are adjacent in the expression DAG, there exists a valid data transfer between the registers, or that both are identical if one is a virtual storage location. $F$, $C$, and $T$ are all domain variables which collectively represent the *extended resource information* (ERI) that specifies: at which functional unit the operation will be executed ($F$); at what cost ($C$), which is the number of execution cycles; and by which machine instruction type ($T$). A combination of a functional unit and machine instruction type is later mapped to a particular machine instruction. Multiple RTs can be combined into the same machine instruction by setting $C$ to 0 when the destination of the result is a virtual storage location and letting the last node in the operation chain account for the required number of execution cycles. The last entity $CS$ is the set of constraints which define the range of values for the domain variables, the constraints defining the dependencies between $D$ and $U_i$, as well as other auxiliary constraints that may be required for the target machine. For example, if the set of RTs matching a node consists of $\{\texttt{c = a + b}, \texttt{a = c + b}\}$, then the corresponding FRT becomes:

$$\langle +, D, [U_1, U_2], F, C, T, \{D \in \{\mathsf{c}, \mathsf{a}\}, U_1 \in \{\mathsf{a}, \mathsf{c}\}, U_2 = \mathsf{b}, D = \mathsf{c} \Rightarrow U_1 = \mathsf{a}\}\rangle$$

For brevity we have omitted several details such as the constraints concerning the ERI. However, according to this model the constraints appears to be limited to within a single FRT, thus hindering support for interdependent machine instructions whose constraints involve multiple FRTs.

This CP model is then solved to optimality by a constraint solver – Bashford and Leupers used ECL$^{i}$PS$^{e}$, which is based on Prolog. However, since optimal covering using FRTs is an NP-complete problem, Bashford and Leupers also applied heuristics to curb the complexity by splitting the expression DAG into smaller pieces at points where intermediate results are shared, and then performing instruction selection on each expression tree in isolation.

Using constraint programming to solve instruction selection seems promising as it, like IP, allows additional constraints of the target machine to be added to the model. In addition, these constraints should be easier to add to a CP model as they need not necessarily be expressed as linear inequalities. At the time of writing, however, the existing techniques for solving integer programming problems are more mature than those of constraint programming, which potentially makes IP solvers more powerful than CP solvers for addressing instruction selection. Having said that, it is still unclear which technique of combinatorial optimization – which also includes SAT – is best suited for instruction selection (and code generation in general).

*Extending grammar rules to contain multiple productions*

Influenced by the earlier work by Leupers, Marwedel, and Bashford, Scharwaechter et al. [224] presented in 2007 another method for handling multiple-output machine instructions by allowing productions in an instruction set grammar to have more than one nonterminal on the left-hand side. Rules having only one LHS nonterminal in their productions are distinguished from rules containing multiple LHS nonterminals by referring to these as *rules* and *complex rules*, respectively. In addition, the single RHS of productions within rules are called *simple patterns* while the multiple RHSs of productions within complex rules are called *split patterns*.[4] We also call a combination of split patterns a *complex pattern*. We illustrate this more clearly below:

$$\underbrace{\text{nt} \rightarrow \overbrace{\text{Op } \text{I}_1 \cdots}^{\text{simple pattern}} \qquad \text{cost} \qquad \text{action}}_{\text{rule}}$$

$$\underbrace{\text{nt}_1 \ \text{nt}_2 \cdots \rightarrow \overbrace{\overbrace{\text{Op}_1 \ \text{I}_{1,1} \cdots}^{\text{split pattern}} , \overbrace{\text{Op}_2 \ \text{I}_{2,1} \cdots}^{\text{split pattern}}}^{\text{complex pattern}} , \cdots \qquad \text{cost} \qquad \text{action}}_{\text{complex rule}}$$

[4] In their paper Scharwaechter et al. call these *simple* and *split rules*, but to conform with the terminology we established on page 27 I chose to call them patterns instead of rules.

The approach of pattern matching is then the same as we have seen before: match simple and split patterns, and then recombine into complex patterns. The pattern selector checks whether it is worth to apply a complex pattern to cover a certain set of nodes, or if the nodes should instead be covered using simple patterns; selecting a complex pattern may have repercussions on the pattern candidates available for the rest of the expression DAG as the intermediate results of nodes within the complex pattern can no longer be reused for other patterns and may thus need to be duplicated, which incurs additional overhead. Consequently, if the cost saved by replacing a set of simple patterns by a single complex pattern is greater than the cost of incurred duplication, then the complex pattern will be selected.

After this step the pattern selector may have selected complex patterns which overlap, thus causing conflicts which must be resolved. This is addressed by formulating a *maximum weighted, independent set* (MWIS) problem, in which a set of nodes is selected from an undirected, weighted graph such that no selected nodes are adjacent. In addition, the sum of the weights of the selected nodes must be maximal. We will discuss this idea in more detail in Section 4.5. Each complex pattern forms a node in the MWIS graph and an edge is drawn between two nodes if the two patterns overlap. The weight is then calculated as the negated sum of the cost of the split patterns in the complex patterns (the paper is ambiguous on how these split pattern costs are calculated). Since the MWIS problem is known to be NP-complete, Scharwaechter et al. employed a greedy heuristic called Gwmin2 by Sakai et al. [221]. Lastly, split patterns which have not been merged into complex patterns are replaced by simple patterns before emitting the code.

Scharwaechter et al. implemented their approach by extending Olive into a system called Cburg, and ran some experiments by generating assembly code for a MIPS architecture. The code generated by Cburg – which exploited complex instructions – was then compared to assembly code which only used simple instructions. The results indicate that Cburg exhibited near-linear complexity and improved performance and code size by almost 25% and 22%, respectively. Ahn et al. [2] later broadened this work by including scheduling dependency conflicts between complex patterns, and incorporating a feedback loop with the register allocator to facilitate register allocation.

In both approaches of Scharwaechter et al. and Ahn et al., however, complex rules can only consist of disconnected simple patterns (i.e. there is no sharing of nodes between the simple patterns). In a 2011 paper – which is a revised and extended version of [224] – Youn et al. [268] address this problem by introducing index subscripts for the operand specification of the complex rules. However, the subscripts are restricted to the input nodes of the pattern, thus still hindering support for completely arbitrary DAG patterns.

*Splitting patterns at marked output nodes*

In the approaches discussed so far, it is assumed that the output of all DAG pattern occur at the root nodes. In 2001 Arnold and Corporaal [18, 19, 20] came up with an approach where the output can be marked explicitly. The DAG pattern will then be split such that each output node receives its own tree pattern, called *partial patterns*; an example is given in Figure 4.5.

Using an $\mathcal{O}(nm)$ algorithm the tree patterns are matched over the expression DAG. After matching an algorithm attempts to merge appropriate combinations of partial pattern instances into full pattern instances. This is done by maintaining an array for each partial pattern instance that maps the pattern nodes to the covered nodes in the input DAG, and then checking whether two partial patterns belong to the same original DAG pattern and do not clash. In other words, no two pattern nodes that correspond to the same pattern node in the original DAG pattern may cover different nodes in the expression DAG. Pattern selection is then done using conventional dynamic programming techniques.

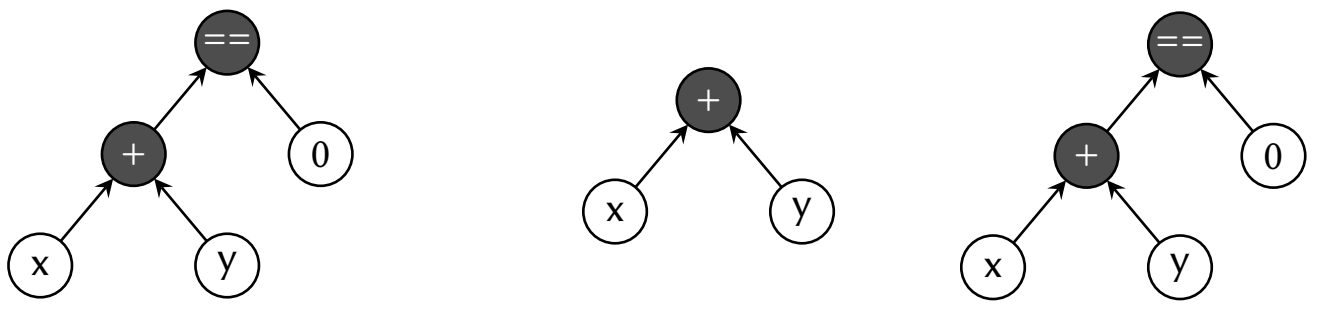

(a) DAG pattern. (b) Tree patterns after splitting.

Figure 4.5: Splitting of a DAG pattern of an add instruction which also sets a condition flag if the result is equal to 0. The darkly shaded nodes indicate the output nodes.

Farfeleder et al. [89] proposed a similar approach by applying an extended version of Lburg for pattern matching followed by a second pass in an attempt to recombine matched patterns that originate from patterns of multiple-output instructions. However, the second pass consists of ad-hoc routines which are not automatically derived from the machine description.

*Exhaustive pattern selection with semantic-preserving transformations*

Hoover and Zadeck [141] developed a design called Toast (Tailored Optimization and Semantic Translation) with the ultimate goal of automating the generation of entire compiler frameworks – including instruction scheduling and register allocation – from a declarative machine description. There instruction selection is done using semantic-preserving transformations during pattern selection to make better use of the instruction set. For example, although $x * 2$ is semantically equivalent to $x \ll 1$ – meaning that $x$ is logically shifted 1 bit to the right – most instruction selectors will fail to select machine instructions implementing the latter when the former appears in the expression DAG as they are syntactically and structurally different.

The approach works as follows: first the frontend emits expression DAGs consisting of *semantic primitives*, which are also used to describe the machine instructions. The expression DAG is then semantically matched using single-output patterns derived from the machine instructions. Semantic matchings are called *toe prints* and are found by a *semantic comparator* which applies syntactic matching – i.e. are the nodes of the same type, which is implemented as a straight-forward $\mathcal{O}(nm)$ pattern matcher – in combination with semantic-preserving transformations for when syntactic matching fails. A transformation is only applied if it will lead to a syntactic match later on, resulting in a bounded exhaustive search for all possible toe prints. Once all toe prints have been found, they are combined into *foot prints* which correspond to the full effects of a machine instruction. These can therefore include only one toe print (as with single-output instructions) or several (as with multi-output instructions). However, the paper lacks details on how exactly this is done. Lastly, all combinations of foot prints are considered in pursuit of the one which leads to the most effective implementation of the expression DAG. To curb the search space, however, this process only considers combinations where each selected foot print syntactically matches at least one semantic primitive in the expression DAG, and only “trivial amounts” of the expression DAG, such as constants, may be included in more than one foot print.

However, due to the exhaustive nature of the approach, in its current form it appears to be unpractical for generating assembly code for all but the smallest input programs; in a prototype implemented by the authors, it was reported that almost $10^{70}$ “implied instruction matches” were found for one of the test cases, and it is unclear how many of them are actually useful.

## 4.5 Reducing pattern selection to a MIS problem

Numerous approaches reduce the pattern selection problem to a *maximum independent set* (MIS) problem; we have already seen this idea in application by Scharwaechter et al. and Ahn et al. (see Section 4.4.1), and we will encounter more such approaches in this section as well as in the next chapter. We therefore proceed with describing this technique in greater detail.

For patterns that overlap one or more nodes in the expression DAG, a corresponding *conflict* or *interference graph* can be formed. An example is given in Figure 4.6. In the matched DAG, shown in (a), we see that pattern $p_1$ overlaps with $p_2$, and pattern $p_4$ overlaps with $p_2$ and $p_3$. For each pattern we form a node in a conflict graph $C$, shown in (b), and draw an edge between the nodes of any two patterns that overlap. By selecting the largest set of vertices from $C$ such that no selected nodes are adjacent in $C$, we obtain a set of patterns such that each node in the expression DAG is completely covered and no two patterns overlap. This problem is, as expected, known to be NP-complete.

To achieve optimal pattern selection we can attach the pattern cost as a weight to each node in the conflict graph, thus augmenting the MIS problem to a *maximum weighted independent set* (MWIS) problem, i.e. to select the largest set of vertices $S$ such that it satisfies MIS and also maximizes $\sum_{s \in S} weight(s)$. Since MWIS has a maximizing goal, we simply assign each weight as the negated cost of that pattern. Note that this technique is not limited to selection of DAG patterns but can just as well be used for graph patterns.

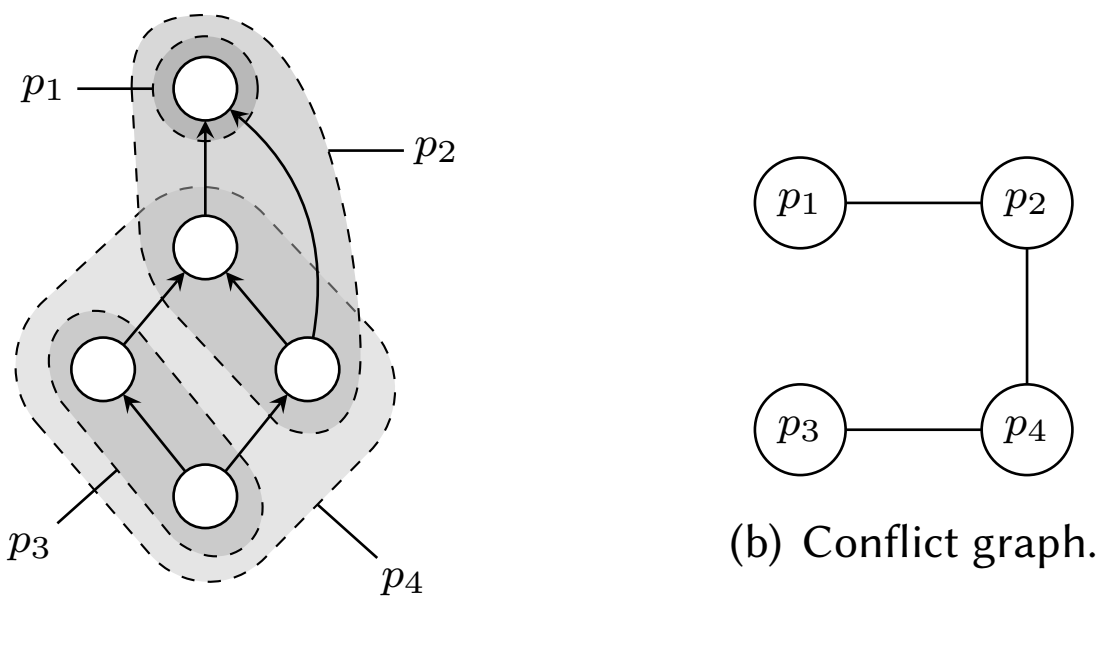

(a) Matched DAG.

(b) Conflict graph.

Figure 4.6: Example of a matched DAG and its corresponding conflict graph.

### 4.5.1 Approaches using MIS and MWIS

In 2001 Kastner et al. [151] developed an approach which targets instruction set generation for hybrid reconfigurable systems. Hence the patterns are not given as input but are generated as part of the problem itself (this is also discussed in the next chapter). Once a pattern set has been generated, a generic subgraph isomorphism algorithm from [75] is used to find all patterns matchings on the expression DAG. Subgraph isomorphism will be described in detail in the next chapter. This is followed by instantiating a corresponding MIS problem which is solved heuristically. Since the matching algorithm does not require the input to be a DAG, the approach can be extended to input of any graph-based format. In an extended version of the paper, Kastner et al. [150] improves the pattern matching algorithm for fast rejection of dissimilar subgraphs. However, no guarantees are made for optimal pattern selection.

*Handling echo instructions*

Brisk et al. [37] also used the idea of MIS for performing instruction selection on target machines with *echo instructions*. Echo instructions are small markers in the assembly code that refer back to earlier portions for re-execution. The assembly code can therefore essentially be compressed using the same idea applied in the LZ77 algorithm [274], where a string can be compressed by replacing common substrings that appear earlier in the string with pointers, and can be reconstructed through simple copy-pasting. Since echo instructions do not incur a branch or a procedure call, the result is reduced code size with zero overhead in terms of performance.

Consequently, the pattern set for instruction selection is not fixed and must be determined as a precursor to pattern matching. Patterns are formed by clustering together adjacent nodes in the expression DAG. The pattern which yields the highest decrease in code size is then selected and the expression DAG updated by replacing recurring pattern instances with new nodes to indicate use of echo instructions. This process repeats until no new pattern better than some user-defined value criteria can be found. Like Kastner et al., the design of Brisk et al. finds these pattern using subgraph isomorphism but applies a different algorithm called VF2 [60]. Although this algorithm exhibits $\mathcal{O}(nn!)$ in worst case, the authors reported that it ran efficiently for most expression DAGs in the experiments.

## 4.6 Other DAG-based techniques

### 4.6.1 Extending means-end analysis to DAGs

20 years after [194] and Cattell et al. published their top-down tree covering approaches (see Section 3.4.1 in the previous chapter), Yu and Hu [270, 271] rediscovered means-end analysis as a method of instruction selection and also made two major improvements. First, the new design handles expression DAGs and DAG patterns whereas those of [194] and Cattell et al. are limited to trees. Second, it combines means-end analysis with *hierarchical planning* [220] which is a search strategy that recognizes that many problems can be arranged in an hierarchical manner to allow larger and more complex problem instances to be solved. This also enables exhaustive exploration of the search space whilst avoiding situations of dead ends and infinite looping that straight-forward implementations of means-end analysis may suffer from ([194] and Cattell et al. both circumvented this problem by enforcing a cut-off when reaching a certain depth in the search space).

The technique exhibits a worst time execution that is exponential to the depth of the search, but Yu and Hu claim in their paper that a depth of 3 is sufficient to yield results of equal quality to that of hand-written assembly code. However, it is unclear whether complex machine instructions with for example disjoint-output and internal-loop characteristics can be handled by this approach.

### 4.6.2 DAG-like trellis diagrams

In 1998 Hanono and Devadas [130, 131] proposed a technique that is similar to Wess's idea of using trellis diagrams – which we discussed in Section 3.7.5 – which they implemented in a system called Aviv. The instruction selector takes an expression DAG as input and duplicates each operation node according to the number of functional units in the target machine on which that operation can run. Special *split* and *transfer nodes* are inserted before and after each duplicated operation node to allow the data flow to diverge and then reconverge before passing to the next operation node in the expression DAG. The use of transfer nodes also allow the cost of transferring data from one functional unit to another to be taken into account. Similarly to the trellis diagram, instruction selection is then reduced to finding a path from the leaf nodes in the expression DAG to its root node. Unlike the optimal dynamic programming-oriented approach of Wess, Hanono and Devadas applied a greedy heuristic which starts from the root node and makes it way towards the leaves.

However, like Wess's design this technique assumes a one-to-one mapping between the expression nodes and the available machine instructions; in fact, the main goal of Aviv was to generate efficient assembly code for VLIW (Very Long Instruction Word) architectures, where focus lies on combining as many operations as possible into the same instruction bundle.

## 4.7 Summary

In this chapter we have investigated several methods which rely on the principle of DAG covering, which is a general form of tree covering. Operating on DAGs instead of trees has several advantages: most importantly, common subexpressions and a largest set of machine instructions – including multiple-output and disjoint-output instructions – can be modeled directly and exploited during instruction selection, leading to improved performance and reduced code size. Consequently, techniques based on DAG covering are today one of the most applied technique for instruction selection in modern compilers.

However, the ultimate cost of transitioning from trees to DAGs is that optimal pattern selection can no longer be achieved in linear time as that problem is NP-complete. At the same time, DAGs are not expressive enough to allow the proper modeling of all aspects in the input program and machine instruction patterns. For example, statements such as `for` loops incur loop-back edges in the expression graphs, thus restricting DAG covering to basic block scope. This obviously also excludes the modeling of internal-loop instructions. Another disadvantage is that optimization opportunities of storing program variables and temporaries in different forms and at different locations across the function. We will see an example in the next chapter where being able to do this increases the performance of the final assembly code.

In the next chapter we will discuss the last and most general principle of instruction selection – graph covering – which addresses some of the aforementioned deficiencies of DAG covering.

# 5

# Graph-Based Approaches

Even though DAG covering, which we examined in the previous chapter, is a more general and powerful form of tree covering, it is still not enough for handling all aspects of the input program and machine instructions. For example, control flow incurred by `for` loop statements cannot be modeled as part of the expressions since that requires cycles, which violates the definition of DAGs. By removing this restriction, we end up with the most general form of covering which is *graph covering*.

## 5.1 The principle

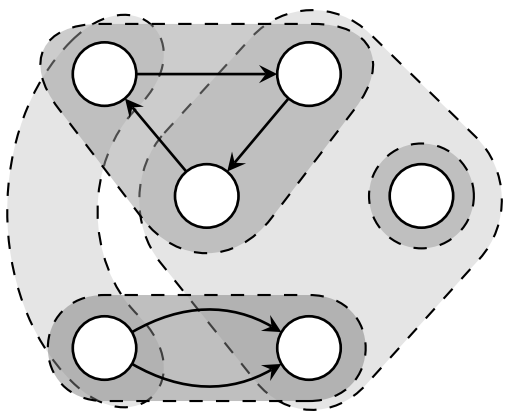

Just like DAG covering is a more general form of tree covering, so is graph covering a more general form of DAG covering. By allowing both the statements of the input program to assume any generic graphical shape, instruction selection can be extended from its previously local scope – i.e. covering a single expression tree or DAG at a time – to a global scope which encompasses all expressions within entire functions as graphs are capable of modeling control flow. This is known as *global instruction selection* and in theory it enables the instruction selector to move operations across the boundaries of basic blocks in order to make better use of complex machine instructions with large patterns that would otherwise not be applicable. For modern target machines where power consumption and heat emission are becoming increasingly important factors, this is one key method to addressing these concerns as using fewer, large instructions is generally more power-efficient than applying many, small instructions.

However, by transitioning from DAGs patterns to graph patterns, even pattern matching becomes NP-complete and thus we can no longer resort to pattern matching techniques of trees and DAGs; Figure 5.1 shows the varying time complexities of pattern matching and optimal pattern selection when using different graphical shapes. Instead we need to apply generic subgraph isomorphism algorithms, and therefore begin with looking at a few such algorithms.

| | Pattern matching | Optimal pattern selection |
|---|---|---|
| Trees | Linear | Linear |
| DAGs | Quadratic | Non-polynomial |
| Graphs | Non-polynomial | Non-polynomial |

Figure 5.1: Time complexities of pattern matching and optimal pattern selection when using different graphical shapes.

## 5.2 Pattern matching using subgraph isomorphism

The problem of subgraph isomorphism is to detect whether an arbitrary graph $G_a$ can be turned, twisted, or mirrored such that it forms a subgraph of another graph $G_b$. In such cases one says that $G_a$ is an *isomorphic subgraph* to $G_b$, and deciding this is known to be an NP-complete [58]. As subgraph isomorphism is found in many other fields, a vast amount of research has been devoted to this problem (see for example [60, 87, 88, 109, 129, 139, 163, 234, 248]). In this section we will look at Ullmann's algorithm and another commonly used algorithm called VF2.

It should be noted that these pattern matching techniques only check for identical structure and not semantics. For example, the expressions $a * (b + c)$ and $a * b + a * c$ are functionally equivalent but will yield differently structured graphs and will thus not match one another. Arora et al. [21] proposed a method where the graphs are normalized prior to pattern matching, but it is limited to arithmetic expression DAGs and does not guarantee that all matchings will be found.

### 5.2.1 Ullmann's algorithm

One of the first – and most well-known – methods to deciding subgraph isomorphism is an algorithm developed by Ullmann [248]. In his paper, published 1976, Ullmann reduces the problem of determining whether a graph $G_a = (V_a, E_a)$ is subgraph isomorphic to another graph $G_b = (V_b, E_b)$ to finding a Boolean $|V_a| \times |V_b|$ matrix $\mathbf{M}$ such that the following conditions holds:

$$\mathbf{C} = \mathbf{M}(\mathbf{MB})^T$$
$$\forall i, j, 1 \le i \le |V_a|, 1 \le j \le |V_b| : (a_{ij} = 1) \Rightarrow (c_{ij} = 1)$$

$\mathbf{A}$ and $\mathbf{B}$ are the respective adjacency matrices of $G_a$ and $G_b$, and $a_{ij}$ and $c_{ij}$ are elements of $\mathbf{A}$ and $\mathbf{C}$. When these conditions hold, every row in $\mathbf{M}$ will contain exactly one 1, and every column in $\mathbf{M}$ will contain at most one 1.

Finding $\mathbf{M}$ can be done with brute-force by initializing every element $m_{ij}$ with 1 and then iteratively pruning away 1's until a solution is found. To reduce this search space Ullmann developed a procedure that eliminates some of the 1's which for sure will not appear in any solution. According to [60], however, even with this improvement the worst-case time complexity of the algorithm is $\mathcal{O}(n!n^2)$.

### 5.2.2 The VF2 algorithm

In 2001 paper Cordella et al. [60] proposed an algorithm, called VF2, which has been used in several DAG and graph covering-based instruction selectors. In particular, it has been applied to assist *instruction set extraction* (ISE), which is sometimes known as *instruction set selection*. Examples of such methods are given in [10, 53, 192], and Galuzzi and Bertels [110] have done a recent survey.

In broad outline, the VF2 algorithm recursively constructs a mapping set of $(n, m)$ pairs, where $n \in G_a$ and $m \in G_b$. The core of the algorithm is the set of rules that check whether a new candidate pair should be allowed to be included in the mapping set. The rules entail a set of *syntactic feasibility checks*, implemented by $F_{\text{syn}}$, followed by a *semantic feasibility check*, implemented by $F_{\text{sem}}$. Without going into any details, we define $F_{\text{syn}}$ as

$$F_{\text{syn}}(s, n, m) = R_{\text{pred}} \vee R_{\text{succ}} \vee R_{\text{in}} \vee R_{\text{out}} \vee R_{\text{new}}$$

where $n \in G_a$, $m \in G_b$ – thus constituting the candidate pair – and $s$ represents the current (partial) mapping set. The first two rules – $R_{\text{pred}}$ and $R_{\text{succ}}$ – ensure that the new mapping set $s'$ is consistent with the structure of $G_a$ and $G_b$. The remaining three rules are used to prune the search space: $R_{\text{in}}$ and $R_{\text{out}}$ looks ahead one step in the search process by ensuring that there will still exist enough unmapped nodes in $G_b$ that can be mapped to the remaining nodes in $G_a$; and similarly $R_{\text{new}}$ looks ahead two steps. The exact definitions of these rules are explained in the referenced paper, and they can be modified with minimal effort to enforce *graph isomorphism* instead of *subgraph* isomorphism; in the former, the structure of $G_a$ and $G_b$ must be rigid and cannot be twisted and turned until they match. $F_{\text{sem}}$ is then a framework that can easily customized to add additional checks regarding the nodes in the candidate pair.

In the worst case this algorithm exhibits $\mathcal{O}(n!n)$ time complexity, but the best case time complexity – which is polynomial – makes it practical for pattern matching over very large graphs; it has been reported that the VF2 algorithm has been successfully used on graphs containing thousands of nodes [60].

## 5.3 Graph-based pattern selection techniques

The main distinction between DAG covering and graph covering is that, in the latter, the problem of pattern matching is much more complex as it is NP-complete. The techniques for addressing pattern selection, however, are often general enough to be applicable for either principle. In this section we will look at more pattern selection techniques, which typically made their first appearance in instruction selection via approaches based on graph covering.

### 5.3.1 Unate and binate covering

The first technique we will discuss is a method where the pattern selection problem is transformed into an equivalent *unate* or *binate covering* problem. The concepts behind the two are identical with the exception of one detail. Although binate covering-based techniques appeared first, we will begin with explaining unate covering as binate covering is an extension of unate covering.

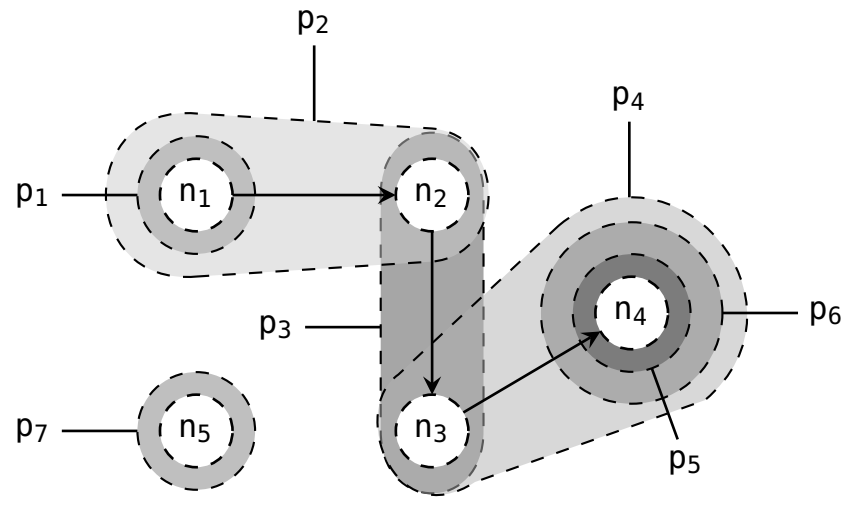

(a) Graph to cover.

| | $p_1$ | $p_2$ | $p_3$ | $p_4$ | $p_5$ | $p_6$ | $p_7$ |
|---|---|---|---|---|---|---|---|
| $n_1$ | 1* | 1* | 0 | 0 | 0 | 0 | 0 |
| $n_2$ | 0 | 1* | 1 | 0 | 0 | 0 | 0 |
| $n_3$ | 0 | 0 | 1 | 1* | 0 | 0 | 0 |
| $n_4$ | 0 | 0 | 0 | 1* | 1 | 1 | 0 |
| $n_5$ | 0 | 0 | 0 | 0 | 0 | 0 | 1* |

(b) Unate covering matrix.

Figure 5.2: Example of unate covering. Unmarked 1s in the matrix represent potential but not selected covers, while the 1s marked with a star (1*) indicate the selected cover which is optimal (assuming all patterns have the same unit cost).

### *Unate covering*

The idea of unate covering is to create a Boolean matrix $\mathbf{M}$ where each row represents a node in the expression graph, and each column represents an instance of a pattern that matches one or more expression nodes. If we denote $m_{ij}$ as row $i$ and column $j$ in $\mathbf{M}$, then $m_{ij} = 1$ indicates that node $i$ is covered by pattern $j$. Hence the problem of pattern selection is thus reduced to finding a valid configuration of $\mathbf{M}$ such that the sum of every row is at least 1. This problem, which is also illustrated in Figure 5.2, is obviously NP-complete, but there exist several efficient techniques for solving it heuristically (e.g. [61, 123]).

However, unate covering alone does not incorporate all constraints of pattern selection as some patterns require – and omit – the selection of other patterns to yield correct assembly code. For example, assume that pattern $p_3$ in Figure 5.2 requires that pattern $p_6$ is selected to cover $n_4$ instead of pattern $p_5$. In instruction set grammars this is enforced via nonterminals, but for unate covering we have no means of expressing this constraint. We therefore turn to binate covering where this is possible.

### *Binate covering*

We first rewrite the Boolean matrix from the unate covering problem into Boolean formulas consisting of conjunctions of uncomplemented disjunctions. We refer to each disjunction – which is a combination of literals and ∨ operations – as a *clause*. Clauses can then in turn be combined with ∧ operations. The Boolean matrix in (b) of Figure 5.2 can thus be rewritten as

$$f = (p_1 \vee p_2) \wedge (p_2 \vee p_3) \wedge (p_3 \vee p_4) \wedge (p_4 \vee p_5 \vee p_6) \wedge p_7.$$

Unlike unate covering, binate covering allows *complemented* literals to appear in the clauses, which we write as $\overline{x}$. The aforementioned constraint regarding the compulsory selection of $p_6$ if $p_4$ is selected can now be expressed as

$$\overline{p_4} \vee p_6$$

which is then added to the Boolean formula $f$. This is called an *implication clause* as it is logically equivalent to $p_4 \Rightarrow p_6$.

*Approaches based on unate and binate covering*

According to [56, 177, 178], Rudell [218] pioneered the use of binate covering in his 1989 PhD thesis by applying it to solve DAG covering as a part of VLSI synthesis. Liao et al. [177, 178] later adapted it in 1995 to instruction selection in a method that optimizes code size for one-register target machines. To curb the search space, Liao et al.'s approach solves the pattern selection problem in two iterations: in the first iteration the costs of necessary data transfers are ignored. After patterns have been selected the nodes covered by the same pattern are collapsed into single nodes, and a second binate covering problem is constructed to minimize the costs of data transfers. Although these two problems can be solved simultaneously, Liao et al. chose not to as the number of necessary clauses becomes extremely large. Cong et al. [56] recently applied binate covering as part of solving application-specific instruction generation for configurable processor architectures.

Unate covering has also been successfully applied by Clark et al. [54] in generating assembly code for acyclic computation accelerators, which can be partially customized in order to increase performance of the currently executed program. Presumably, the target machines are homogeneous enough that implication clauses are unnecessary. This approach was later extended by Hormati et al. [142] to reduce inter-connect and data-centered latencies of accelerator designs. In 2009 Martin et al. [183] also applied the idea of unate covering in solving a similar problem concerning reconfigurable processor extensions, but modeled it as a constraint programming problem which also incorporated the constraints of instruction scheduling. In addition, they used CP in a separate procedure to find all applicable pattern matchings.

### 5.3.2 CP-based techniques

Although the aforementioned approach by Martin et al. uses constraint programming, the constraints of pattern selection are simply a direct encoding of unate covering and thus not specific for CP per se. When Floch et al. [97] later adapted the CP model to handle processors with reconfigurable cell fabric, they replaced the method of pattern selection with constraints that are radically different from that of unate covering. In addition, unlike the pioneering CP approach by Bashford and Leupers [28] which we saw earlier in the previous chapter, the design applies the more conventional form of pattern matching and pattern selection instead of breaking down the patterns into RTs and then selecting machine instructions that combine as many RTs as possible.

In the CP model by Floch et al., the requirement that every expression node must be covered by exactly one pattern is enforced via a $\textsc{Count}(i, var, val)$ constraint, which dictates that in the set of domain variables $var$, exactly $val$ number of variables must assume the value $i$.[1] $val$ can be a fixed number or represent another domain variable. Let us assume that every node in the expression graph has an associated matchset containing all the patterns that may cover that node, and that each pattern $m$ appearing in the matchset has been assigned a unique integer value $i_m$. We introduce a domain variable $match_n$ for each expression node $n$ to represent the pattern selected to cover $n$. For each pattern $m$ we also introduce a domain variable $nodecount_m \in \{0, size(m)\}$, where $size(m)$ is the number of pattern nodes in $m$, and define $mset_m = \bigcup_{n \in nodes(m)} match_n$ as the set of $match_n$ variables in which pattern $m$ may appear. With this we can express that every expression node must be covered exactly once as

$$\forall m \in M : \textsc{Count}(i_m, mset_m, nodecount_m)$$

where $M$ is the total set of matching patterns. This constraint is more restrictive than that of unate covering and yields more propagation which reduces the search space. To identify which patterns have been selected we simply check whether $nodecount_m > 0$ for every pattern $m$.

[1] Remember that a domain variable has an initial set of possible values that it may assume, and a solution to a CP model is a configuration of single-value assignments to all domain variables such that it is valid for every constraint appearing in the model.

The model was also further extended by Arslan and Kuchcinski [22] to accommodate VLIW architectures and disjoint-output instructions. This is essentially done by splitting each disjoint-output instruction into multiple subinstructions, each modeled by a disjoint pattern. A generic subgraph isomorphism algorithm is used to find all pattern matchings and pattern selection is then modeled as an instance of the CP model with the additional constraints that the subinstructions are scheduled such that they can be replaced by the original disjoint-output machine instruction. This technique thereby differs from previous approaches that we have seen before (e.g. Scharwaechter et al. [224], Ahn et al. [2], Arnold and Corporaal [18, 19, 20]) where partial patterns are recombined into complex pattern instances prior to pattern selection. The design assumes that there are no loops in the expression graph, but it can consist of multiple disconnected graphs.

However, in all CP models by Martin et al., Floch et al., and Arslan and Kuchcinski, the target machine is assumed to have a homogeneous register architecture as they do not model the necessary data transfers between different register classes.

### 5.3.3 PBQP-based techniques

In 2003 Eckstein et al. [77] presented an approach which uses a modeling method called PBQP during pattern selection, and by taking an SSA graph as input it is one of the few techniques that addresses *global* instruction selection. We will therefore discuss this approach in greater detail, but first we must explain several concepts which are new to us. We begin with describing SSA.

#### *Rewriting input programs into SSA form*

SSA stands for *static single assignment* and is a form of program representation which is well-explained in most compiler textbooks (I recommend [59]). In essence, SSA restricts each variable or temporary in a function to be defined only *once*. The effect of this is that the *live range* of each variable (i.e. the length within the code that the value of the variable may not be destroyed) is contiguous, which in turn means that each variable corresponds to a single value. As a consequence many optimization routines can be simplified, and several modern compilers are based on this form (e.g. LLVM and GCC).

However, the *define-only-once* restriction causes problems for variables whose value can come from more than one source, which occurs when `IF` statements and `for` loops are involved. For example, the following C function (which computes the factorial)

```
1  function factorial(int n) {
2    init:
3      int i = 0;
4      int f = 1;
5    loop:
6      if (i < 10) goto end;
7      f = f * i;
8      i = i + 1;
9      goto loop;
10   end:
11     return f;
12 }
```

is not in SSA form as `f` and `i` are defined multiple times (first at lines 2 and 3, and then at lines 6 and 7, respectively). Such situations are resolved through the use of *$\phi$-functions*, which allow variables to be defined using one of several values that each originate from a distinct basic block. By declaring additional variables and inserting $\phi$-functions at the beginning of the `loop` block, we can rewrite the example above into SSA form:

```
1  function factorial(int n) {
2    init:
3      int i_1 = 0;
4      int f_1 = 1;
5    loop:
6      int i_2 = φ(i_1, i_3);
7      int f_2 = φ(f_1, f_3);
8      if (i_2 < 10) goto end;
9      int f_3 = f_2 * i_2;
10     int i_3 = i_2 + 1;
11     goto loop;
12   end:
13     return f_2;
14 }
```

The problem of $\phi$-functions is that now we have a problem of deciding which value to use. For example, on line 6 in the SSA code above – and likewise for line 7 – we should use the value $i_1$ for defining $i_2$ on the first loop iteration, and then the value $i_3$ for the next following iterations. However, the details on how this is handled is out of scope for our purpose as we will be mostly interested in the *SSA graph*, which is akin to an expression graph. We will see an example of this shortly.

*An example where global instruction selection is beneficial*

Eckstein et al. recognized that limiting instruction selection to local scope can lessen code quality of fixed-point arithmetic assembly code generated for dedicated digital signal processors. A common idiosyncrasy of DSPs is that their fixed-point multiplication units will often leave the result shifted one bit to the left. Hence, if a value is computed by accumulating values from fixed-point multiplications, it should remain in shifted mode until all fixed-point multiplications have been performed or else the accumulated value will needlessly be shifted back and forth.

In (a) of Figure 5.3 we see an example of a C function where this can occur. A fixed-point value `s` is computed as the scalar product of two fixed-point arrays `a` and `b`. For efficient

```c
1 int f(short* a, short* b, int n) {
2     int s = 0;
3     for (int i = 0; i < n; i++) {
4         s = abs(s) + a[i] * b[i];
5     }
6     return s;
7 }
```

(a) C function.

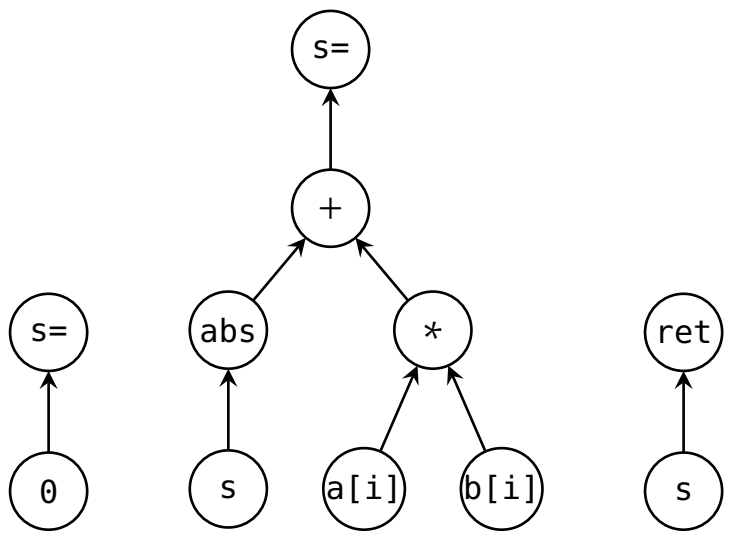

(b) Corresponding expression trees.

Figure 5.3: A fixed-point C function for which a local-scope instruction selector most likely would produce inefficient assembly code (from [77]).

```c
int f(short* a, short* b, int n) {
  init:
    int s1 = 0;
    int i1 = 0;
  loop:
    int i2 = φ(i1, i3);
    int s2 = φ(s1, s3);
    if (i2 < n) goto end;
    s3 = abs(s2) + a[i2] * b[i2];
    int i3 = i2 + 1;
  end:
    return s2;
}
```

(a) C function in SSA form.

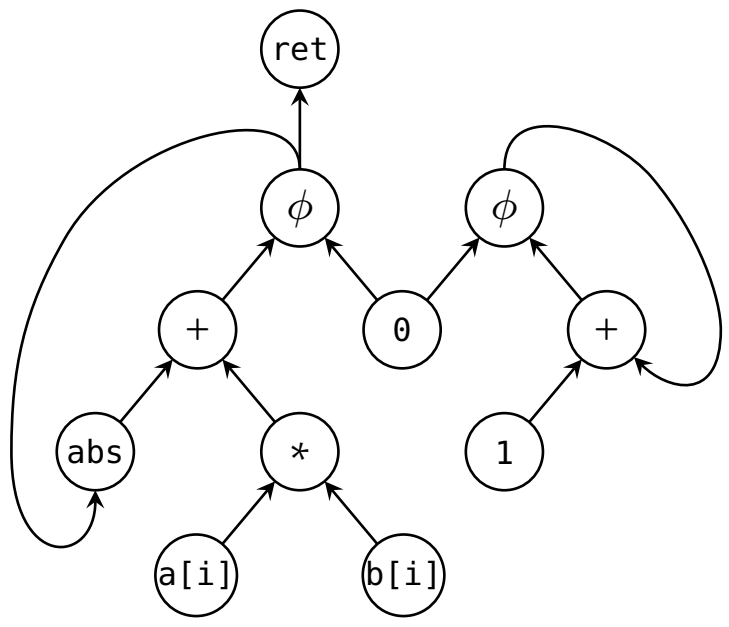

(b) Corresponding SSA graph.

Figure 5.4: Same as Figure 5.3 but in SSA form (from [77]).

execution on a DSP, the value of s should be stored in shifted mode within the `for` loop (lines 3–5), and only shifted back before the return (line 6). In an instruction set grammar this can be modeled by introducing an additional nonterminal which represents the shifted mode, and then add rules for transitioning from shifted mode to normal mode by emitting the appropriate shift instructions.

However, efficient use of such modes is difficult to achieve if the instruction selector only considers one basic block at a time (i.e. is limited to local scope) since the definitions and uses of s occur in different expression trees (see (b) of Figure 5.3). By contrast, in the SSA graph all definitions and uses appear in the same graph. In Figure 5.4 we see the same function in SSA form, together with its corresponding SSA graph. Note that the SSA graph does not contain any nodes representing control flow, which will have to be dealt with separately. In addition, in order to arrange the generated assembly code into basic blocks the instruction selector has to take the original IR code as additional input.

*Solving pattern selection with PBQP*

In the approach by Eckstein et al., the pattern selection problem is solved by reducing it to a *partitioned Boolean quadratic problem* (PBQP). PBQP was first introduced Scholz and Eckstein [225] in 2002 for solving register allocation, and is an extension of the *quadratic assignment problem* (QAP) which is a fundamental combinatorial optimization problem. Both QAP and PBQP are NP-complete, and Eckstein et al. therefore developed their own heuristic solver which is also described in the referenced 2003 paper. We will explain PBQP by building the model bottom-up, starting with the definitions.

To begin with, this design assumes that the machine instructions are given as a linear-form grammar – this was introduced on page 51 in Chapter 3 – where each rule is either a base rule or a chain rule. For each node $n$ in the SSA graph we define a Boolean vector $\mathbf{r}_n$ with a length equal to the number of base rules that match the node. Since every production is in linear form a pattern match only requires that the that the operator of the base rule's tree pattern matches the operator of a node in the SSA graph.

The costs of selecting a matching base rule is given as another vector $\mathbf{c}_n$ of equal length, where each element is the rule cost weighted against the estimated relative execution frequency of the node operation. This is needed to give higher priority to low-cost instructions for operations that reside in loops as these will have a greater impact on performance. With this we can define a cost function $f$ of covering the SSA graph as

$$f = \sum_{1 \leq n \leq |N|} \mathbf{r}_n^T \cdot \mathbf{c}_n$$

where $|N|$ is the number of nodes. We call this the accumulated *base cost* as it gives the total cost of applying base rules to cover the SSA graph, and the goal is to cover each SSA node exactly once (i.e. $\forall 1 \leq n \leq |N| : \mathbf{r}_n^T \cdot \mathbf{1} = 1$) such that the value of $f$ is minimized.

However, this does not necessarily produce a valid covering as there is no connection *between* the base rules, meaning that a selected base rule may reduce to a different nonterminal than what is required by another selected base rule. This is addressed by introducing a cost matrix $\mathbf{C}_{nm}$ for every pair of nodes $n$ and $m$ in the SSA graph where there exists a directed edge from $m$ to $n$. An element $c_{ij}$ in $\mathbf{C}_{nm}$ then reflects the cost of selecting rule $i$ for node $n$ and rule $j$ for node $m$, and the value is set as follows:

1. If rule $j$ reduces $m$ to the nonterminal expected at a certain position on the right-hand side in the production of rule $i$, then $c_{ij} = 0$.
2. If the nonterminal produced by rule $j$ can be reduced to the expected nonterminal via a series of chain rules, then $c_{ij} = \sum c_k$ where $c_k$ denotes the cost of an applied chain rule $k$.
3. Otherwise $c_{ij} = \infty$, which prevents selection of this rule combination.

These chain costs are calculated by computing the transitive closure for all chain rules. For this Eckstein et al. appears to have used the Floyd-Warshall algorithm [98], and Schäfer and Scholz [223] later discovered a method that computes the lowest cost for each $c_{ij}$ by finding the optimal sequence of chain rules. Lastly, the costs are weighted according to the execution frequency.

We now augment $f$ by adding the accumulated *chain cost*, resulting in

$$f = \sum_{1 \leq n < m \leq |N|} \mathbf{r}_n^T \cdot \mathbf{C}_{nm} \cdot \mathbf{r}_m + \sum_{1 \leq n \leq |N|} \mathbf{r}_n^T \cdot \mathbf{c}_n$$

which is solved using a heuristic PBQP solver (also developed by Eckstein et al.).

The astute reader may at this point have noticed that this scheme assumes that the SSA graph is not a *multigraph* – i.e. a graph where more than one edge can exist between the same pair of nodes – which prevents direct modeling of expressions such as `y = x + x`. Fortunately, the need for such edges can be removed by introducing new temporaries and linking these via value copies (see inlined code).

```
t = x;
y = x + t;
```

Eckstein et al. tested their implementation on a selected set of DSP problems which highlight the need for global instruction selection; the results indicate that the approach improved performance by 40–60% on average – and at most 82% for one problem – compared to a traditional tree-based instruction selector. The high gain in performance comes from the limitation of tree-based approaches where the program variables must be preassigned a nonterminal, which may have a detrimental effect on code quality; if chosen poorly additional assembly code must be inserted to undo these decisions, thus reducing performance. Unfortunately, the design by Eckstein et al. is restricted to tree patterns which hinders exploitation of many common target machine features such as multi-output instructions.

*Extending PBQP to DAG patterns*

In 2008 Ebner et al. [76] extended the PBQP model by Eckstein et al. to handle DAG patterns. By replacing the default instruction selector in LLVM 2.1 – a greedy DAG rewriter – the performance of the generated assembly code improved by an average of 13% for a set of selected problems compiled for the ARMv5 processor, with negligible impact on compilation time.

First the grammar format is extended to allow a rule to contain multiple productions – we refer to these as *complex rules* – which is done in a similar fashion as that of Scharwaechter et al. [224] (see page 72 in the previous chapter). We say that each production within a complex rule constitute a *proxy rule*. The PBQP problem is then augmented to accommodate the selection of complex rules, which essentially entails introducing new vectors and matrices that decide whether a complex rule is selected together with constraints that enforce that all corresponding proxy rules are also selected. We also need to prevent combinations of complex rules which incur cyclic data dependencies.

With more than one cost matrix it becomes necessary to be able to distinguish one from another. We say that all base and proxy rules belong to the $\mathcal{B}$ category and all complex rules belong to the $\mathcal{C}$ category. A cost matrix is then written as $\mathbf{C}^{\mathcal{X}\to\mathcal{Y}}$ which indicates that it concerns the costs of transitioning from category $\mathcal{X}$ to $\mathcal{Y}$. As an example, the cost matrix which we previously used to compute the accumulated chain cost is henceforth written as $\mathbf{C}_{nm}^{\mathcal{B}\to\mathcal{B}}$ since it only concerns the base rules. We can now proceed with augmenting the PBQP model.

First, for every node $n$ in the SSA graph the $\mathbf{r}_n$ vector is extended with the proxy rules that match at $n$. If two or more proxy rules derived from different complex rules are identical the length of the vector only increases by 1 element. Second, we create an *instance* of a complex rule for every permutation of distinct nodes where matched proxy rules can be combined into such a complex rule. Each instance gives rise to a 2-element decision vector $\mathbf{d}_i$ which indicates whether the complex rule instance $i$ is selected or not (i.e. a 1 in the first element indicates *not* *selected* and a 1 in the second element indicates *selected*).[2] We then accumulate the costs of the selected complex rules like we did for the base rules by adding

$$\sum_{1 \leq i \leq |I|} \mathbf{d}_i^T \cdot \mathbf{c}_i^{\mathcal{C}}$$

to $f$, where $I$ is the number of complex rule instances and $\mathbf{c}_i^{\mathcal{C}}$ is a 2-element cost vector whose elements are either 0 or the cost of the complex rule (depending on the arrangement of $\mathbf{d}_i$).

As previously stated, in order to select a complex rule instance $i$ all proxy rules of $i$ must also be selected. We enforce this via a cost matrix $\mathbf{C}_{ni}^{\mathcal{B} \to \mathcal{C}}$ where $n$ is a particular node in the SSA graph and $i$ is a particular instance of a complex rule. An element $c_{mj}$ in $\mathbf{C}_{ni}^{\mathcal{B} \to \mathcal{C}}$ is then set as follows:

- If $j$ represents that $i$ is not selected, then $c_{mj} = 0$.
- If $m$ is a base or proxy rule not associated with the complex rule of $i$, then $c_{mj} = 0$.
- Otherwise $c_{mj} = \infty$.

Hence by adding

$$\sum_{\substack{1 \leq n \leq |N| \\ 1 \leq i \leq |I|}} \mathbf{r}_n^T \cdot \mathbf{C}_{ni}^{\mathcal{B} \to \mathcal{C}} \cdot \mathbf{d}_i$$

to $f$ we force the selection of the necessary proxy if $\mathbf{d}_i$ is set as selected.

However, if the cost of all proxy rules is 0 then solutions are allowed where all proxy rules of a complex rule are selected but the instance itself is not selected. Ebner et al. solved this by setting a high cost $M$ to all proxy rules and modifying the cost of all complex rules to $\mathit{cost}(i) - |i|M$ where $|i|$ is the number of proxy rules required by the complex rule $i$. This offsets the artificial cost of the selected proxy patterns, thus bringing down the total cost of the selected proxy rules and complex pattern to $\mathit{cost}(i)$.

Lastly, we need a cost matrix $\mathbf{C}_{ij}^{\mathcal{C} \to \mathcal{C}}$ that prevents two complex pattern instances $i$ and $j$ to be selected simultaneously if they overlap or if such a combination incurs a cyclic data dependency. This is enforced by setting the elements in $\mathbf{C}_{ij}^{\mathcal{C} \to \mathcal{C}}$ that correspond to such situations to $\infty$, otherwise to 0, and augmenting $f$ accordingly. Hence the complete definition of $f$ becomes:

$$\begin{aligned} f = & \sum_{1 \leq i < j \leq |I|} \mathbf{d}_i^T \cdot \mathbf{C}_{ij}^{\mathcal{C} \to \mathcal{C}} \cdot \mathbf{d}_j + \sum_{\substack{1 \leq n \leq |N| \\ 1 \leq i \leq |I|}} \mathbf{r}_n^T \cdot \mathbf{C}_{ni}^{\mathcal{B} \to \mathcal{C}} \cdot \mathbf{d}_i + \sum_{1 \leq i \leq |I|} \mathbf{d}_i^T \cdot \mathbf{c}_i^{\mathcal{C}} \\ & + \sum_{1 \leq n < m \leq |N|} \mathbf{r}_n^T \cdot \mathbf{C}_{nm}^{\mathcal{B} \to \mathcal{B}} \cdot \mathbf{r}_m + \sum_{1 \leq n \leq |N|} \mathbf{r}_n^T \cdot \mathbf{c}_n^{\mathcal{B}} \end{aligned}$$

[2]Here one might wonder: Why not just use a Boolean variable instead of a 2-element vector? The reason has to do with the fact that the PBQP model must consist of only matrices and vectors, and for all vectors the sum of it must always be exactly 1.

*Using rewrite rules instead of grammar rules*

Another technique based on PBQP was presented in 2010 by Buchwald and Zwinkau [39]. Unlike Eckstein et al. and Ebner et al., Buchwald and Zwinkau approached instruction selection as a formal graph transformation problem – for which much previous work already exists – and expressed the machine instructions as rewrite rules instead of grammar rules. By using a formal foundation the instruction selector can be verified that it can handle all possible input programs using an automated tool; if this check fails the tool can provide the necessary rewrite rules that are missing from the instruction set.

Buchwald and Zwinkau's approach works as follows: first the SSA graph is converted into a DAG by splitting $\phi$ nodes into two nodes which effectively break the cycles. After having found all applicable rewrite rules for the SSA DAG – which is done using traditional pattern matching – a corresponding PBQP instance is formed and solved as before using a PBQP solver.

Buchwald and Zwinkau also discovered and addressed flaws in the PBQP solver by Eckstein et al. which may fail to find a solution in certain situations due to inadequate propagation of information. However, Buchwald and Zwinkau cautioned that their own implementation does not scale well when the number of overlapping patterns grows.

## 5.4 Approaches based on processor modeling

In the previous chapter we saw a microcode generation approach which models the entire data paths of the target processor instead of just the available instruction set. Here we will look at techniques which rely on the same modeling scheme but address the more traditional problem of instruction selection.

### 5.4.1 Chess

In 1990 Lanneer et al. [166] developed an approach which later was adopted by Van Praet et al. [249, 250] in their implementation of Chess, a well-known compiler targeting DSPs and ASIPs (Application-Specific Instruction-set Processors) which was the outcome of a European project.

When comparing Chess to MSSQ – which we read about in Section 4.3.5 – there are two striking differences. First, in MSSQ the data paths of the processor are given by a manually written description file, whereas in Chess the data paths are automatically derived from an nML-based specification [90, 91] which requires less manual effort.

Second, the method of bundling – i.e. combining operations for parallel execution – is different: in MSSQ the instruction selector relies on DAG covering by attempting to find subgraphs in the data path-modeling graph for covering expression trees, and after pattern selection a subsequent routine attempts to schedule the selected instructions in parallel. In contrast Chess takes a more incremental approach by formulating a *chaining graph*, where each node represents an operation in the expression graph (which is allowed to contain cycles). The nodes are annotated with the functional units on which the operation can be executed. On a DSP processor the functional units are commonly grouped into *functional building blocks* (FBBs), and an edge is added to the chaining graph for every pair of nodes that could potentially be executed on the same FBB (this is referred to as *chaining*). An heuristic algorithm then attempts to collapse nodes in the chaining graph by selecting an edge, replacing the two nodes with a new node, and then removing edges which can no longer be executed on the same FBB as the operations of the new node. This process iterates until no more nodes can be collapsed, where every remaining node in the chaining graph constitute a bundle. The same authors later extended this approach in [249] to considering selection between *all* possible bundles via branch-and-bound search, and allowing the same operations in the expression graph to appear in multiple bundles.

By forming bundles incrementally through the chaining graph, the approach of Van Praet et al. is capable of modeling entire functions as expression graphs and bundle operations across basic block boundaries. Its integrated nature also allows efficient assembly code to be generated for complex architectures, making it suitable for DSPs and ASIPs where the data paths are typically very irregular, and since it is graph-based it appears to be extendable to model machine instructions with internal loops. However, it is not clear how interdependent machine instructions are handled in this approach.

### 5.4.2 Simulated annealing

Another, albeit unusual, technique was proposed by Visser [251]. Like MSSQ and Chess, Visser's design is an integrated code generation technique, but applies the theories of *simulated annealing* [158] to improve the solutions. In brief terms, an initial solution is found by randomly mapping each node in the expression graph to a node in the hardware graph which models the processors, and a schedule is found using a heuristic list scheduler. A fitness function is then applied to judge the effectiveness of the solution – the exact details are omitted in the referenced paper – and various modifications are then performed in order to find more effective, neighboring solutions. A proof-of-concept prototype was developed and tested on a simple input program, but it appears no further research has been conducted on this idea.

## 5.5 Summary

In this chapter we have considered a number of instruction selection techniques that rely on graph covering. Compared to designs that operate solely on trees or DAGs, graph-based instruction selectors are the most powerful as both the input program and the machine instruction patterns can take arbitrary graphical shapes. This also enables global instruction selection as the expression graph can model an entire function of the input program instead of just single basic blocks, thus allowing for the generation of more efficient assembly code (as showcased by Eckstein et al.).

However, with graphs the pattern matching also becomes an NP-complete problem, thus requiring optimal graph covering to consist of *two* NP-complete problems instead of "just" one for DAG covering. Consequently, it is most likely that we will only see such approaches be applied in compilers where its users can afford very long compilation times (e.g. in embedded systems with extremely high demands on performance, code size, power consumption, or a combination thereof).

# 6

# Conclusions

In this report we have discussed, examined, and assessed a vast majority of the various approaches to instruction selection.[1] Starting with monolithic instruction selector – which were created ad-hoc and by hand – we have seen how the field advanced into retargetable macro-expanding designs, which later became overruled by the more powerful principle of tree covering. With the introduction of formal methodologies, it became possible of automatically generating the instruction selector from a declarative specification which describes the target machine. Tree covering in turn evolved into the more general forms which are DAG and graph covering at the expense of increased complexity. As optimal DAG and graph covering are both NP-complete problems, most such approaches apply heuristics to curb the search space. Simultaneously, developments were made to combine macro expansion with peephole optimization, which has proven a very effective methodology and is currently in use in GCC. An accessible overview of the studied approaches is given in Appendix A, and Appendix B contains a diagram which illustrates how the research has progressed over time.

## 6.1 Is instruction selection a closed problem?

Despite the tremendous advancements that have been made over the past 40 years, the instruction selection techniques constituting the current state-of-the-art still have several significant shortcomings; most notably, no approach – at least to my knowledge – is capable of modeling machine instructions with internal-loop characteristics. Today this impact is mitigated by augmenting the instruction selector with customized routines which detect when particular machine instructions can be exploited, but this is an error-prone and tedious task to implement. More flexible solutions are to use *compiler intrinsics* – which can be seen as additional node types in the expression graph that express more complicated operations such as $\sqrt{x}$ – or to make calls to target-specific library functions – which are implemented directly in assembly – to execute particular operations. However, neither is ideal as extending the compiler with additional intrinsics to take advantage of new machine instructions typically requires a signifi-cant amount of manual labor, and library functions must be rewritten by hand for each target machine.

[1] My ambition was to exhaust the entire body of literature, but this proved difficult as it is not clear where to draw the line as there exist many related problems – such as microcode generation and instruction set selection – that share many aspects with instruction selection. In addition, I have no doubts that at least one or two papers managed to escape my scrutiny.

In addition, as we stated in the introduction all three aspects of code generation must be taken into account simultaneously in order to generate truly optimal assembly code. Optimal instruction selection in isolation is futile for several reasons: for example, making efficient use of condition codes (or status flags as they are also called) is impossible without taking the instruction schedule into consideration as one must make sure that the flags are not prematurely overwritten by another instruction. The same goes for VLIW architectures where multiple machine instructions can be bundled and executed in parallel. Another problem is *rematerialization* which means that instead of retaining a value in register until its next use, the value is recomputed. If the target machine has few registers, or the cost of spilling is prohibitively high, this can be an effective method of improving code quality. However, rematerialization is only beneficial if it actually assist register allocation, or if it can be done with low overhead (e.g. by exploiting parallel execution). The bond between the instruction selection and register allocation becomes even tighter for target machines with multiple register classes that require a special set of instructions for accessing each register class.

Having said this, most contemporary techniques only consider instruction selection in isolation, and it is often unclear whether they can be fully and efficiently integrated with instruction scheduling and register allocation.

## 6.2 Where to go from here?

These problems notwithstanding, the techniques based on methods from combinatorial optimization – i.e. integer programming, constraint programming, SAT, etc.; a few such examples covered in this report include Wilson et al. [263], Bashford and Leupers [28], Bednarski and Kessler [29], and Floch et al. [97] – have shown considerable promise: first, the underlying modeling mechanisms facilitate an integrated code generation approach; second, auxiliary constraints can often be added to the already existing model, thus allowing interdependent machine instructions to be exploited without having to resort to hand-written assembly code or customized optimization routines; and third, recent advancements in solver technology have made these kinds of techniques viable options for code generation (see for example Castañeda Lozano et al. [43]). However, current implementations are still orders-of-magnitude slower than their heuristic counterparts, indicating that these ideas are in need of further research.

To conclude, although the field has come far since its initial attempts in the 1960s, instruction selection is still – despite common belief – an evasive problem. Moreover, with the current trend being integrated code generation and increasingly more complicated target machines, it appears that instruction selection is in greater need of increased understanding than ever before.

# 7

# Bibliography

[1] Adl-Tabatabai, A.-R., Langdale, G., Lucco, S., and Wahbe, R. "Efficient and Language-Independent Mobile Programs". In: *Proceedings of the ACM SIGPLAN 1996 Conference on Programming Language Design and Implementation*. PLDI '96. Philadelphia, Pennsylvania, USA: ACM, 1996, pp. 127–136. ISBN: 0-89791-795-2.

[2] Ahn, M., Youn, J. M., Choi, Y., Cho, D., and Paek, Y. "Iterative Algorithm for Compound Instruction Selection with Register Coalescing". In: *Proceedings of the 12th Euromicro Conference on Digital System Design, Architectures, Methods and Tools*. DSD '09. IEEE Press, 2009, pp. 513–520. ISBN: 978-0-7695-3782-5.

[3] Aho, A. V. and Johnson, S. C. "Optimal Code Generation for Expression Trees". In: *Journal of the ACM* 23.3 (1976), pp. 488–501. ISSN: 0004-5411.

[4] Aho, A. V., Johnson, S. C., and Ullman, J. D. "Code Generation for Expressions with Common Subexpressions". In: *Proceedings of the 3rd ACM SIGACT-SIGPLAN symposium on Principles on Programming Languages*. POPL '76. Atlanta, Georgia, USA: ACM, 1976, pp. 19–31.

[5] Aho, A. V. and Corasick, M. J. "Efficient String Matching: An Aid to Bibliographic Search". In: *Communications of the ACM* 18.6 (1975), pp. 333–340. ISSN: 0001-0782.

[6] Aho, A. V. and Ganapathi, M. "Efficient Tree Pattern Matching: An Aid to Code Generation". In: *Proceedings of the 12th ACM SIGACT-SIGPLAN symposium on Principles of Programming Languages*. POPL '85. New Orleans, Louisiana, USA: ACM, 1985, pp. 334–340. ISBN: 0-89791-147-4.

[7] Aho, A. V., Ganapathi, M., and Tjiang, S. W. K. "Code Generation Using Tree Matching and Dynamic Programming". In: *ACM Trans. Program. Lang. Syst.* 11.4 (1989), pp. 491–516. ISSN: 0164-0925.

[8] Aho, V. A., Sethi, R., and Ullman, J. D. *Compilers: Principles, Techniques, and Tools*. 2nd ed. Addison-Wesley, 2006. ISBN: 978-0321486813.

[9] Aigrain, P., Graham, S. L., Henry, R. R., McKusick, M. K., and Pelegrí-Llopart, E. "Experience with a Graham-Glanville Style Code Generator". In: *Proceedings of the 1984 SIGPLAN symposium on Compiler Construction*. SIGPLAN '84. Montreal, Canada: ACM, 1984, pp. 13–24. ISBN: 0-89791-139-3.

[10] Almer, O., Bennett, R., Böhm, I., Murray, A., Qu, X., Zuluaga, M., Franke, B., and Topham, N. "An End-to-End Design Flow for Automated Instruction Set Extension and Complex Instruction Selection based on GCC". In: *1st International Workshop on GCC Research Opportunities*. GROW '09. Paphos, Cyprus, 2009, pp. 49–60.

[11] Ammann, U., Nori, K. V., Jensen, K., and Nägeli, H. *The PASCAL (P) Compiler Implementation Notes*. Tech. rep. Eidgenössische Technishe Hochschule, Zürich, Switzerland: Instituts für Informatik, 1974.

[12] Ammann, U. *On Code Generation in a PASCAL Compiler*. Tech. rep. Eidgenössische Technishe Hochschule, Zürich, Switzerland: Instituts für Informatik, 1977.

[13] Anderson, J. P. "A Note on Some Compiling Algorithms". In: *Communications of the ACM* 7.3 (1964), pp. 149–150. ISSN: 0001-0782.

[14] Appel, A., Davidson, J., and Ramsey, N. *The Zephyr Compiler Infrastructure*. Tech. rep. 1998.

[15] Appel, A. W. and Palsberg, J. *Modern Compiler Implementation in Java*. 2nd ed. Cambridge University Press, 2002. ISBN: 0-521-82060-X.

[16] Araujo, G. and Malik, S. "Optimal Code Generation for Embedded Memory Non-Homogeneous Register Architectures". In: *Proceedings of the 8th International symposium on System Synthesis*. ISSS '95. Cannes, France, 1995, pp. 36–41. ISBN: 0-89791-771-5.

[17] Araujo, G., Malik, S., and Lee, M. T.-C. "Using Register-Transfer Paths in Code Generation for Heterogeneous Memory-Register Architectures". In: *Proceedings of the 33rd annual Design Automation Conference*. DAC '96. Las Vegas, Nevada, USA: ACM, 1996, pp. 591–596. ISBN: 0-89791-779-0.

[18] Arnold, M. *Matching and Covering with Multiple-Output Patterns*. Tech. rep. 1-68340-44. Delft, The Netherlands: Delft University of Technology, 1999.

[19] Arnold, M. and Corporaal, H. "Automatic Detection of Recurring Operation Patterns". In: *Proceedings of the seventh International Workshop on Hardware/Software Codesign*. CODES '99. Rome, Italy: ACM, 1999, pp. 22–26. ISBN: 1-58113-132-1.

[20] Arnold, M. and Corporaal, H. "Designing Domain-Specific Processors". In: *Proceedings of the 9th International symposium on Hardware/Software Codesign*. CODES '01. Copenhagen, Denmark: ACM, 2001, pp. 61–66. ISBN: 1-58113-364-2.

[21] Arora, N., Chandramohan, K., Pothineni, N., and Kumar, A. "Instruction Selection in ASIP Synthesis Using Functional Matching". In: *Proceedings of the 23rd International Conference on VLSI Design*. VLSID '10. 2010, pp. 146–151.

[22] Arslan, M. A. and Kuchcinski, K. "Instruction Selection and Scheduling for DSP Kernels on Custom Architectures". In: *Proceedings of the 16th EUROMICRO Conference on Digital System Design*. DSD '13. Santanader, Cantabria, Spain: IEEE Press, Sept. 4–6, 2013.

[23] Associated Compiler Experts, ACE. *COSY Compilers: Overview of Construction and Operation*. CoSy-8004-construct. 2003.

[24] Auslander, M. and Hopkins, M. “An Overview of the PL.8 Compiler”. In: *Proceedings of the 1982 SIGPLAN symposium on Compiler Construction*. SIGPLAN ’82. Boston, Massachusetts, USA: ACM, 1982, pp. 22–31. ISBN: 0-89791-074-5.

[25] Bailey, M. W. and Davidson, J. W. “Automatic Detection and Diagnosis of Faults in Generated Code for Procedure Calls”. In: *IEEE Transactions on Software Engineering* 29.11 (2003), pp. 1031–1042. ISSN: 0098-5589.

[26] Balachandran, A., Dhamdhere, D. M., and Biswas, S. “Efficient Retargetable Code Generation Using Bottom-Up Tree Pattern Matching”. In: *Computer Languages* 15.3 (1990), pp. 127–140. ISSN: 0096-0551.

[27] Balakrishnan, M., Bhatt, P. C. P., and Madan, B. B. “An Efficient Retargetable Microcode Generator”. In: *Proceedings of the 19th annual workshop on Microprogramming*. MICRO 19. New York, New York, USA: ACM, 1986, pp. 44–53. ISBN: 0-8186-0736-X.

[28] Bashford, S. and Leupers, R. “Constraint Driven Code Selection for Fixed-Point DSPs”. In: *Proceedings of the 36th annual ACM/IEEE Design Automation Conference*. DAC ’99. New Orleans, Louisiana, United States: ACM, 1999, pp. 817–822. ISBN: 1-58113-109-7.

[29] Bednarski, A. and Kessler, C. W. “Optimal Integrated VLIW Code Generation with Integer Linear Programming”. In: *Proceedings of the 12th International Euro-Par Conference*. Vol. 4128. Lecture Notes in Computer Science. Dresden, Germany: Springer, 2006, pp. 461–472.

[30] Bendersky, E. *A Deeper Look into the LLVM Code Generator: Part 1*. Feb. 25, 2013. URL: `http://eli.thegreenplace.net/2013/02/25/a-deeper-look-into-the-llvm-code-generator-part-1/` (visited on 05/10/2013).

[31] Borchardt, B. “Code Selection by Tree Series Transducers”. In: *Proceedings of the 9th International Conference on Implementation and Application of Automata*. CIAA ’04. Sophia Antipolis, France: Springer, 2004, pp. 57–67. ISBN: 978-3-540-24318-2.

[32] Boulytchev, D. “BURS-Based Instruction Set Selection”. In: *Proceedings of the 6th International Andrei Ershov memorial Conference on Perspectives of Systems Informatics*. PSI ’06. Novosibirsk, Russia: Springer, 2007, pp. 431–437. ISBN: 978-3-540-70880-3.

[33] Boulytchev, D. and Lomov, D. “An Empirical Study of Retargetable Compilers”. In: *Proceedings of the 4th International Andrei Ershov Memorial Conference on Perspectives of System Informatics (PSI ’01)*. Ed. by Bjørner, D., Broy, M., and Zamulin, A. V. Vol. 2244. Lecture Notes in Computer Science. Springer Berlin Heidelberg, 2001, pp. 328–335. ISBN: 978-3-540-43075-9.

[34] Brandner, F. “Completeness of Automatically Generated Instruction Selectors”. In: *Proceedings of the 21st IEEE International Conference on Application-specific Systems Architectures and Processors*. ASAP ’10. 2010, pp. 175–182.

[35] Brandner, F., Ebner, D., and Krall, A. “Compiler Generation from Structural Architecture Descriptions”. In: *Proceedings of the 2007 International Conference on Compilers, Architecture, and Synthesis for Embedded Systems*. CASES ’07. Salzburg, Austria: ACM, 2007, pp. 13–22. ISBN: 978-1-59593-826-8.

[36] Bravenboer, M. and Visser, E. “Rewriting Strategies for Instruction Selection”. In: *Proceedings of the 13th International Conference on Rewriting Techniques and Applications (RTA ’02)*. Ed. by Tison, S. Vol. 2378. Lecture Notes in Computer Science. Springer, 2002, pp. 237–251. ISBN: 978-3-540-43916-5.

[37] Brisk, P., Nahapetian, A., and Sarrafzadeh, M. “Instruction Selection for Compilers That Target Architectures with Echo Instructions”. In: *Proceedings of the 8th International Workshop on Software and Compilers for Embedded Systems (SCOPES ’04)*. Ed. by Schepers, H. Vol. 3199. Lecture Notes in Computer Science. Springer Berlin Heidelberg, 2004, pp. 229–243. ISBN: 978-3-540-23035-9.

[38] Bruno, J. and Sethi, R. “Code Generation for a One-Register Machine”. In: *Journal of the ACM* 23.3 (1976), pp. 502–510. ISSN: 0004-5411.

[39] Buchwald, S. and Zwinkau, A. “Instruction Selection by Graph Transformation”. In: *Proceedings of the 2010 International Conference on Compilers, Architectures and Synthesis for Embedded Systems*. CASES ’10. Scottsdale, Arizona, USA: ACM, 2010, pp. 31–40. ISBN: 978-1-60558-903-9.

[40] Cai, J., Paige, R., and Tarjan, R. “More Efficient Bottom-Up Multi-Pattern Matching in Trees”. In: *Theoretical Computer Science* 106.1 (1992), pp. 21–60. ISSN: 0304-3975.

[41] Canalda, P., Cognard, L., Despland, A., Jourdan, M., Mazaud, M., Parigot, D., Thomasset, F., and Voluceau, D. de. *PAGODE: A Realistic Back-End Generator*. Tech. rep. Rocquencourt, France: INRIA, 1995.

[42] Cao, Z., Dong, Y., and Wang, S. “Compiler Backend Generation for Application Specific Instruction Set Processors”. In: *Proceedings of the 9th Asian Symposium on Programming Languages and Systems (APLAS ’11)*. Ed. by Yang, H. Vol. 7078. Lecture Notes in Computer Science. Springer Berlin Heidelberg, 2011, pp. 121–136. ISBN: 978-3-642-25317-1.

[43] Castañeda Lozano, R., Carlsson, M., Drejhammar, F., and Schulte, C. “Constraint-Based Register Allocation and Instruction Scheduling”. In: *Proceedings of the 18th International Conference on the Principles and Practice of Constraint Programming (CP ’12)*. Ed. by Milano, M. Vol. 7514. Lecture Notes in Computer Science. Springer Berlin Heidelberg, 2012, pp. 750–766. ISBN: 978-3-642-33557-0.

[44] Cattell, R. G. “Automatic Derivation of Code Generators from Machine Descriptions”. In: *ACM Transactions on Programming Languages and Systems* 2.2 (1980), pp. 173–190. ISSN: 0164-0925.

[45] Cattell, R. G. G. *A Survey and Critique of Some Models of Code Generation*. Tech. rep. Pittsburgh, Pennsylvania, USA: School of Computer Science, Carnegie Mellon University, 1979.

[46] Cattell, R. G. G. “Formalization and Automatic Derivation of Code Generators”. AAI7815197. PhD thesis. Pittsburgh, Pennsylvania, USA: Carnegie Mellon University, 1978.

[47] Cattell, R. G., Newcomer, J. M., and Leverett, B. W. "Code Generation in a Machine-Independent Compiler". In: *Proceedings of the 1979 SIGPLAN symposium on Compiler construction*. SIGPLAN '79. Denver, Colorado, USA: ACM, 1979, pp. 65–75. ISBN: 0-89791-002-8.

[48] Ceruzzi, P. E. *A History of Modern Computing*. 2nd ed. The MIT Press, 2003. ISBN: 978-0262532037.

[49] Chase, D. R. "An Improvement to Bottom-Up Tree Pattern Matching". In: *Proceedings of the 14th ACM SIGACT-SIGPLAN symposium on Principles of programming languages*. POPL '87. Munich, West Germany: ACM, 1987, pp. 168–177. ISBN: 0-89791-215-2.

[50] Chen, D., Batson, R. G., and Dang, Y. *Applied Integer Programming: Modeling and Solution*. Wiley, 2010. ISBN: 978-0-470-37306-4.

[51] Chen, T., Lai, F., and Shang, R. "A Simple Tree Pattern Matching Algorithm for Code Generator". In: *Proceedings of the 19th annual International Conference on Computer Software and Applications*. COMPSAC '95. Dallas, Texas, USA: IEEE Press, 1995, pp. 162–167.

[52] Christopher, T. W., Hatcher, P. J., and Kukuk, R. C. "Using Dynamic Programming to Generate Optimized Code in a Graham-Glanville Style Code Generator". In: *Proceedings of the 1984 SIGPLAN symposium on Compiler construction*. SIGPLAN '84. Montreal, Canada: ACM, 1984, pp. 25–36. ISBN: 0-89791-139-3.

[53] Clark, N., Zhong, H., and Mahlke, S. "Processor Acceleration Through Automated Instruction Set Customization". In: *Proceedings of the 36th annual IEEE/ACM International Symposium on Microarchitecture*. MICRO 36. IEEE Press, 2003, pp. 129–140. ISBN: 0-7695-2043-X.

[54] Clark, N., Hormati, A., Mahlke, S., and Yehia, S. "Scalable Subgraph Mapping for Acyclic Computation Accelerators". In: *Proceedings of the 2006 International Conference on Compilers, Architecture and Synthesis for Embedded Systems*. CASES '06. Seoul, Korea: ACM, 2006, pp. 147–157. ISBN: 1-59593-543-6.

[55] Cole, R. and Hariharan, R. "Tree Pattern Matching and Subset Matching in Randomized $O(n \log^3 m)$ Time". In: *Proceedings of the 29th annual ACM symposium on Theory of Computing*. STOC '97. El Paso, Texas, USA: ACM, 1997, pp. 66–75. ISBN: 0-89791-888-6.

[56] Cong, J., Fan, Y., Han, G., and Zhang, Z. "Application-Specific Instruction Generation for Configurable Processor Architectures". In: *Proceedings of the 2004 ACM/SIGDA 12th International symposium on Field Programmable Gate Arrays*. FPGA '04. Monterey, California, USA: ACM, 2004, pp. 183–189. ISBN: 1-58113-829-6.

[57] Conway, M. E. "Proposal for an UNCOL". In: *Communications of the ACM* 1.10 (1958), pp. 5–8. ISSN: 0001-0782.

[58] Cook, S. A. "The Complexity of Theorem-Proving Procedures". In: *Proceedings of the 3rd annual ACM Symposium on Theory Of Computing*. STOC '71. Shaker Heights, Ohia, USA: ACM, 1971, pp. 151–158.

[59] Cooper, K. D. and Torczon, L. *Engineering a Compiler*. Morgan Kaufmann, 2004. ISBN: 1-55860-699-8.

[60] Cordella, L. P., Foggia, P., Sansone, C., and Vento, M. "An Improved Algorithm for Matching Large Graphs". In: *Proceedings of the 3rd IAPR-TC15 Workshop on Graph-based Representations in Pattern Recognition*. 2001, pp. 149–159.

[61] Cordone, R., Ferrandi, F., Sciuto, D., and Wolfler Calvo, R. "An Efficient Heuristic Approach to Solve the Unate Covering Problem". In: *Proceedings of the Conference and exhibition on Design, Automation and Test in Europe*. DATE '00. 2000, pp. 364–371.

[62] Cormen, T. H., Leiserson, C. E., Rivest, R. L., and Stein, C. *Introduction to Algorithms*. 3rd ed. The MIT Press, 2009. ISBN: 978-0262033848.

[63] Davidson, J. W. and Fraser, C. W. "Code Selection Through Object Code Optimization". In: *Transactions on Programming Languages and Systems* 6.4 (1984), pp. 505–526. ISSN: 0164-0925.

[64] Davidson, J. W. and Fraser, C. W. "Eliminating Redundant Object Code". In: *Proceedings of the 9th ACM SIGPLAN-SIGACT symposium on Principles of Programming Languages*. POPL '82. Albuquerque, New Mexico, USA: ACM, 1982, pp. 128–132. ISBN: 0-89791-065-6.

[65] Davidson, J. W. and Fraser, C. W. "The Design and Application of a Retargetable Peephole Optimizer". In: *Transactions on Programming Languages and Systems* 2.2 (1980), pp. 191–202. ISSN: 0164-0925.

[66] Despland, A., Mazaud, M., and Rakotozafy, R. "Code Generator Generation Based on Template-Driven Target Term Rewriting". In: *Rewriting Techniques and Applications*. Bordeaux, France: Springer-Verlag, 1987, pp. 105–120. ISBN: 0-387-17220-3.

[67] Despland, A., Mazaud, M., and Rakotozafy, R. "Using Rewriting Techniques to Produce Code Generators and Proving Them Correct". In: *Science of Computer Programming* 15.1 (1990), pp. 15–54. ISSN: 0167-6423.

[68] Deutsch, L. P. and Schiffman, A. M. "Efficient Implementation of the Smalltalk-80 System". In: *Proceedings of the 11th ACM SIGACT-SIGPLAN symposium on Principles of Programming Languages*. POPL '84. Salt Lake City, Utah, USA: ACM, 1984, pp. 297–302. ISBN: 0-89791-125-3.

[69] Dias, J. and Ramsey, N. "Automatically Generating Instruction Selectors Using Declarative Machine Descriptions". In: *Proceedings of the 37th annual ACM SIGPLAN-SIGACT symposium on Principles of Programming Languages*. POPL '10. Madrid, Spain: ACM, 2010, pp. 403–416. ISBN: 978-1-60558-479-9.

[70] Dias, J. and Ramsey, N. "Converting Intermediate Code to Assembly Code Using Declarative Machine Descriptions". In: *Proceedings of the 15th International Conference on Compiler Construction*. CC '06. Vienna, Austria: Springer-Verlag, 2006, pp. 217–231. ISBN: 3-540-33050-X, 978-3-540-33050-9.

[71] Dold, A., Gaul, T., Vialard, V., and Zimmermann, W. "ASM-Based Mechanized Verification of Compiler Back-Ends". In: *Proceedings of the 5th International Workshop on Abstract State Machines*. Ed. by Glässer, U. and Schmitt, P. H. Magdeburg, Germany, 1998, pp. 50–67.

[72] Donegan, M. K. "An Approach to the Automatic Generation of Code Generators". AAI7321548. PhD thesis. Houston, Texas, USA: Rice University, 1973.

[73] Dubiner, M., Galil, Z., and Magen, E. "Faster Tree Pattern Matching". In: *Journal of the ACM* 41.2 (1994), pp. 205–213. ISSN: 0004-5411.

[74] Earley, J. "An Efficient Context-Free Parsing Algorithm". In: *Communications of the ACM* 13.2 (1970), pp. 94–102. ISSN: 0001-0782.

[75] Ebeling, C. and Zaijicek, O. "Validating VLSI Circuit Layout by Wirelist Comparison". In: *Proceedings of the International Conference on Computer-Aided Design*. ICCAD '83. 1983, pp. 172–173.

[76] Ebner, D., Brandner, F., Scholz, B., Krall, A., Wiedermann, P., and Kadlec, A. "Generalized Instruction Selection Using SSA-Graphs". In: *Proceedings of the 2008 ACM SIGPLAN-SIGBED Conference on Languages, Compilers, and Tools for Embedded Systems*. LCTES '08. Tucson, Arizona, USA: ACM, 2008, pp. 31–40. ISBN: 978-1-60558-104-0.

[77] Eckstein, E., König, O., and Scholz, B. "Code Instruction Selection Based on SSA-Graphs". In: *Proceedings of the 7th International Workhops on Software and Compilers for Embedded Systems (SCOPES '03)*. Ed. by Krall, A. Vol. 2826. Lecture Notes in Computer Science. Springer Berlin Heidelberg, 2003, pp. 49–65. ISBN: 978-3-540-20145-8.

[78] Elson, M. and Rake, S. T. "Code-Generation Technique for Large-Language Compilers". In: *IBM Systems Journal* 9.3 (1970), pp. 166–188. ISSN: 0018-8670.

[79] Emmelmann, H., Schröer, F.-W., and Landwehr, R. "BEG: A Generator for Efficient Back Ends". In: *Proceedings of the ACM SIGPLAN 1989 Conference on Programming Language Design and Implementation*. PLDI '89. Portland, Oregon, USA: ACM, 1989, pp. 227–237. ISBN: 0-89791-306-X.

[80] Emmelmann, H. "Testing Completeness of Code Selector Specifications". In: *Proceedings of the 4th International Conference on Compiler Construction*. CC '92. Springer-Verlag, 1992, pp. 163–175. ISBN: 3-540-55984-1.

[81] Engelfriet, J., Fülöp, Z., and Vogler, H. "Bottom-Up and Top-Down Tree Series Transformations". In: *Journal of Automata, Languages and Combinatorics* 7.1 (July 2001), pp. 11–70. ISSN: 1430-189X.

[82] Engler, D. R. "VCODE: A Retargetable, Extensible, Very Fast Dynamic Code Generation System". In: *Proceedings of the ACM SIGPLAN 1996 Conference on Programming Language Design and Implementation*. PLDI '96. Philadelphia, Pennsylvania, USA: ACM, 1996, pp. 160–170. ISBN: 0-89791-795-2.

[83] Engler, D. R. and Proebsting, T. A. "DCG: An Efficient, Retargetable Dynamic Code Generation System". In: *Proceedings of the sixth International Conference on Architectural Support for Programming Languages and Operating Systems*. ASPLOS VI. San Jose, California, USA: ACM, 1994, pp. 263–272. ISBN: 0-89791-660-3.

[84] Eriksson, M. V., Skoog, O., and Kessler, C. W. "Optimal vs. Heuristic Integrated Code Generation for Clustered VLIW Architectures". In: *Proceedings of the 11th International Workshop on Software & Compilers for Embedded Systems*. SCOPES '08. Munich, Germany: ACM, 2008, pp. 11–20.

[85] Ertl, M. A. "Optimal Code Selection in DAGs". In: *Proceedings of the 26th ACM SIGPLAN-SIGACT symposium on Principles Of Programming Languages*. POPL '99. San Antonio, Texas, USA: ACM, 1999, pp. 242–249. ISBN: 1-58113-095-3.

[86] Ertl, M. A., Casey, K., and Gregg, D. "Fast and Flexible Instruction Selection with On-Demand Tree-Parsing Automata". In: *Proceedings of the 2006 ACM SIGPLAN Conference on Programming language design and implementation*. PLDI '06. Ottawa, Ontario, Canada: ACM, 2006, pp. 52–60. ISBN: 1-59593-320-4.

[87] Fan, W., Li, J., Ma, S., Tang, N., Wu, Y., and Wu, Y. "Graph Pattern Matching: From Intractable to Polynomial Time". In: *Proceedings of the VLDB Endowment* 3.1-2 (2010), pp. 264–275. ISSN: 2150-8097.

[88] Fan, W., Li, J., Luo, J., Tan, Z., Wang, X., and Wu, Y. "Incremental Graph Pattern Matching". In: *Proceedings of the 2011 ACM SIGMOD International Conference on Management Of Data*. SIGMOD '11. Athens, Greece: ACM, 2011, pp. 925–936. ISBN: 978-1-4503-0661-4.

[89] Farfeleder, S., Krall, A., Steiner, E., and Brandner, F. "Effective Compiler Generation by Architecture Description". In: *Proceedings of the 2006 ACM SIGPLAN/SIGBED Conference on Language, Compilers, and Tool support for Embedded Systems*. LCTES '06. Ottawa, Ontario, Canada: ACM, 2006, pp. 145–152. ISBN: 1-59593-362-X.

[90] Fauth, A., Van Praet, J., and Freericks, M. "Describing Instruction Set Processors Using nML". In: *Proceedings of the 1995 European Conference on Design and Test*. EDTC '95. IEEE Press, 1995, pp. 503–507. ISBN: 0-8186-7039-8.

[91] Fauth, A., Freericks, M., and Knoll, A. "Generation of Hardware Machine Models from Instruction Set Descriptions". In: *Proceedings of the IEEE Workshop on VLSI Signal Processing*. VI' 93. 1993, pp. 242–250.

[92] Fauth, A., Hommel, G., Knoll, A., and Müller, C. "Global Code Selection of Directed Acyclic Graphs". In: *Proceedings of the 5th International Conference on Compiler Construction*. CC '94. Springer-Verlag, 1994, pp. 128–142. ISBN: 3-540-57877-3.

[93] Feldman, J. and Gries, D. "Translator Writing Systems". In: *Communications of the ACM* 11.2 (1968), pp. 77–113. ISSN: 0001-0782.

[94] Ferdinand, C., Seidl, H., and Wilhelm, R. "Tree Automata for Code Selection". In: *Acta Informatica* 31.9 (1994), pp. 741–760. ISSN: 0001-5903.

[95] Fernández, M. and Ramsey, N. "Automatic Checking of Instruction Specifications". In: *Proceedings of the 19th International Conference on Software Engineering*. ICSE '97. Boston, Massachusetts, USA: ACM, 1997, pp. 326–336. ISBN: 0-89791-914-9.

[96] Fischer, C. N., Cytron, R. K., and LeBlanc, R. J. J. *Crafting a Compiler*. Pearson, 2009. ISBN: 978-0138017859.

[97] Floch, A., Wolinski, C., and Kuchcinski, K. "Combined Scheduling and Instruction Selection for Processors with Reconfigurable Cell Fabric". In: *Proceedings of the 21st International Conference on Application-specific Systems Architectures and Processors*. ASAP '10. 2010, pp. 167–174.

[98] Floyd, R. W. "Algorithm 97: Shortest Path". In: *Communications of the ACM* 5.6 (1962), p. 345. ISSN: 0001-0782.

[99] Floyd, R. W. "An Algorithm for Coding Efficient Arithmetic Operations". In: *Communications of the ACM* 4.1 (1961), pp. 42–51. ISSN: 0001-0782.

[100] Fraser, C. W. "A Language for Writing Code Generators". In: *Proceedings of the ACM SIGPLAN 1989 Conference on Programming Language Design and Implementation*. PLDI '89. Portland, Oregon, USA: ACM, 1989, pp. 238–245. ISBN: 0-89791-306-X.

[101] Fraser, C. W. and Wendt, A. L. "Automatic Generation of Fast Optimizing Code Generators". In: *Proceedings of the ACM SIGPLAN 1988 Conference on Programming Language Design and Implementation*. PLDI '88. Atlanta, Georgia, USA: ACM, 1988, pp. 79–84. ISBN: 0-89791-269-1.

[102] Fraser, C. W. "A Compact, Machine-Independent Peephole Optimizer". In: *Proceedings of the 6th ACM SIGACT-SIGPLAN symposium on Principles of Programming Languages*. POPL '79. San Antonio, Texas, USA: ACM, 1979, pp. 1–6.

[103] Fraser, C. W. and Proebsting, T. A. "Finite-State Code Generation". In: *Proceedings of the ACM SIGPLAN 1999 Conference on Programming Language Design and Implementation*. PLDI '99. Atlanta, Georgia, USA: ACM, 1999, pp. 270–280. ISBN: 1-58113-094-5.

[104] Fraser, C. W., Hanson, D. R., and Proebsting, T. A. "Engineering a Simple, Efficient Code-Generator Generator". In: *ACM Letters on Programming Languages and Systems* 1.3 (1992), pp. 213–226. ISSN: 1057-4514.

[105] Fraser, C. W., Henry, R. R., and Proebsting, T. A. "BURG – Fast Optimal Instruction Selection and Tree Parsing". In: *SIGPLAN Notices* 27.4 (1992), pp. 68–76. ISSN: 0362-1340.

[106] Fraser, C. W. "A Knowledge-Based Code Generator Generator". In: *Proceedings of the Symposium on Artificial Intelligence and Programming Languages*. ACM, 1977, pp. 126–129.

[107] Fraser, C. W. "Automatic Generation of Code Generators". PhD thesis. New Haven, Connecticut, USA: Yale University, 1977.

[108] Fröhlich, S., Gotschlich, M., Krebelder, U., and Wess, B. "Dynamic Trellis Diagrams for Optimized DSP Code Generation". In: *Proceedings of the IEEE International symposium on Circuits and Systems*. ISCAS '99. IEEE Press, 1999, pp. 492–495.

[109] Gallagher, B. *The State of the Art in Graph-Based Pattern Matching*. Tech. rep. UCRL-TR-220300. Livermore, California, USA: Lawrence Livermore National Laboratory, Mar. 31, 2006.

[110] Galuzzi, C. and Bertels, K. "The Instruction-Set Extension Problem: A Survey". In: *Reconfigurable Computing: Architectures, Tools and Applications (ARC '08)*. Ed. by Woods, R., Compton, K., Bouganis, C., and Diniz, P. C. Vol. 4943. Lecture Notes in Computer Science. Springer Berlin Heidelberg, 2008, pp. 209–220. ISBN: 978-3-540-78609-2.

[111] Ganapathi, M. "Prolog Based Retargetable Code Generation". In: *Computer Languages* 14.3 (1989), pp. 193–204. ISSN: 0096-0551.

[112] Ganapathi, M. "Retargetable Code Generation and Optimization Using Attribute Grammars". AAI8107834. PhD thesis. Madison, Wisconsin, USA: The University of Wisconsin–Madison, 1980.

[113] Ganapathi, M. and Fischer, C. N. "Affix Grammar Driven Code Generation". In: *ACM Transactions on Programming Languages and Systems* 7.4 (1985), pp. 560–599. ISSN: 0164-0925.

[114] Ganapathi, M. and Fischer, C. N. "Description-Driven Code Generation Using Attribute Grammars". In: *Proceedings of the 9th ACM SIGPLAN-SIGACT symposium on Principles of programming languages*. POPL '82. Albuquerque, New Mexico, USA: ACM, 1982, pp. 108–119. ISBN: 0-89791-065-6.

[115] Ganapathi, M. and Fischer, C. N. *Instruction Selection by Attributed Parsing*. Tech. rep. No. 84-256. Stanford, California, USA: Stanford University, 1984.

[116] Ganapathi, M., Fischer, C. N., and Hennessy, J. L. "Retargetable Compiler Code Generation". In: *ACM Computing Surveys* 14.4 (1982), pp. 573–592. ISSN: 0360-0300.

[117] Gecseg, F. and Steinby, M. *Tree Automata*. Akademiai Kiado, 1984. ISBN: 978-96305317-02.

[118] Genin, D., De Moortel, J., Desmet, D., and Velde, E. Van de. "System design, Optimization and Intelligent Code Generation for Standard Digital Signal Processors". In: *Proceedings of the IEEE International Symposium on Circuits and Systems*. ISCAS '90. 1989, pp. 565–569.

[119] Giegerich, R. "A Formal Framework for the Derivation of Machine-Specific Optimizers". In: *Transactions on Programming Languages and Systems* 5.3 (1983), pp. 478–498. ISSN: 0164-0925.

[120] Giegerich, R. and Schmal, K. "Code Selection Techniques: Pattern Matching, Tree Parsing, and Inversion of Derivors". In: *Proceedings of the 2nd European Symposium on Programming*. Ed. by Ganzinger, H. Vol. 300. ESOP '88. Nancy, France: Springer, 1998, pp. 247–268. ISBN: 978-3-540-19027-1.

[121] Glanville, R. S. and Graham, S. L. "A New Method for Compiler Code Generation". In: *Proceedings of the 5th ACM SIGACT-SIGPLAN symposium on Principles of programming languages*. POPL '78. Tucson, Arizona: Springer, 1978, pp. 231–254.

[122] Goldberg, D. E. *Genetic Algorithms in Search, Optimization, and Machine Learning*. Addison-Wesley, 1989. ISBN: 978-0201157673.

[123] Goldberg, E. I., Carloni, L. P., Villa, T., Brayton, R. K., and Sangiovanni-Vincentelli, A. L. "Negative Thinking in Branch-and-Bound: The Case of Unate Covering". In: *Transactions of Computer-Aided Design of Integrated Ciruits and Systems* 19.3 (2006), pp. 281–294. ISSN: 0278-0070.

[124] Gough, K. J. *Bottom-Up Tree Rewriting Tool MBURG*. Tech. rep. Brisbane, Australia: Faculty of Information Technology, Queensland University of Technology, July 18, 1995.

[125] Gough, K. J. and Ledermann, J. "Optimal Code-Selection using MBURG". In: *Proceedings of the 20th Australasian Computer Science Conference*. ACSC '97. Sydney, Australia, 1997.

[126] Graham, S. L. "Table-Driven Code Generation". In: *Computer* 13.8 (1980). IEEE Press, pp. 25–34. ISSN: 0018-9162.

[127] Graham, S. L., Henry, R. R., and Schulman, R. A. "An Experiment in Table Driven Code Generation". In: *Proceedings of the 1982 SIGPLAN symposium on Compiler construction*. SIGPLAN '82. Boston, Massachusetts, USA: ACM, 1982, pp. 32–43. ISBN: 0-89791-074-5.

[128] Granlund, T. and Kenner, R. "Eliminating Branches Using a Superoptimizer and the GNU C Compiler". In: *Proceedings of the ACM SIGPLAN 1992 Conference on Programming Language Design and Implementation*. PLDI '92. San Francisco, California, USA: ACM, 1992, pp. 341–352. ISBN: 0-89791-475-9.

[129] Guo, Y., Smit, G. J., Broersma, H., and Heysters, P. M. "A Graph Covering Algorithm for a Coarse Grain Reconfigurable System". In: *Proceedings of the 2003 ACM SIGPLAN Conference on Language, Compiler, and Tool for Embedded Systems*. LCTES '03. San Diego, California, USA: ACM, 2003, pp. 199–208. ISBN: 1-58113-647-1.

[130] Hanono, S. and Devadas, S. "Instruction Selection, Resource Allocation, and Scheduling in the AVIV Retargetable Code Generator". In: *Proceedings of the 35th annual Design Automation Conference*. DAC '98. San Francisco, California, USA: ACM, 1998, pp. 510–515. ISBN: 0-89791-964-5.

[131] Hanono, S. Z. "AVIV: A Retargetable Code Generator for Embedded Processors". AAI7815197. PhD thesis. Cambridge, Massachusetts, USA: Massachusetts Institute of Technology, 1999.

[132] Hanson, D. R. and Fraser, C. W. *A Retargetable C Compiler: Design and Implementation*. Addison-Wesley, 1995. ISBN: 978-0805316704.

[133] Harrison, W. H. "A New Strategy for Code Generation the General-Purpose Optimizing Compiler". In: *Transactions Software Engineering* 5.4 (1979), pp. 367–373. ISSN: 0098-5589.

[134] Hatcher, P. J. and Christopher, T. W. "High-Quality Code Generation via Bottom-Up Tree Pattern Matching". In: *Proceedings of the 13th ACM SIGACT-SIGPLAN symposium on Principles of Programming Languages*. POPL '86. St. Petersburg Beach, Florida, USA: ACM, 1986, pp. 119–130.

[135] Hatcher, P. "The Equational Specification of Efficient Compiler Code Generation". In: *Computer Languages* 16.1 (1991), pp. 81–95. ISSN: 0096-0551.

[136] Hatcher, P. and Tuller, J. W. "Efficient Retargetable Compiler Code Generation". In: *Proceedings for the International Conference on Computer Languages*. Miami Beach, Florida, USA: IEEE Press, 1988, pp. 25–30. ISBN: 0-8186-0874-9.

[137] Henry, R. R. *Encoding Optimal Pattern Selection in Table-Driven Bottom-Up Tree-Pattern Matcher*. Tech. rep. 89-02-04. Seattle, Washington, USA: University of Washington, 1989.

[138] Henry, R. R. "Graham-Glanville Code Generators". UCB/CSD-84-184. PhD thesis. Berkeley, California, USA: EECS Department, University of California, May 1984.

[139] Hino, T., Suzuki, Y., Uchida, T., and Itokawa, Y. "Polynomial Time Pattern Matching Algorithm for Ordered Graph Patterns". In: *Proceedings of the 22nd International Conference on Inductive Logic Programming*. ILP '12. Dubrovnik, Croatia, 2012.

[140] Hoffmann, C. M. and O'Donnell, M. J. "Pattern Matching in Trees". In: *Journal of the ACM* 29.1 (1982), pp. 68–95. ISSN: 0004-5411.

[141] Hoover, R. and Zadeck, K. "Generating Machine Specific Optimizing Compilers". In: *Proceedings of the 23rd ACM SIGPLAN-SIGACT symposium on Principles of Programming Languages*. POPL '96. St. Petersburg Beach, Florida, USA: ACM, 1996, pp. 219–229. ISBN: 0-89791-769-3.

[142] Hormati, A., Clark, N., and Mahlke, S. "Exploiting Narrow Accelerators with Data-Centric Subgraph Mapping". In: *Proceedings of the International Symposium on Code Generation and Optimization*. CGI '07. 2007, pp. 341–353.

[143] Johnson, S. C. "A Portable Compiler: Theory and Practice". In: *Proceedings of the 5th ACM SIGACT-SIGPLAN symposium on Principles of Programming Languages*. POPL '78. Tucson, Arizona, USA: ACM, 1978, pp. 97–104.

[144] Johnson, S. C. "A Tour Through the Portable C Compiler". In: *Unix Programmer's Manual*. 7th ed. 1981.

[145] Joshi, R., Nelson, G., and Randall, K. "Denali: A Goal-Directed Superoptimizer". In: *Proceedings of the ACM SIGPLAN 2002 Conference on Programming Language Design and Implementation*. PLDI '02. Berlin, Germany: ACM, 2002, pp. 304–314. ISBN: 1-58113-463-0.

[146] Joshi, R., Nelson, G., and Zhou, Y. "Denali: A Practical Algorithm for Generating Optimal Code". In: *Transactions on Programming Languages and Systems* 28.6 (2006), pp. 967–989. ISSN: 0164-0925.

[147] Kang, K. "A Study on Generating an Efficient Bottom-Up Tree Rewrite Machine for JBURG". In: *Proceedings of the International Conference on Computational Science and Its Applications (ICCSA '04)*. Ed. by Laganá, A., Gavrilova, M., Kumar, V., Mun, Y., Tan, C., and Gervasi, O. Vol. 3043. Lecture Notes in Computer Science. Assisi, Italy: Springer Berlin Heidelberg, 2004, pp. 65–72. ISBN: 978-3-540-22054-1.

[148] Kang, K. and Choe, K. *On the Automatic Generation of Instruction Selector Using Bottom-Up Tree Pattern Matching*. Tech. rep. CS/TR-95-93. Daejeon, South Korea: Korea Advanced Institute of Science and Technology, 1995.

[149] Karp, R. M., Miller, R. E., and Rosenberg, A. L. "Rapid Identification of Repeated Patterns in Strings, Trees and Arrays". In: *Proceedings of the 4th Annual ACM Symposium on Theory of Computing*. STOC '72. Denver, Colorado, USA: ACM, 1972, pp. 125–136.

[150] Kastner, R., Kaplan, A., Memik, S. O., and Bozorgzadeh, E. "Instruction Generation for Hybrid Reconfigurable Systems". In: *Transactions on Design Automation of Electronic Systems* 7.4 (2002), pp. 605–627. ISSN: 1084-4309.

[151] Kastner, R., Ogrenci-Memik, S., Bozorgzadeh, E., and Sarrafzadeh, M. "Instruction Generation for Hybrid Reconfigurable Systems". In: *Proceedings of the International Conference on Computer-Aided Design*. ICCAD '01. San Jose, California, USA, Nov. 2001, pp. 127–130.

[152] Kessler, C. W. and Bednarski, A. "A Dynamic Programming Approach to Optimal Integrated Code Generation". In: *Proceedings of the Conference on Languages, Compilers, and Tools for Embedded Systems*. LCTES '01. ACM, 2001, pp. 165–174.

[153] Kessler, C. W. and Bednarski, A. "Optimal Integrated Code Generation for Clustered VLIW Architectures". In: *Proceedings of the ACM SIGPLAN joint Conference on Languages, Compilers, and Tools for Embedded Systems and Software and Compilers for Embedded Systems*. LCTES '02/SCOPES '02. ACM, 2002, pp. 102–111.

[154] Kessler, P. B. "Discovering Machine-Specific Code Improvements". In: *Proceedings of the 1986 SIGPLAN symposium on Compiler Construction*. SIGPLAN '86. Palo Alto, California, USA: ACM, 1986, pp. 249–254. ISBN: 0-89791-197-0.

[155] Kessler, R. R. "PEEP: An Architectural Description Driven Peephole Optimizer". In: *Proceedings of the 1984 SIGPLAN symposium on Compiler Construction*. SIGPLAN '84. Montreal, Canada: ACM, 1984, pp. 106–110. ISBN: 0-89791-139-3.

[156] Keutzer, K. "DAGON: Technology Binding and Local Optimization by DAG Matching". In: *Proceedings of the 24th ACM/IEEE Design Automation Conference*. DAC '87. ACM, 1987, pp. 341–347.

[157] Khedker, U. "Workshop on Essential Abstractions in GCC". Lecture. GCC Resource Center, Department of Computer Science and Engineering, IIT Bombay. Bombay, India, June 30–July 3, 2012.

[158] Kirkpatrick, S., Gelatt, C. D., and Vecchi, M. P. "Optimization by Simulated Annealing". In: *Science* 220.4598 (1983), pp. 671–680.

[159] Knuth, D. E. "Semantics of Context-Free Languages". In: *Mathematical Systems Theory* 2.2 (1968), pp. 127–145.

[160] Knuth, D. E., Morris, J. H. J., and Pratt, V. R. "Fast Pattern Matching in Strings". In: *SIAM Journal of Computing* 6.2 (1977), pp. 323–350. ISSN: 0097-5397.

[161] Koes, D. R. and Goldstein, S. C. "Near-Optimal Instruction Selection on DAGs". In: *Proceedings of the 6th annual IEEE/ACM International symposium on Code Generation and Optimization*. CGO '08. Boston, Massachusetts, USA: ACM, 2008, pp. 45–54. ISBN: 978-1-59593-978-4.

[162] Kreuzer, W., Gotschlich, W., and B, W. "REDACO: A Retargetable Data Flow Graph Compiler for Digital Signal Processors". In: *Proceedings of the International Conference on Signal Processing Applications and Technology*. ICSPAT '96. 1996.

[163] Krissinel, E. B. and Henrick, K. "Common Subgraph Isomorphism Detection by Backtracking Search". In: *Software–Practice & Experience* 34.6 (2004), pp. 591–607. ISSN: 0038-0644.

[164] Krumme, D. W. and Ackley, D. H. "A Practical Method for Code Generation Based on Exhaustive Search". In: *Proceedings of the 1982 SIGPLAN symposium on Compiler Construction*. SIGPLAN '82. Boston, Massachusetts, USA: ACM, 1982, pp. 185–196. ISBN: 0-89791-074-5.

[165] Langevin, M. and Cerny, E. "An automata-theoretic approach to local microcode generation". In: *Proceedings of the 4th European Conference on Design Automation = with the European Event in ASIC Design*. EDAC '93. 1993, pp. 94–98.

[166] Lanneer, D., Catthoor, F., Goossens, G., Pauwels, M., Van Meerbergen, J., and De Man, H. "Open-Ended System for High-Level Synthesis of Flexible Signal Processors". In: *Proceedings of the Conference on European Design Automation*. EURO-DAC '90. Glasgow, Scotland: IEEE Press, 1990, pp. 272–276. ISBN: 0-8186-2024-2.

[167] Lattner, C. and Adve, V. "LLVM: A Compilation Framework for Lifelong Program Analysis & Transformation". In: *Proceedings of the International Symposium on Code Generation and Optimization: Feedback-Directed and Runtime Optimization*. CGO '04. Palo Alto, California: IEEE Computer Society, 2004, pp. 75–86. ISBN: 0-7695-2102-9.

[168] Leupers, R. and Marwedel, P. "Instruction Selection for Embedded DSPs with Complex Instructions". In: *Proceedings of the Conference on European Design Automation*. EURO-DAC '96/EURO-VHDL '96. Geneva, Switzerland: IEEE Press, 1996, pp. 200–205. ISBN: 0-8186-7573-X.

[169] Leupers, R. "Code Generation for Embedded Processors". In: *Proceedings of the 13th International Symposium on System Synthesis*. ISSS '00. Madrid, Spain: IEEE Press, 2000, pp. 173–178. ISBN: 1-58113-267-0.

[170] Leupers, R. "Code Selection for Media Processors with SIMD Instructions". In: *Proceedings of the Conference on Design, Automation and Test in Europe*. DATE '00. Paris, France: ACM, 2000, pp. 4–8. ISBN: 1-58113-244-1.

[171] Leupers, R. and Bashford, S. "Graph-Based Code Selection Techniques for Embedded Processors". In: *Transactions on Design Automation of Electronic Systems* 5 (4 2000), pp. 794–814. ISSN: 1084-4309.

[172] Leupers, R. and Marwedel, P. "Retargetable Code Generation Based on Structural Processor Description". In: *Design Automation for Embedded Systems* 3.1 (1998), pp. 75–108. ISSN: 0929-5585.

[173] Leupers, R. and Marwedel, P. *Retargetable Compiler Technology for Embedded Systems*. Kluwer Academic Publishers, 2001. ISBN: 0-7923-7578-5.

[174] Leupers, R. and Marwedel, P. "Retargetable Generation of Code Selectors from HDL Processor Models". In: *Proceedings of the 1997 European Design and Test Conference*. EDTC '97. Paris, France: IEEE Press, 1997, pp. 140–144.

[175] Leupers, R. and Marwedel, P. "Time-Constrained Code Compaction for DSPs". In: *Proceedings of the 8th International Symposium on System Synthesis*. ISSS '95. Cannes, France: ACM, 1995, pp. 54–59. ISBN: 0-89791-771-5.

[176] Leverett, B. W., Cattell, R. G. G., Hobbs, S. O., Newcomer, J. M., Reiner, A. H., Schatz, B. R., and Wulf, W. A. "An Overview of the Production-Quality Compiler-Compiler Project". In: *Computer* 13.8 (1980). IEEE Press, pp. 38–49. ISSN: 0018-9162.

[177] Liao, S., Keutzer, K., Tjiang, S., and Devadas, S. "A New Viewpoint on Code Generation for Directed Acyclic Graphs". In: *ACM Transactions on Design Automation of Electronic Systems* 3.1 (1998), pp. 51–75. ISSN: 1084-4309.

[178] Liao, S., Devadas, S., Keutzer, K., and Tjiang, S. "Instruction Selection Using Binate Covering for Code Size Optimization". In: *Proceedings of the 1995 IEEE/ACM International Conference on Computer-Aided Design*. ICCAD '95. San Jose, California, USA: IEEE Press, 1995, pp. 393–399. ISBN: 0-8186-7213-7.

[179] Liem, C., May, T., and Paulin, P. "Instruction-Set Matching and Selection for DSP and ASIP Code Generation". In: *Proceedings of 1994 European Design and Test Conference (EDAC '94/ETC '94/EUROASIC '94)*. IEEE Press, 1994, pp. 31–37.

[180] Lowry, E. S. and Medlock, C. W. "Object Code Optimization". In: *Communications of the ACM* 12.1 (1969), pp. 13–22. ISSN: 0001-0782.

[181] Madhavan, M., Shankar, P., Rai, S., and Ramakrishna, U. "Extending Graham-Glanville Techniques for Optimal Code Generation". In: *ACM Trans. Program. Lang. Syst.* 22.6 (2000), pp. 973–1001. ISSN: 0164-0925.

[182] Mahmood, M., Mavaddat, F., and Elmastry, M. "Experiments with an Efficient Heuristic Algorithm for Local Microcode Generation". In: *Proceedings of the IEEE International Conference on Computer Design: VLSI in Computers and Processors*. ICCD '90. 1990, pp. 319–323.

[183] Martin, K., Wolinski, C., Kuchcinski, K., Floch, A., and Charot, F. "Constraint-Driven Instructions Selection and Application Scheduling in the DURASE System". In: *Proceedings of the 20th IEEE International Conference on Application-specific Systems, Architectures and Processors*. ASAP '09. IEEE Press, 2009, pp. 145–152. ISBN: 978-0-7695-3732-0.

[184] Marwedel, P. "Code Generation for Core Processors". In: *Proceedings of the 1997 Design Automation Conference*. DAC '97. Anaheim, California, USA: IEEE Press, 1997, pp. 232–237. ISBN: 0-7803-4093-0.

[185] Marwedel, P. "The MIMOLA Design System: Tools for the Design of Digital Processors". In: *Proceedings of the 21st Design Automation Conference*. DAC '84. Albuquerque, New Mexico, USA: IEEE Press, 1984, pp. 587–593. ISBN: 0-8186-0542-1.

[186] Marwedel, P. "Tree-Based Mapping of Algorithms to Predefined Structures". In: *Proceedings of the 1993 IEEE/ACM International Conference on Computer-Aided Design*. ICCAD '93. Santa Clara, California, USA: IEEE Press, 1993, pp. 586–593. ISBN: 0-8186-4490-7.

[187] Massalin, H. "Superoptimizer: A Look at the Smallest Program". In: *Proceedings of the 2nd International Conference on Architectual Support for Programming Languages and Operating Systems*. ASPLOS II. Palo Alto, California, USA: IEEE Press, 1987, pp. 122–126. ISBN: 0-8186-0805-6.

[188] McKeeman, W. M. "Peephole Optimization". In: *Communications of the ACM* 8.7 (July 1965), pp. 443–444. ISSN: 0001-0782.

[189] Miller, P. L. "Automatic Creation of a Code Generator from a Machine Description". MA thesis. Cambridge, Massachusetts, USA: Massachusetts Institute of Technology, 1971.

[190] Muchnick, S. *Advanced Compiler Design & Implementation*. Morgan Kaufmann, 1997. ISBN: 978-1558603202.

[191] Müller, C. *Code Selection from Directed Acyclid Graphs in the Context of Domain Specific Digital Signal Processors*. Tech. rep. Berlin, Germany: Humboldt-Universität, 1994.

[192] Murray, A. and Franke, B. "Compiling for Automatically Generated Instruction Set Extensions". In: *Proceedings of the 10th International Symposium on Code Generation and Optimization*. CGO '12. San Jose, California, USA: ACM, 2012, pp. 13–22. ISBN: 978-1-4503-1206-6.

[193] Nakata, I. "On Compiling Algorithms for Arithmetic Expressions". In: *Communications of the ACM* 10.8 (1967), pp. 492–494. ISSN: 0001-0782.

[194] Newcomer, J. M. "Machine-Independent Generation of Optimal Local Code". Order number: AAI7521781. PhD thesis. Pittsburgh, Pennsylvania, USA: Carnegie Mellon University, 1975.

[195] Newell, A and Simon, H. A. "The Simulation of Human Thought". In: *Program on Current Trends in Psychology*. The University of Pittsburgh, Pennsylvania, USA, 1959.

[196] Nowak, L. and Marwedel, P. "Verification of Hardware Descriptions by Retargetable Code Generation". In: *Proceedings of the 26th ACM/IEEE Design Automation Conference*. DAC '89. Las Vegas, Nevada, USA: ACM, 1989, pp. 441–447. ISBN: 0-89791-310-8.

[197] Nymeyer, A. and Katoen, J.-P. "Code Generation Based on Formal Bottom-Up Rewrite Systems Theory and Heuristic Search". In: *Acta Informatica* 34.4 (1997), pp. 597–635.

[198] Nymeyer, A., Katoen, J.-P., Westra, Y., and Alblas, H. "Code Generation = A* + BURS". In: *Proceedings of the 6th International Conference on Compiler Construction (CC '06)*. Ed. by Gyimóthy, T. Vol. 1060. Lecture Notes in Computer Science. Springer Berlin Heidelberg, 1996, pp. 160–176. ISBN: 978-3-540-61053-3.

[199] O'Donnell, M. J. *Equational Logic as a Programming Language*. The MIT Press, 1985. ISBN: 978-0262150286.

[200] Orgass, R. J. and Waite, W. M. "A Base for a Mobile Programming System". In: *Communications of the ACM* 12.9 (1969), pp. 507–510. ISSN: 0001-0782.

[201] Paulin, P. G., Liem, C., May, T. C., and Sutarwala, S. "DSP Design Tool Requirements for Embedded Systems: A Telecommunications Industrial Perspective". In: *Journal of VLSI Signal Processing Systems for Signal, Image and Video Technology* 9.1–2 (1995), pp. 23–47. ISSN: 0922-5773.

[202] Pelegrí-Llopart, E. and Graham, S. L. "Optimal Code Generation for Expression Trees: An Application of BURS Theory". In: *Proceedings of the 15th ACM SIGPLAN-SIGACT symposium on Principles of Programming Languages*. POPL '88. San Diego, California, USA: ACM, 1988, pp. 294–308. ISBN: 0-89791-252-7.

[203] Pennello, T. J. "Very Fast LR Parsing". In: *Proceedings of the 1986 SIGPLAN symposium on Compiler Construction*. SIGPLAN '86. Palo Alto, California, USA: ACM, 1986, pp. 145–151. ISBN: 0-89791-197-0.

[204] Perkins, D. R. and Sites, R. L. "Machine-Independent PASCAL Code Optimization". In: *Proceedings of the 1979 SIGPLAN symposium on Compiler Construction*. SIGPLAN '79. Denver, Colorado, USA: ACM, 1979, pp. 201–207. ISBN: 0-89791-002-8.

[205] Pierre, P. P., Liem, C., May, T., and Sutarwala, S. "CODESYN: A Retargetable Code Synthesis System". In: *Proceedings of the 7th International Symposium on High Level Synthesis*. San Diego, California, USA, 1994, pp. 94–95.

[206] Proebsting, T. A. *Least-Cost Instruction Selection in DAGs is NP-Complete*. 1995. URL: `http://web.archive.org/web/20081012050644/http://research.microsoft.com/~toddpro/papers/proof.htm` (visited on 04/23/2013).

[207] Proebsting, T. A. "Simple and Efficient BURS Table Generation". In: *Proceedings of the ACM SIGPLAN 1992 Conference on Programming Language Design and Implementation*. PLDI '92. San Francisco, California, USA: ACM, 1992, pp. 331–340. ISBN: 0-89791-475-9.

[208] Proebsting, T. A. "BURS Automata Generation". In: *ACM Transactions on Programming Language Systems* 17.3 (1995), pp. 461–486. ISSN: 0164-0925.

[209] Proebsting, T. A. and Whaley, B. R. "One-Pass, Optimal Tree Parsing – With or Without Trees". In: *Proceedings of the 6th International Conference on Compiler Construction (CC '06)*. Ed. by Gyimóthy, T. Vol. 1060. Lecture Notes in Computer Science. Springer Berlin Heidelberg, 1996, pp. 294–306. ISBN: 978-3-540-61053-3.

[210] Purdom Jr., P. W. and Brown, C. A. "Fast Many-to-One Matching Algorithms". In: *Proceedings of the 1st International Conference on Rewriting Techniques and Applications*. Dijon, France: Springer, 1985, pp. 407–416. ISBN: 0-387-15976-2.

[211] Ramesh, R. and Ramakrishnan, I. V. "Nonlinear Pattern Matching in Trees". In: *Journal of the ACM* 39.2 (1992), pp. 295–316. ISSN: 0004-5411.

[212] Ramsey, N. and Davidson, J. W. "Machine Descriptions to Build Tools for Embedded Systems". In: *Proceedings of the ACM SIGPLAN Workshop on Languages, Compilers, and Tools for Embedded Systems*. LCTES '98. Springer-Verlag, 1998, pp. 176–192. ISBN: 3-540-65075-X.

[213] Ramsey, N. and Dias, J. "Resourceable, Retargetable, Modular Instruction Selection Using a Machine-Independent, Type-Based Tiling of Low-Level Intermediate Code". In: *Proceedings of the 38th annual ACM SIGPLAN-SIGACT symposium on Principles of Programming Languages*. POPL '11. Austin, Texas, USA: ACM, 2011, pp. 575–586. ISBN: 978-1-4503-0490-0.

[214] Redziejowski, R. R. "On Arithmetic Expressions and Trees". In: *Communications of the ACM* 12.2 (1969), pp. 81–84. ISSN: 0001-0782.

[215] Reiser, J. F. "Compiling Three-Address Code for C Programs". In: *The Bell System Technical Journal* 60.2 (1981), pp. 159–166.

[216] Ripken, K. *Formale Beschreibung von Maschinen, Implementierungen und Optimierender Maschinencodeerzeugung aus Attributierten Programmgraphen*. Tech. rep. TUM-INFO-7731. Munich, Germany: Institut für Informatik, Technical University of Munich, July 1977.

[217] Rossi, F., Van Beek, P., and Walsh, T., eds. *Handbook of Constraint Programming*. Elsevier, 2006. ISBN: 978-0-444-52726-4.

[218] Rudell, R. L. "Logic Synthesis for VLSI Design". AAI9006491. PhD thesis. 1989.

[219] Russell, S. J. and Norvig, P. *Artificial Intelligence: A Modern Approach*. 3rd ed. Pearson Education, 2010. ISBN: 0-13-604259-7.

[220] Sacerdoti, E. D. "Planning in a Hierarchy of Abstraction Spaces". In: *Proceedings of the 3rd International Joint Conference on Artificial Intelligence*. Ed. by Nilsson, N. J. IJCAI '73. Stanford, California, USA: William Kaufmann, 1973, pp. 412–422.

[221] Sakai, S., Togasaki, M., and Yamazaki, K. "A Note on Greedy Algorithms for the Maximum Weghted Independent Set Problem". In: *Discrete Applied Mathematics* 126.2-3 (2003), pp. 313–322. ISSN: 0166-218X.

[222] Sarkar, V., Serrano, M. J., and Simons, B. B. "Register-Sensitive Selection, Duplication, and Sequencing of Instructions". In: *Proceedings of the 15th International Conference on Supercomputing*. ICS '01. Sorrento, Italy: ACM, 2001, pp. 277–288. ISBN: 1-58113-410-X.

[223] Schäfer, S. and Scholz, B. "Optimal Chain Rule Placement for Instruction Selection Based on SSA Graphs". In: *Proceedings of the 10th International Workshop on Software and Compilers for Embedded Systems*. SCOPES '07. Nice, France: ACM, 2007, pp. 91–100.

[224] Scharwaechter, H., Youn, J. M., Leupers, R., Paek, Y., Ascheid, G., and Meyr, H. "A Code-Generator Generator for Multi-Output Instructions". In: *Proceedings of the 5th IEEE/ACM International Conference on Hardware/Software Codesign and System Synthesis*. CODES+ISSS '07. Salzburg, Austria: ACM, 2007, pp. 131–136. ISBN: 978-1-59593-824-4.

[225] Scholz, B. and Eckstein, E. "Register Allocation for Irregular Architectures". In: *Proceedings of the joint Conference on Languages, Compilers and Tools for Embedded Systems and Software and Compilers for Embedded Systems*. LCTES/SCOPES '02. Berlin, Germany: ACM, 2002, pp. 139–148. ISBN: 1-58113-527-0.

[226] Schrijver, A. *Theory of Linear and Integer Programming*. Wiley, 1998. ISBN: 978-04719823-26.

[227] Schulte, C. *Private correspondence*. Apr. 2013.

[228] Sethi, R. and Ullman, J. D. "The Generation of Optimal Code for Arithmetic Expressions". In: *Journal of the ACM* 17.4 (1970), pp. 715–28. ISSN: 0004-5411.

[229] Shamir, R. and Tsur, D. "Faster Subtree Isomorphism". In: *Journal of Algorithms* 33.2 (1999), pp. 267–280. ISSN: 0196-6774.

[230] Shankar, P., Gantait, A., Yuvaraj, A. R., and Madhavan, M. "A New Algorithm for Linear Regular Tree Pattern Matching". In: *Theoretical Computer Science* 242.1-2 (2000), pp. 125–142. ISSN: 0304-3975.

[231] Shu, J., Wilson, T. C., and Banerji, D. K. "Instruction-Set Matching and GA-based Selection for Embedded-Processor Code Generation". In: *Proceedings of the 9th International Conference on VLSI Design: VLSI in Mobile Communication*. VLSID '96. IEEE Press, 1996, pp. 73–76. ISBN: 0-8186-7228-5.

[232] Simoneaux, D. C. "High-Level Language Compiling for User-Defineable Architectures". PhD thesis. Monterey, California, USA: Naval Postgraduate School, 1975.

[233] Snyder, A. "A Portable Compiler for the Language C". MA thesis. Cambridge, Massachusetts, USA, 1975.

[234] Sorlin, S. and Solnon, C. "A Global Constraint for Graph Isomorphism Problems". In: *Proceedings of the 1st International Conference on Integration of AI and OR Techniques in Constraint Programming for Combinatorial Optimization Problems (CPAIOR '04)*. Ed. by Régin, J.-C. and Rueher, M. Vol. 3011. Lecture Notes in Computer Science. Springer Berlin Heidelberg, 2004, pp. 287–301. ISBN: 978-3-540-21836-4.

[235] Stallman, R. *Internals of GNU CC*. Version 1.21. Apr. 24, 1988. URL: `http://trinity.engr.uconn.edu/~vamsik/internals.pdf` (visited on 05/29/2013).

[236] Strong, J., Wegstein, J., Tritter, A., Olsztyn, J., Mock, O., and Steel, T. "The Problem of Programming Communication with Changing Machines: A Proposed Solution". In: *Communications of the ACM* 1.8 (1958), pp. 12–18. ISSN: 0001-0782.

[237] Sudarsanam, A., Malik, S., and Fujita, M. "A Retargetable Compilation Methodology for Embedded Digital Signal Processors Using a Machine-Dependent Code Optimization Library". In: *Design Automation for Embedded Systems* 4.2–3 (1999), pp. 187–206. ISSN: 0929-5585.

[238] Tanaka, H., Kobayashi, S., Takeuchi, Y., Sakanushi, K., and Imai, M. "A Code Selection Method for SIMD Processors with PACK Instructions". In: *Proceedings of the 7th International Workshop on Software and Compilers for Embedded Systems (SCOPES '03)*. Ed. by Krall, A. Vol. 2826. Lecture Notes in Computer Science. Springer Berlin Heidelberg, 2003, pp. 66–80. ISBN: 978-3-540-20145-8.

[239] Tanenbaum, A. S., Staveren, H. van, Keizer, E. G., and Stevenson, J. W. "A Practical Tool Kit for Making Portable Compilers". In: *Communications of the ACM* 26.9 (1983), pp. 654–660. ISSN: 0001-0782.

[240] *HBURG*. URL: http://www.bytelabs.org/hburg.html.

[241] *IBM CPLEX Optimizer*. URL: http://www-01.ibm.com/software/commerce/optimization/cplex-optimizer/.

[242] Tirrell, A. K. "A Study of the Application of Compiler Techniques to the Generation of Micro-Code". In: *Proceedings of the SIGPLAN/SIGMICRO interface meeting*. Harriman, New York, USA: ACM, 1973, pp. 67–85.

[243] *JBURG*. URL: http://jburg.sourceforge.net/.

[244] Tjiang, S. W. K. *An Olive Twig*. Tech. rep. Synopsys Inc., 1993.

[245] Tjiang, S. W. K. *Twig Reference Manual*. Tech. rep. Murray Hill, New Jersey, USA: AT&T Bell Laboratories, 1986.

[246] *OCAMLBURG*. URL: http://www.cminusminus.org/tools.html.

[247] *TMS320C55x DSP Mnemonic Instruction Set Reference Guide*. SPRU374G. Texas Instruments. Oct. 2002.

[248] Ullmann, J. R. "An Algorithm for Subgraph Isomorphism". In: *J. ACM* 23.1 (1976), pp. 31–42. ISSN: 0004-5411.

[249] Van Praet, J., Lanneer, D., Geurts, W., and Goossens, G. "Processor Modeling and Code Selection for Retargetable Compilation". In: *ACM Transactions on Design Automation of Electronic Systems* 6.3 (2001), pp. 277–307. ISSN: 1084-4309.

[250] Van Praet, J., Goossens, G., Lanneer, D., and De Man, H. "Instruction Set Definition and Instruction Selection for ASIPs". In: *Proceedings of the 7th International Symposium on Systems Synthesis*. ISSS '94. Niagra-on-the-Lake, Ontario, Canada: IEEE Press, 1994, pp. 11–16. ISBN: 0-8186-5785-5.

[251] Visser, B.-S. "A Framework for Retargetable Code Generation Using Simulated Annealing". In: *Proceedings of the 25th EUROMICRO '99 Conference on Informatics: Theory and Practice for the New Millenium*. 1999, pp. 1458–1462. ISBN: 0-7695-0321-7.

[252] Visser, E. "A Survey of Strategies in Rule-Based Program Transformation Systems". In: *Journal of Symbolic Computation* 40.1 (2005), pp. 831–873.

[253] Visser, E. "Stratego: A Language for Program Transformation Based on Rewriting Strategies - System Description of Stratego 0.5". In: *Rewriting Techniques and Applications*. Vol. 2051. RTA '01. Springer, 2001, pp. 357–361.

[254] Živojnović, V., Martínez Velarde, J., Schläger, C., and Meyr, H. "DSPstone: A DSP-Oriented Benchmarking Methodology". In: *Proceedings of the International Conference on Signal Processing Applications and Technology*. ICSPAT '94. Dallas, Texas, USA, Oct. 1994.

[255] Wasilew, S. G. "A Compiler Writing System with Optimization Capabilities for Complex Object Order Structures". AAI7232604. PhD thesis. Evanston, Illinois, USA: Northwestern University, 1972.

[256] Weingart, S. W. "An Efficient and Systematic Method of Compiler Code Generation". AAI7329501. PhD thesis. New Haven, Connecticut, USA: Yale University, 1973.

[257] Weisgerber, B. and Wilhelm, R. "Two Tree Pattern Matchers for Code Selection". In: *Proceedings of the 2nd CCHSC Workshop on Compiler Compilers and High Speed Compilation*. Springer, 1989, pp. 215–229. ISBN: 3-540-51364-7.

[258] Wendt, A. L. "Fast Code Generation Using Automatically-Generated Decision Trees". In: *Proceedings of the ACM SIGPLAN 1990 Conference on Programming Language Design and Implementation*. PLDI '90. White Plains, New York, USA: ACM, 1990, pp. 9–15. ISBN: 0-89791-364-7.

[259] Wess, B. "Automatic Instruction Code Generation Based on Trellis Diagrams". In: *Proceedings of the IEEE International symposium on Circuits and Systems*. Vol. 2. ISCAS '92. IEEE Press, 1992, pp. 645–648.

[260] Wess, B. "Code Generation Based on Trellis Diagrams". In: *Code Generation for Embedded Processors*. Ed. by Marwedel, P. and Goossens, G. 1995. Chap. 11, pp. 188–202.

[261] Wilcox, T. R. "Generating Machine Code for High-Level Programming Languages". AAI7209959. PhD thesis. Ithaca, New York, USA: Cornell University, 1971.

[262] Wilhelm, R. and Maurer, D. *Compiler Design*. Addison-Wesley, 1995. ISBN: 978-0201422-900.

[263] Wilson, T., Grewal, G., Halley, B., and Banerji, D. "An Integrated Approach to Retargetable Code Generation". In: *Proceedings of the 7th International Symposium on High-Level Synthesis*. ISSS '94. Niagra-on-the-Lake, Ontario, Canada: IEEE Press, 1994, pp. 70–75. ISBN: 0-8186-5785-5.

[264] Wulf, W. A., Johnsson, R. K., Weinstock, C. B., Hobbs, S. O., and Geschke, C. M. *The Design of an Optimizing Compiler*. Elsevier Science Inc., 1975. ISBN: 0444001581.

[265] Wuu, H.-T. L., and Yang, W. "A Simple Tree Pattern-Matching Algorithm". In: *In Proceedings of the Workshop on Algorithms and Theory of Computation*. 2000.

[266] Yang, W. “A Fast General Parser for Automatic Code Generation”. In: *Proceedings of the 2nd Russia-Taiwan Conference on Methods and Tools of Parallel Programming multicomputers*. MTPP ’10. Vladivostok, Russia: Springer, 2010, pp. 30–39. ISBN: 978-3-642-14821-7.

[267] Yates, J. S. and Schwartz, R. A. “Dynamic Programming and Industrial-Strength Instruction Selection: Code Generation by Tiring, but not Exhaustive, Search”. In: *SIGPLAN Notices* 23.10 (1988), pp. 131–140. ISSN: 0362-1340.

[268] Youn, J. M., Lee, J., Paek, Y., Lee, J., Scharwaechter, H., and Leupers, R. “Fast Graph-Based Instruction Selection for Multi-Output Instructions”. In: *Software—Practice & Experience* 41.6 (2011), pp. 717–736. ISSN: 0038-0644.

[269] Young, R. “The Coder: A Program Module for Code Generation in High Level Language Compilers”. MA thesis. Urbana-Champaign, Illinois, USA: Computer Science Department, University of Illinois, 1974.

[270] Yu, K. H. and Hu, Y. H. “Artificial Intelligence in Scheduling and Instruction Selection for Digital Signal Processors”. In: *Applied Artificial Intelligence* 8.3 (1994), pp. 377–392.

[271] Yu, K. H. and Hu, Y. H. “Efficient Scheduling and Instruction Selection for Programmable Digital Signal Processors”. In: *Transactions on Signal Processing* 42.12 (1994), pp. 3549–3552. ISSN: 1053-587X.

[272] Zimmermann, G. “The MIMOLA Design System: A Computer Aided Digital Processor Design Method”. In: *Proceedings of the 16th Design Automation Conference*. DAC ’79. San Diego, California, USA: IEEE Press, 1979, pp. 53–58.

[273] Zimmermann, W. and Gaul, T. “On the Construction of Correct Compiler Back-Ends: An ASM-Approach”. In: *Journal of Universal Computer Science* 3.5 (May 28, 1997), pp. 504–567.

[274] Ziv, J and A., L. “A Universal Algorithm for Sequential Data Compression”. In: *Transactions on Information Theory* 23.3 (1977), pp. 337–343.

# A

# List of Approaches

| Approach | Pub. | Prin. | Sc. | Opt. | SO | MO | DO | IL | In. | Implementations |
|---|---|---|---|---|---|---|---|---|---|---|
| Lowry and Medlock [180] | 1969 | ME | L | | ✓ | | | | | |
| Orgass and Waite [200] | 1969 | ME | L | | ✓ | | | | | Simcmp |
| Elson and Rake [78] | 1970 | ME | L | | ✓ | | | | | |
| Miller [189] | 1971 | ME | L | | ✓ | | | | | Dmacs |
| Wilcox [261] | 1971 | ME | L | | ✓ | | | | | |
| Wasilew [255] | 1972 | TC | L | | ✓ | | | | | |
| Donegan [72] | 1973 | ME | L | | ✓ | | | | | |
| Tirrell [242] | 1973 | ME | L | | ✓ | | | | | |
| Weingart [256] | 1973 | TC | L | | ✓ | | | | | |
| Young [269] | 1974 | ME | L | | ✓ | | | | | |
| Newcomer [194] | 1975 | TC | L | | ✓ | | | | | |
| Simoneaux [232] | 1975 | ME | L | | ✓ | | | | | |
| Snyder [233] | 1975 | ME | L | | ✓ | | | | | |
| Fraser [106, 107] | 1977 | ME | L | | ✓ | | | | | |
| Ripken [216] | 1977 | TC | L | ✓ | ✓ | | | | | |
| Glanville and Graham [121] | 1978 | TC | L | | ✓ | | | | | |
| Johnson [143] | 1978 | TC | L | | ✓ | | | | | PCC |
| Cattell et al. [44, 47, 176] | 1980 | TC | L | | ✓ | | | | | PQCC |
| Ganapathi and Fischer [112, 113, 114, 115] | 1982 | TC | L | | ✓ | | | | | |
| Krumme and Ackley [164] | 1982 | ME | L | | ✓ | | | | | |
| Christopher et al. [52] | 1984 | TC | L | ✓ | ✓ | | | | | |
| Davidson and Fraser [63] | 1984 | $ME^+$ | L | | ✓ | ✓ | ✓ | | | GCC, ACK, Zephyr/VPO |
| Henry [138] | 1984 | TC | L | ✓ | ✓ | | | | | |
| Aho et al. [6, 7, 245] | 1985 | TC | L | ✓ | ✓ | | | | | Twig |
| Hatcher and Christopher [134] | 1986 | TC | L | ✓ | ✓ | | | | | |
| Fraser and Wendt [101] | 1988 | $ME^+$ | L | | ✓ | | | | | |
| Giegerich and Schmal [120] | 1988 | TC | L | | ✓ | | | | | |
| Hatcher and Tuller [136] | 1988 | TC | L | ✓ | ✓ | | | | | UNH-Codegen |
| Pelegrí-Llopart and Graham [202] | 1988 | TC | L | ✓ | ✓ | | | | | |
| Yates and Schwartz [267] | 1988 | TC | L | ✓ | ✓ | | | | | |
| Emmelmann et al. [79] | 1989 | ME | L | ✓ | ✓ | | | | | BEG, CoSy |

Pub.: date of publication. Prin.: fundamental principle on which the approach is based [macro expansion (ME); macro expansion with peephole optimization ($ME^+$); tree covering (TC); trellis diagrams (TD) – is actually sorted under TC in the report; DAG covering (DC); or graph covering (GC)]. Sc.: scope of instruction selection [local (L) – isolated to a single basic block; or global (G) – considers an entire function]. Opt.: whether the approach claims to be optimal. Supported machine instruction characteristics: single-output (SO), multi-output (MO), disjoint-output (DO), internal-loop (IL), and interdependent instructions (In.).

| Approach | Pub. | Prin. | Sc. | Opt. | SO | MO | DO | IL | In. | Implementations |
|---|---|---|---|---|---|---|---|---|---|---|
| Ganapathi [111] | 1989 | TC | L | | ✓ | ✓ | | | | |
| Genin et al. [118] | 1989 | $ME^+$ | L | | ✓ | ✓ | | | | |
| Nowak and Marwedel [196] | 1989 | DC | L | | ✓ | ✓ | ✓ | | | MSSC |
| Balachandran et al. [26] | 1990 | TC | L | ✓ | ✓ | | | | | |
| Despland et al. [41, 66, 67] | 1990 | TC | L | | ✓ | | | | | Pagode |
| Wendt [258] | 1990 | $ME^+$ | L | | ✓ | ✓ | | | | |
| Hatcher [135] | 1991 | TC | L | ✓ | ✓ | | | | | UCG |
| Fraser et al. [104] | 1992 | TC | L | ✓ | ✓ | | | | | Iburg, Record, Redaco |
| Proebsting [105, 207, 208, 209] | 1992 | TC | L | ✓ | ✓ | | | | | Burg, Hburg, Wburg |
| Wess [259, 260] | 1992 | TD | L | ✓ | ✓ | | | | | |
| Marwedel [186] | 1993 | DC | L | | ✓ | ✓ | ✓ | | | MSSV |
| Tjiang [244] | 1993 | TC | L | ✓ | ✓ | | | | | Olive |
| Engler and Proebsting [83] | 1994 | TC | L | ✓ | ✓ | | | | | DCG |
| Fauth et al. [92, 191] | 1994 | DC | L | | ✓ | ✓ | | | | CBC |
| Ferdinand et al. [94] | 1994 | TC | L | ✓ | ✓ | | | | | |
| Liem et al. [179, 201, 205] | 1994 | DC | L | | ✓ | | | | | CodeSyn |
| Van Praet et al. [166, 249, 250] | 1994 | GC | G | ✓ | ✓ | ✓ | ✓ | | | Chess |
| Wilson et al. [263] | 1994 | DC | L | ✓ | ✓ | ✓ | | | ✓ | |
| Yu and Hu [270, 271] | 1994 | DC | L | ✓ | ✓ | ✓ | | | | |
| Hanson and Fraser [132] | 1995 | TC | L | ✓ | ✓ | | | | | Lburg, LCC |
| Liao et al. [177, 178] | 1995 | GC | L | ✓ | ✓ | ✓ | | | | |
| Hoover and Zadeck [141] | 1996 | DC | L | | ✓ | ✓ | ✓ | | | |
| Leupers and Marwedel [168, 175] | 1996 | DC | L | | ✓ | ✓ | ✓ | | | |
| Nymeyer et al. [197, 198] | 1996 | TC | L | | ✓ | | | | | |
| Shu et al. [231] | 1996 | TC | L | | ✓ | | | | | |
| Gough and Ledermann [124, 125] | 1997 | TC | L | ✓ | ✓ | | | | | Mburg |
| Hanono and Devadas [130, 131] | 1998 | TD | L | | ✓ | | | | | Aviv |
| Leupers and Marwedel [172] | 1998 | DC | L | | ✓ | ✓ | ✓ | | | MSSQ |
| Bashford and Leupers [28] | 1999 | DC | L | | ✓ | ✓ | ✓ | | | |
| Ertl [85] | 1999 | DC | L | ✓ | ✓ | ✓ | | | | Dburg |
| Fraser and Proebsting [103] | 1999 | ME | L | | ✓ | | | | | Gburg |
| Fröhlich et al. [108] | 1999 | TD | L | ✓ | ✓ | | | | | |
| Visser [251] | 1999 | GC | G | | ✓ | ✓ | ✓ | | | |
| Leupers [170] | 2000 | DC | L | | ✓ | ✓ | ✓ | | ✓ | |
| Madhavan et al. [181] | 2000 | TC | L | ✓ | ✓ | | | | | |
| Arnold and Corporaal [18, 19, 20] | 2001 | DC | L | | ✓ | ✓ | | | | |
| Kastner et al. [151] | 2001 | DC | L | | ✓ | ✓ | | | | |
| Sarkar et al. [222] | 2001 | DC | L | | ✓ | | | | | |
| Bravenboer and Visser [36] | 2002 | TC | L | ✓ | ✓ | | | | | |
| Eckstein et al. [77] | 2003 | GC | G | | ✓ | | | | | |
| Tanaka et al. [238] | 2003 | DC | L | | ✓ | ✓ | ✓ | | | |
| Borchardt [31] | 2004 | TC | L | ✓ | ✓ | | | | | |
| Brisk et al. [37] | 2004 | DC | L | | ✓ | ✓ | | | | |
| Cong et al. [56] | 2004 | GC | L | | ✓ | ✓ | ✓ | | | |
| Lattner and Adve [167] | 2004 | DC | L | | ✓ | | | | | LLVM |
| Bednarski and Kessler [29] | 2006 | DC | L | ✓ | ✓ | ✓ | | | ✓ | Optimist |

Pub.: date of publication. Prin.: fundamental principle on which the approach is based [macro expansion (ME); macro expansion with peephole optimization ($ME^+$); tree covering (TC); trellis diagrams (TD) – is actually sorted under TC in the report; DAG covering (DC); or graph covering (GC)]. Sc.: scope of instruction selection [local (L) – isolated to a single basic block; or global (G) – considers an entire function]. Opt.: whether the approach claims to be optimal. Supported machine instruction characteristics: single-output (SO), multi-output (MO), disjoint-output (DO), internal-loop (IL), and interdependent instructions (In.).

| Approach | Pub. | Prin. | Sc. | Opt. | SO | MO | DO | IL | In. | Implementations |
|---|---|---|---|---|---|---|---|---|---|---|
| Clark et al. [54] | 2006 | GC | L | | ✓ | ✓ | ✓ | | | |
| Dias and Ramsey [70] | 2006 | $ME^+$ | L | | ✓ | ✓ | | | | |
| Ertl et al. [86] | 2006 | TC | L | ✓ | ✓ | | | | | |
| Farfeleder et al. [89] | 2006 | DC | L | | ✓ | ✓ | | | | |
| Hormati et al. [142] | 2007 | GC | L | | ✓ | ✓ | ✓ | | | |
| Scharwaechter et al. [224] | 2007 | DC | L | | ✓ | ✓ | ✓ | | | Cburg |
| Ebner et al. [76] | 2008 | GC | G | | ✓ | ✓ | ✓ | | | |
| Koes and Goldstein [161] | 2008 | DC | L | | ✓ | ✓ | | | | Noltis |
| Ahn et al. [2] | 2009 | DC | L | | ✓ | ✓ | ✓ | | | |
| Martin et al. [183] | 2009 | GC | G | ✓ | ✓ | ✓ | ✓ | | ✓ | |
| Buchwald and Zwinkau [39] | 2010 | GC | G | | ✓ | | | | | |
| Dias and Ramsey [69, 213] | 2010 | $ME^+$ | L | | ✓ | ✓ | | | | |
| Floch et al. [97] | 2010 | GC | G | ✓ | ✓ | ✓ | ✓ | | | |
| Yang [266] | 2010 | TC | L | ✓ | ✓ | | | | | |
| Arslan and Kuchcinski [22] | 2013 | GC | G | ✓ | ✓ | ✓ | ✓ | | ✓ | |

Pub.: date of publication. Prin.: fundamental principle on which the approach is based [macro expansion (ME); macro expansion with peephole optimization ($ME^+$); tree covering (TC); trellis diagrams (TD) – is actually sorted under TC in the report; DAG covering (DC); or graph covering (GC)]. Sc.: scope of instruction selection [local (L) – isolated to a single basic block; or global (G) – considers an entire function]. Opt.: whether the approach claims to be optimal. Supported machine instruction characteristics: single-output (SO), multi-output (MO), disjoint-output (DO), internal-loop (IL), and interdependent instructions (In.).

# B

# Publication Diagram

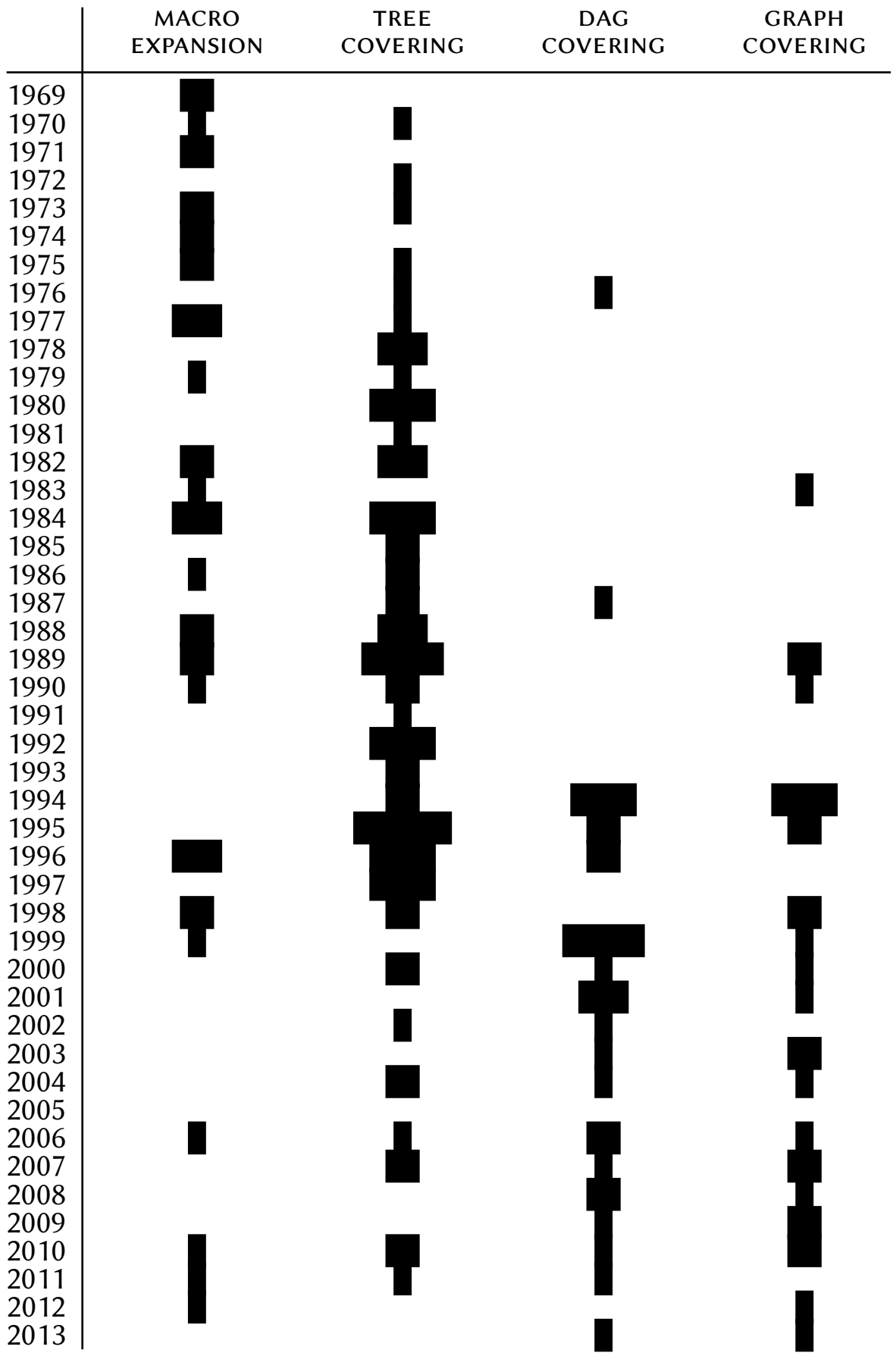

Figure B.1: Diagram that illustrates a publication timeline, categorised after the respective, fundamental principles of instruction selection. The width of the bars indicate the relative number of publications at a given year.

# C

# Graph Definitions

A *graph* is defined as a tuple $G = \langle N, E \rangle$ where $N$ is a set of *nodes* (also known as *vertices*) and $E$ is a set of *edges*, each consisting of a pair of nodes $n, m \in N$. A graph is *undirected* if its edges have no direction, and *directed* if they do. We write a directed edge from a node $n$ to another node $m$ as $n \to m$, and say that such an edge is *outgoing* with respect to node $n$, and *ingoing* with respect to node $m$. We also introduce the following functions:

$$\begin{aligned} \mathit{src} &: E \to N & \mathit{dst} &: E \to N \\ \mathit{src}(n \to m) &= n & \mathit{dst}(n \to m) &= m \end{aligned}$$

Edges for which $\mathit{src}(e) = \mathit{dst}(e)$ are known as a *loop edges* (or simply *loops*). If there exists more than one edge between a pair of nodes then the graph is a *multigraph*; otherwise it is a *simple* graph.

A list of edges that describe how to get from one node to another is known as a *path*. More formally we define a path between two nodes $n$ and $m$ is an ordered list of edges $p = (e_1, \ldots, e_n)$ such that, for a directed graph:

$$e_1, \ldots, e_l \in E, n \in e_1, m \in e_l, \forall 1 \leq i < l : \mathit{dst}(e_i) = \mathit{src}(e_{i+1})$$

Paths for undirected graphs are similarly defined and will thus be skipped. A path for which $\mathit{src}(e_1) = \mathit{dst}(e_l)$ is known as a *cycle*. Two nodes $n$ and $m$, $n \neq m$, are said to be *connected* if there exists a path from $n$ to $m$, and if the path is of length 1 then $n$ and $m$ are also *adjacent*. An undirected graph is *connected* if there exists a path for every distinct pair of nodes. For completeness, a directed graph is *strongly connected* if there exists a path from $n$ to $m$ and a path from $m$ to $n$ for every pair of distinct nodes $n$ and $m$. Likewise, a directed graph is only *weakly connected* if replacing all edges with undirected edges yields a connected undirected graph (see Figure C.1 for example).

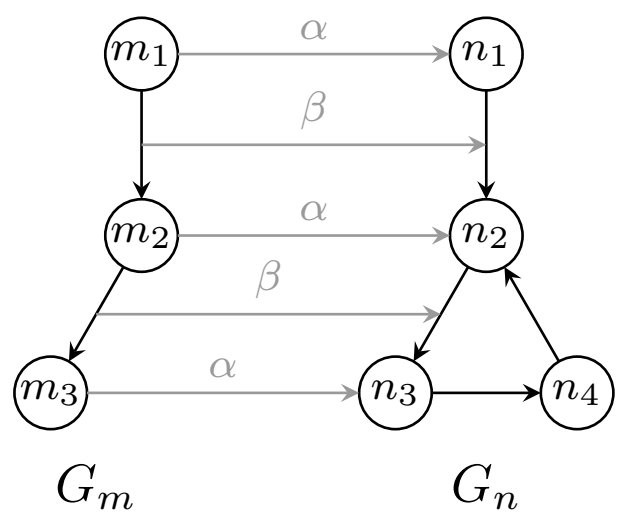

Figure C.1: Example of two simple, directed graphs $G_m$ and $G_n$. Through the mapping functions $\alpha$ and $\beta$ we see that $G_m$ is an isomorphic subgraph of $G_n$. Moreover, $G_n$ has a strongly connected component – consisting of $n_2$, $n_3$, and $n_4$, which form a cycle – while the entire $G_m$ is only weakly connected.

A simple graph that is connected, contains no cycles, and has exactly one path between any two nodes is called a *tree*, and a set of trees constitutes a *forest*. Nodes in a tree who are adjacent with exactly one other node are known as *leaves*, and depending on the direction of the edges wherein the leaves appear, there exists a node with no outgoing (or ingoing) edges which is called the *root* of the tree. In this report we will always draw trees with the root appearing at the top. A directed graph which contains no cycles is known as a *directed acyclic graph* or *DAG* for short.

A graph $G = \langle N, E \rangle$ is a *subgraph* of another graph $G' = \langle N', E' \rangle$ if $N \subseteq N'$ and $E \subseteq E'$, which is denoted as $G \subseteq G'$. Moreover, $G$ is an *isomorphic subgraph* of $G'$ if there exist two mapping functions $\alpha$ and $\beta$ such that $\{\alpha(n) \mid n \in N\} \subseteq N'$ and $\{\beta(e) \mid e \in E\} \subseteq E'$. An example of this is given in Figure C.1. Likewise a tree $T$ is a *subtree* of another tree $T'$ if $T \subseteq T'$.

Lastly we introduce the notion of *topological sort*, where the nodes of a graph $G = \langle N, E \rangle$ are arranged in an ordered list $s = (n_1, \dots, n_n)$, $\forall 1 \leq i \leq |N| : n_i \in N$, such that for no pair of nodes $n_i$ and $n_j$ where $i < j$ does there exist an edge $e = n_j \to n_i$. In other words, if the edges are added to the list then all edges will go point forward from left to right.

# D

# Taxonomy

The terms and exact meanings – in particular that of optimal instruction selection – often differ from one paper to another, thereby making them difficult to discuss and compare without a common foundation. A taxonomy with a well-defined vocabulary has therefore been established and is used consistently throughout the report. Although this may seem a bit daunting and superfluous at first, most items in the taxonomy are easy to understand and having explicitly defined these terms hopefully mitigates confusions that may otherwise occur.

## D.1 Common terms

Several terms are continuously used to when discussing instruction selection. Most of these are obvious and appear in other, related literature, while others are may require a little explanation.

- *input program* – the program under compilation, and therefore the input to the compiler and, subsequently, to the instruction selector. In the former this refers to the source code, while in the latter it usually refers to the IR code, either the entire code or parts of it (e.g. a function or single basic block) depending on the scope of the instruction selector.
- *target machine* – the hardware platform for which an input program is compiled to run on. Most often this refers to the instruction set architecture (ISA) implemented by its processing unit.
- *instruction selector* – a component or program responsible of implementing and executing the task of instruction selection. If this program is automatically generated from a specification, the term refers to the *generated product* – not the *generator*.
- *frontend* – a component or program responsible of parsing, validating, and translating an input program into equivalent IR code, a representation used internally by the compiler. This may also be referred to as the *compiler frontend*.
- *code generation* – the task of generating assembly code for a given program input by performing instruction selection, instruction scheduling, and register allocation.
- *backend* – a component or program responsible of implementing and executing the task of code generation. This may also be referred to as the *compiler backend* or *code generator backend*.

- *compilation time* – the time required to compile a given input program.
- *pattern matching* – the problem of detecting when and where it is possible to use a certain machine instruction for a given input program.
- *pattern selection* – the problem of deciding which instructions to select from the candidate set found when solving the pattern matching problem.
- *offline cost analysis* – the task of shifting the computation of optimization decisions from the input program compilation phase to the compiler building phase, thereby reducing the time it takes to compile an input program at the cost of increasing the time it takes to generating the compiler.

## D.2 Machine instruction characteristics

A machine instruction exhibits one or more *machine instruction characteristics*. For this study, the following characteristics were identified:

- *single-output* – the instruction only produces a single observable output. In this sense, observable means a value that can be accessed by the program. This includes instructions that perform typical arithmetic operations such as addition and multiplication as well as bit operations such as `OR` and `NOT`. Note that the instruction must produce *only* this value and nothing else in order to be considered a single-output instruction (compare this with the next characteristic).
- *multiple-output* (or *side-effect*) – the instruction produces more than one observable output from the same input. Examples include `divrem` instructions that produces both the quotient as well as the remainder of two terms, but it also arithmetic operation instructions that set *status flags* (also known as *condition* flags or *condition codes*) in addition to computing the arithmetic result. A status flag is a bit that signify additional information about the result (e.g. if there was a carry overflow or the result was `0`) and are therefore often called side effects of the instruction. In reality, however, these bits are no more than additional outputs produced by the instruction and are thus referred to as multiple-output instructions.
- *disjoint-output* – the instruction produces observable output values from disjoint input value sets. This means that if the expression for computing each observable output value formed a directed graph (see Appendix C for definitions), then these graphs would be disjoint. In comparison, multiple-output instructions all form a single graph. Disjoint-output instructions typically include SIMD (single-instruction-multiple-data) and vector instructions which execute the same operations simultaneously on many input values.
- *internal-loop* – the internal behavior of the instruction cannot be described without resorting to the use of loops. For example, the TI TMS320C55x [247] processor provides an `RPT k` instruction that repeats the immediately following instruction $k$ times, where $k$ is an immediate value given as part of the instruction.
- *interdependent* – the instruction enforces additional constraints when appearing in combination with other machine instructions. An example includes an `ADD` instruction from the TMS320C55x instruction set which cannot be combined with an `RPT k` instruction if the addressing mode is set to a specific mode. As seen in this report, this kind of instructions is very difficult to handle by most approaches.

## D.3 Scope

Most IR code is typically arranged as a sequence of statements, executed one after another, with special operations that allow the execution to jump to another position in the code. By grouping together statements such that all jumps only occur to the beginning of a group or at the end of a group, the code can be divided into a set of *basic blocks*. Let us look at an example.

Assume we have the example in Figure D.1 of a C code fragment and its corresponding IR code, where `jmp x` means "jump to label `x`" and `condjmp v x` means "jump to label `x` if `v` is true, otherwise jump the next statement". We see then that:

```c
int b = 0;
for (int i = 0; i < 10; i++) {
  b = b + i;
  if (b == 25) break;
}
int c = b;
```

(a) C code.

```
1  init:
2    b = 0
3    i = 0
4  loop:
5    t1 = b < 10
6    condjmp t end
7    b = b + i
8    t2 = b == 25
9    condjmp t end
10   i = i + 1
11   jmp loop
12 end:
13   c = b
```

(b) Corresponding IR code.

Figure D.1: Basic block example.

- lines 1 through 3 form a basic block, since there there are jumps to line 4;
- lines 4 through 6 form a basic block, since there is a potential jump at line 6;
- lines 7 through 9 form a basic block, for the same reason;
- as does lines 10 through 11; and
- the last basic block is formed by lines 12 through 13.

With this in mind, we can divide instruction selectors into two categories depending on their scope:

- *local* – the instruction selector only considers one basic block at a time.
- *global* – the instruction selector considers more than one block at a time, typically all blocks within an entire function.

The main difference between local and global instruction selection is that the latter is capable of combining statements across blocks in order implement then using a single machine instruction, thus effectively improving code quality.

## D.4 Principles

All techniques reviewed in this report have been fitted into one of four principles. These are only listed and briefly explained here as an entire chapter is dedicated for each.

- *macro expansion* – each IR construct is expanded into one or more machine instructions using macros. This is a simple strategy but generally produces very inefficient code as a machine instruction often can implement more than one IR construct. Consequently, modern instruction selectors that apply this approach also incorporate peephole optimization that combines many machine instructions into single equivalents.
- *tree covering* – the IR code and machine instructions are represented as graphical trees. Each machine instruction gives rise to a pattern which is then matched over the IR tree (this is the pattern matching problem). From the matching set of patterns, a subset is selected such that the entire IR tree is covered at the lowest cost.
- *DAG covering* – the same idea as tree covering but uses directed acyclic graphs (DAGs) instead of trees. Since DAGs are a more general form of trees, DAG covering supersedes tree covering.
- *graph covering* – the same idea as DAG covering but uses generic graphs instead of DAGs. Again, since graphs are a more general form av DAGs, graph covering supersedes DAG covering.

# E

# Document Revisions

2013-10-04

- Merged simulation chapter with macro expansion chapter
- Addressed misunderstandings of several approaches.
- Completely rewrote many parts of the chapters to make the material more accessible and to strengthened the discussion of many approaches
- Revised the drawing of all trees and graphs to put the root at the top instead of at the bottom
- Added additional references
- Added appendix for listing the approaches in a table
- Added appendix for document revisions
- Removed compiler appendix as it became obsolete
- Added KTH-TRITA, ISBN, ISSN, and ISRN number
- Added acknowledgements
- Fixed typographical and grammatical errors
- Added middle-name initial to clearly indicate the surname
- Second public release

2013-06-19

- Rewrote and rearranged many passages
- Added a few more references
- Fixed typographical and grammatical errors
- Changed title slightly
- Added placeholder for KTH-TRITA number
- Added copyright page
- Added list of abbreviations
- First public release

2013-05-30

- First draft

# Index